\documentclass{aa}
\usepackage[T1]{fontenc}
\usepackage[utf8]{inputenc}

\usepackage{graphicx}
\usepackage{natbib}
\usepackage{ulem}
\usepackage{textcomp}
\usepackage{gensymb}
\usepackage{longtable}
\usepackage{threeparttable}
\usepackage{multicol}
\usepackage{multirow}
\usepackage{float}
\usepackage{todonotes}
\usepackage[Symbol]{upgreek}
\usepackage{amsmath}
\usepackage{etoolbox}
\usepackage{txfonts}
\usepackage{url}
\usepackage{xcolor}

\usepackage[most]{tcolorbox}
\usepackage{hyperref}
\usepackage{xcolor}
\newcommand{\enzo}{{\it {\small ENZO}}}

\begin{document}

\title{The evolution of cosmic ray electrons in the cosmic web: seeding by AGN, star formation and shocks}

\author{F. Vazza\inst{1,2,3},  C. Gheller\inst{2}, F Zanetti\inst{1}, M. Tsizh\inst{1,4}, E. Carretti\inst{2},  S. Mtchedlidze\inst{1,5}, M.  Br\"{u}ggen\inst{3}}

\offprints{%
 E-mail: franco.vazza2@unibo.it}

\institute{Dipartimento di Fisica e Astronomia, Universit\'{a} di Bologna, Via Gobetti 93/2, 40122, Bologna, Italy
\and INAF-Istituto di Radioastronomia, Via Gobetti 101, 40129, Bologna, Italy
\and  Hamburger Sternwarte, University of Hamburg, Gojenbergsweg 112, 21029 Hamburg, Germany
\and Astronomical Observatory of Ivan Franko National University of Lviv, Kyryla i Methodia str. 8, Lviv, 79005, Ukraine 
\and
School of Natural Sciences and Medicine, Ilia State University, 3-5 Cholokashvili St., 0194 Tbilisi, Georgia
}
\authorrunning{F. Vazza et al.}
\titlerunning{Relativistic electrons in the cosmic web}

\date{Accepted ???. Received ???; in original form ???}

%\begin{abstract}
\abstract{Several processes in the Universe convert a fraction of gas kinetic energy into the acceleration of relativistic electrons, making them observable at radio wavelengths, or contributing to a dormant reservoir of low-energy cosmic rays in cosmic structures.
We present a new suite of cosmological simulations, {with simple galaxy formation models calibrated to work at a specific spatial resolution},  tailored to study all most important processes of injection of 
relativistic electrons in evolving large-sale structures: accretion and merger shocks, feedback from active galactic nuclei and winds from star forming regions.  We also follow the injection of magnetic fields by active galactic nuclei and star formation, and compute the observational signatures of these mechanisms. {We find that} the injection of cosmic ray electrons by shocks is the most volume filling process, { and that it also} dominates the energy density of fossil relativistic electrons in halos. The combination of the seeding mechanisms studied in this work, regardless of the uncertainties related to physical or numerical uncertainties, is more than enough to fuel large-scale radio emissions with a { large} amount of seed fossil electrons. We derive an approximated formula to predict the number of fossil cosmic ray electrons injected by $z=0$ by the total activity of shocks, AGN and star formation in the volume of halos. { By looking at the maximum possible contribution to the magnetisation of the cosmic web by all our simulated sources, we conclude that galaxy formation-related processes, alone, cannot explain the values of Faraday Rotation of background polarised sources recently detected using LOFAR.}}

\maketitle
\label{firstpage}

 \section{Introduction}\label{sec::intro}

Modern and very sensitive radio observations have detected extended and diffuse emission from the extreme periphery of clusters of galaxies 
\citep{2022Natur.609..911C,2022SciA....8.7623B}, as well as from filaments
\citep{2019Sci...364..981G,2020MNRAS.499L..11B,vern21,2022A&A...660A..81V,vern23}. These observations provide evidence for substantial magnetic fields ($B \sim 10-100$~nG) as well as for relativistic electrons ($E \geq 0.1-1$~GeV) existing on these scales. By an large, such radio-emitting electrons cannot be accelerated in-situ and directly from the thermal pool, but should rather be re-accelerated by large-scale plasma perturbations, either in shocks \citep[e.g.][]{ka12,2013MNRAS.435.1061P,2017NatAs...1E...5V}, or turbulence \citep[e.g.][]{1977ApJ...212....1J,bl07,bj14,2015ApJ...800...60M}.  The existence of pockets of fossil radio plasma, whose synchrotron emissivity gets boosted by compression following accretion phenomena in clusters, is also the leading model to explain the formation of more classical "radio phoenices" \citep[][]{2002Natur.418..301B,2015MNRAS.448.2197D}, and is suggested by other recent observations of faint and diffuse remnant radio emission from the repeated activity of central radio galaxies \citep[e.g.][]{brienza21,2021A&A...650A.170B}.

Several mechanisms can naturally lead to the injection of large amounts of cosmic ray electrons (CRe in the rest of the paper) in large-scale structures, and in this work we focus on the three which are believed to be the most significant. 
First, { outflows and jets from active galactic nuclei} can store a large fraction of their internal energy in the form of non-thermal components (Cosmic Rays
and magnetic fields), although the balance between thermal and non-thermal components may depend on the radio galaxy type, as well as on the surrounding environment \cite[e.g.][]{Volk&Atoyan..ApJ.2000, 2008MNRAS.386.1709C,2018MNRAS.476.1614C,vb24}. 

Second, the ubiquitous process of star formation can also enrich galactic halos with cosmic rays, through the collective inflation of expanding { shocked shells} produced by main sequence stars, red supergiant stars, Wolf-Rayet stars and supernova remnants \citep[][]{2004Ap&SS.289..337D,2017ApJ...847L..13P,2018JKAS...51...37S,2023MNRAS.524.6374M}. 

Third, the growth of cosmic structures is accompanied by the formation of shock waves, which dissipate a large fraction of infall kinetic energy into gas heating  \citep[e.g.][]{1972A&A....20..189S,by08}. 
Strong  ($\mathcal{M} \geq 10-10^2$ where $\mathcal{M}$ is the sonic Mach number) quasi-stationary accretion shocks are expected beyond the virial radius of clusters of galaxies \citep[e.g.][]{2009ApJ...696.1640M,2024NatAs...8.1195V}, as well as shocks connected with the interface between filamentary accretions and the intra-cluster or intra-group medium \citep[e.g.][]{2011JApA...32..577B}, or internal merger shocks violently crossing the innermost regions of clusters \citep[e.g.][]{mv07}.  Basically all modern cosmological simulations agree that the bulk of kinetic energy in the cosmic volume gets dissipated by $2 \leq \mathcal{M} \leq 4$ shocks  \citep[see e.g.,][for a comparison of cosmological simulations]{va11comparison}, and that the ensemble of shocks developed in the cosmic web can steadily refill cosmic structures with Cosmic Rays \citep[e.g.][]{mi00,ry03,pf06,2013MNRAS.428.1643P,scienzo,2016MNRAS.461.4441S}, and also produce levels of radio emission potentially detectable with the next generation of radio interferometers \citep[e.g.][]{va15radio,2020PASA...37....2W,2023arXiv231013734B}. However, the $\mathcal{M} \leq 5$ regime is a one where making predictions of CR acceleration gets difficult, owing to a number of physical and numerical uncertainties \citep[e.g.][for recent discussions]{Bykov19,boss}. 

 Modern numerical simulations have attempted, with various degrees of success, to cover the daunting physical complexities and vast range of scales needed to model the interplay between multi-phase accretion onto supermassive black holes, and their ejected relativistic jets 
\citep[e.g.][for a few recent reviews]{ 2023arXiv230605864P,2023Galax..11...73B}. 
Cosmological simulations have been successfully used to simulate the "radio"-mode feedback (i.e. mediated by kinetic jets dissipating their energy via heating of the gas reservoir undergoing cooling) \citep[e.g.][]{2008ApJ...687L..53P,2010MNRAS.401.1670F,2012MNRAS.420.2662D,2017MNRAS.470.1121T,2017MNRAS.470.4530W,2017Galax...5...35D,2020NatRP...2...42V,2023Galax..11...73B}.  However, to the best of our knowledge, no cosmological simulation has so far included an even simplistic model to co-evolve magnetic fields and cosmic ray electrons injected by radio galaxies, star formation and shocks at the same time. 

Here, we introduce a new suite of cosmological simulations with cosmic-ray electrons, injected at run time by shock waves, star feedback and active galactic nuclei, which also are sources of magnetic fields and allow us a run-time monitoring of how these non-thermal energy components are advected in the cosmic volume. 
Our paper is structured as follows: in the following Section (Sec.~\ref{sec:methods}) we give an overview of the new numerical implementations adopted in our runs. In Sec.~\ref{sec::res} we give our main results, both on the simulation of galaxy properties and on the properties of the cosmic ray electron fluid and magnetic fields on various scales of the cosmic web. A discussion of the main limitations of our approach is given in Sec.~\ref{sec:discussion}, while our main conclusions are summarised in Sec.~\ref{sec:conclusions}.

\begin{figure*}
\begin{center}
\includegraphics[width=0.99\textwidth]{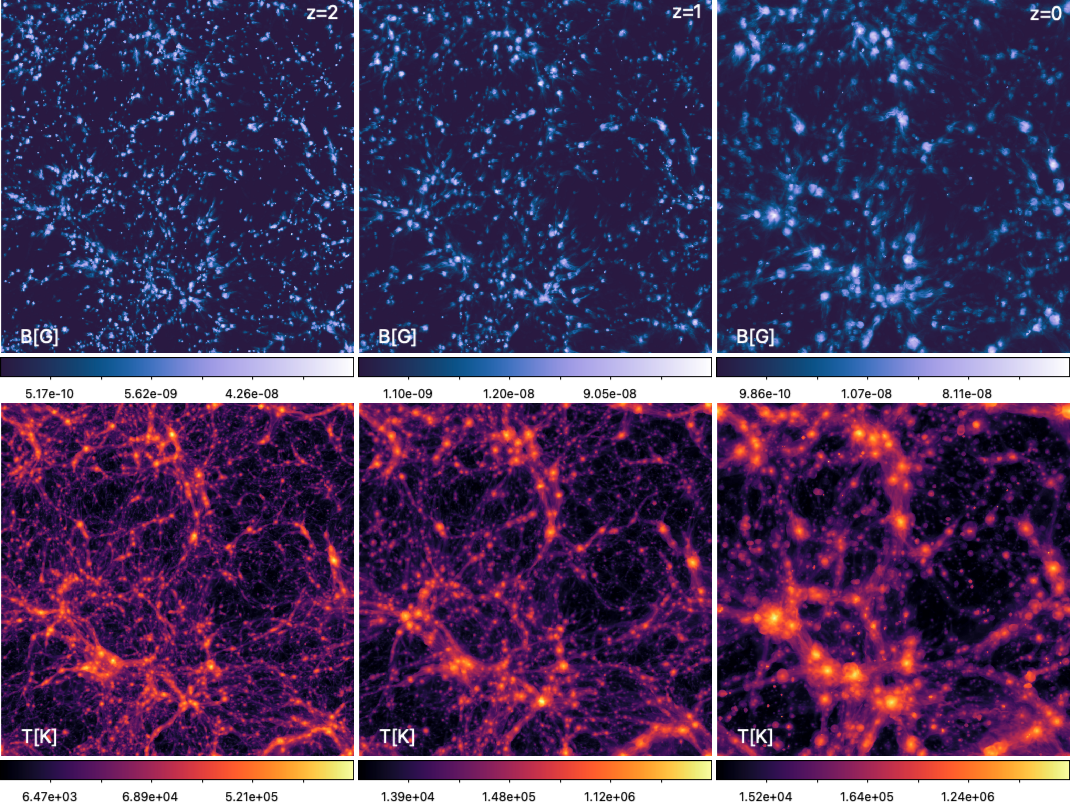}
\caption{Projected mean magnetic field intensity { (top panels, in units of comoving Gauss) and projected mean mass-weighted gas temperature (bottom panels) across the full simulated $42 \rm  Mpc^3$ volume in our B4 run at $z \approx 2$ (left), for the $z \approx 1 $ (centre) and $z \approx 0$ (right) epochs}. Movies of some of our runs can be found at \url{https://www.youtube.com/playlist?list=PL8ecsjnxOKP7LXQICrNBLHZkPgxh0jCzj}.}
\label{fig:map0}
\end{center}
\end{figure*}

\begin{figure*}
\begin{center}
\includegraphics[width=0.99\textwidth]{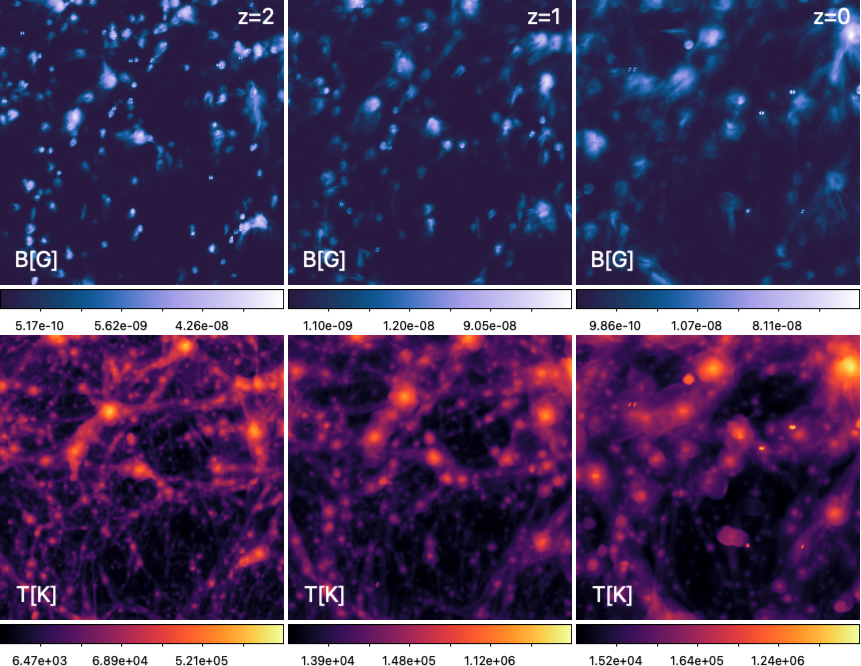}
\caption{Same as Fig.~\ref{fig:map0}, but for a zoomed portion with side $15 \times 15 \rm Mpc^2$  (and $42 \rm  Mpc$ along the line of sight) for our B4 run at three different epochs.}
\label{fig:mapz}
\end{center}
\end{figure*}

\begin{table*}[bht]
    \centering
    \begin{tabular}{cccccccccccccc}

         ID & $m_{\rm *} \rm [M_\odot]$ &  $t_{\rm *}[Myr]$ & $\epsilon_{SF}$&  $\epsilon_{B,SF}$ &$\xi_{SF}$& $\alpha_{B,cold}$& $\alpha_{B,hot}$&$f_{\rm k,cold}$ &$f_{\rm k,hot}$ &$\epsilon_{B,AGN}$ &$\xi_{AGN}$&$B_{\rm 1Mpc}\rm[nG]$ & $\xi_e(\mathcal{M})$\\ \hline
 
         A1& $10^{7}$ & 10 & $10^{-8}$&  0.1 &$10^{-5}$& 1& 10& 0.18& 0.81& 0.1&$10^{-3}$&$10^{-7}$ & Eq.\ref{eq:shocks}\\

 A2& $10^{7}$& 10 &$10^{-8}$& 0.1 &$10^{-5}$& 1& 50& 0.18& 0.81&0.1 &$10^{-3}$&$10^{-7}$ & Eq.\ref{eq:shocks}\\

 B1& $3 \cdot 10^{7}$& 10 &$3 \cdot 10^{-8}$& 0.1 &$10^{-5}$& 1& 10& 0.18& 0.81&0.1 &$10^{-3}$&$10^{-7}$ & Eq.\ref{eq:shocks} \\
 B2& $5 \cdot 10^{7}$&10 & $5\cdot 10^{-8}$& 0.1 &$10^{-5}$& 1& 10& 0.18& 0.81&0.1 &$10^{-3}$&$10^{-7}$ & Eq.\ref{eq:shocks}\\
  B3& $3 \cdot 10^{7}$& 30 &$3\cdot 10^{-8}$& 0.1 &$10^{-5}$& 1& 10& 0.18& 0.81&0.1 &$10^{-3}$&$10^{-7}$ & Eq.\ref{eq:shocks}\\
\textbf{ B4 }& $3 \cdot 10^{7}$& 20 &$3\cdot 10^{-8}$& 0.1 &$10^{-5}$& 1& 10& 0.18& 0.81&0.1 &$10^{-3}$&$10^{-7}$ & Eq.\ref{eq:shocks}\\  
 C1& $3 \cdot 10^{7}$& 20& $3\cdot 10^{-8}$&0.1 & $10^{-5}$& 1& 10& 0.18& 0.81& 0.1 & 
 $10^{-3}$&$0.37$, $k^{-1}$ & Eq.\ref{eq:shocks}\\
 C2& $3 \cdot 10^{7}$& 20& $3 \cdot 10^{-8}$& 0.1 & $10^{-5}$& 1& 10& 0.18& 0.81& 0.1 & $10^{-3}$&$0.37$, $k^{-1}$ & Eq.\ref{eq:shocks},  $\theta_B \geq 45^\circ$\\ \hline

    \end{tabular}
    \caption{Main simulation parameters used in our runs. Each column gives: a) the run ID used in the paper, b) the minimum mass of star forming particles; c)  the minimum timescale for forming star particles; d) the efficiency of stellar feedback (referred to the accreted rest mass energy); e) the efficiency of magnetic feedback (referred to the feedback energy);  f) the acceleration efficiency of  CRe by stellar feedback; g) the used boost parameter in Bondi accretion formula for cold gas; h) the used boost parameter in Bondi accretion formula for hot gas; i) the fraction of feedback energy assigned to kinetic feedback in the cold accretion mode; l) the  fraction of feedback energy assigned to kinetic feedback in the hot accretion mode; m) the efficiency of magnetic feedback energy (referred to the total feedback energy); n)  the acceleration efficiency of  CRe by AGN feedback; o) the  average initial magnetic field amplitude smoothed on a comoving $1 \rm~ Mpc$; p) the assume injection efficiency by shocks.  
    Relevant parameters which were instead kept the same for all runs were: the physical gas density for star formation, $n_{\rm SF}=10^{-3}\rm part/cm^3$, the comoving density for the simulation of SMBH, $\rho_{\rm AGN}=20 \bar{\rho}$ , the assumed fixed temperature in the Bondi-Hoyle accretion rate, $T_{\rm AGN}=10^{5} \rm K$. { We mark in boldface model B4 as the one overall best performing in our suite of runs.}}
    \label{tab:mhd}
\end{table*}

\begin{figure*}
\begin{center}
\includegraphics[width=0.95\textwidth]{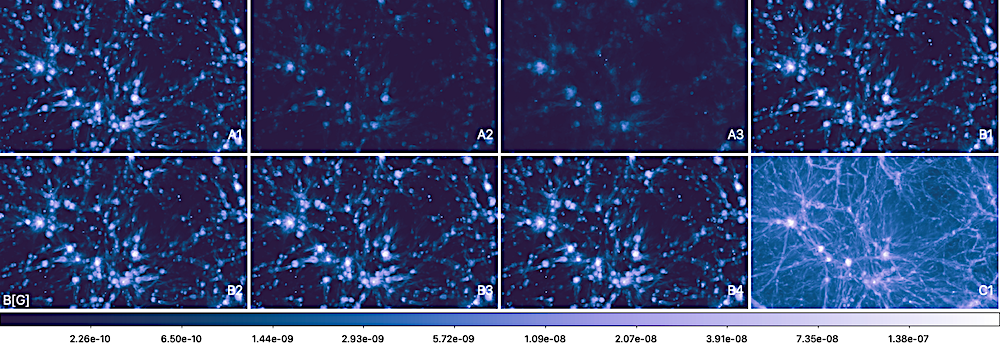}
\includegraphics[width=0.95\textwidth]{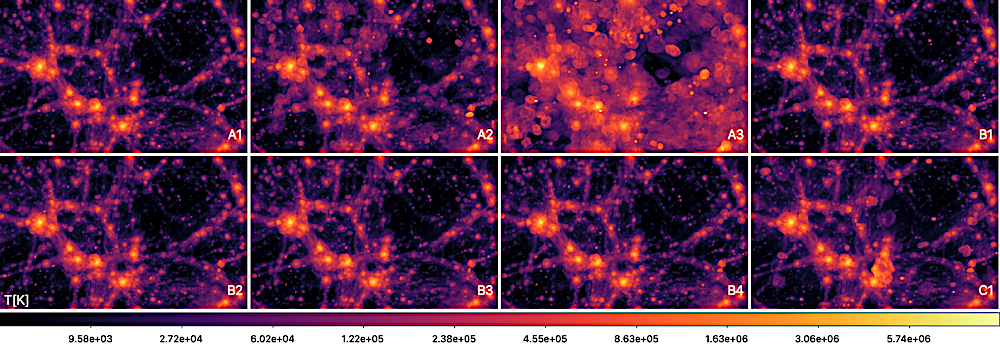}
\caption{Top projected mean (mass weighted) gas temperature (in units of [K]) across the entire simulated volume for our seven runs at $z=0.02$. Bottom: for the same epoch and volume selection, projected mean (mass weighted) magnetic field strength (in units of [G]). The selected volume in all panels is about $21.25 \times 42.5 \rm Mpc^2$ and the lenght along the line of sight is $42.5\rm~ Mpc$.}
\label{fig:mapTemps}
\end{center}
\end{figure*}

\section{Methods}
\label{sec:methods}

\subsection{ENZO cosmological simulations}

We produced new ideal Magneto-Hydro-Dynamical (MHD) cosmological simulations with the Eulerian code  {\enzo} v2.6 \footnote{
\url{www.enzo-project.org}} \citep[][]{enzo14}. These simulations include a number of ad-hoc modifications to inject a passive cosmic-ray electron fluid and to implement source feedback mechanisms for reproducing the population of active radio galaxies.   

This version of {\enzo} allows us to take full advantage of the GPU architecture on the LEONARDO supercluster at CINECA, which ensures that our cosmological simulations are overall $\times 4$ faster then their corresponding CPU-only version, on the problem being analysed.  For this project we used the GPU-accelerated version of  {\enzo} \citep[][]{wang10}, which allowed us to run our simulations employing  4 GPU per node and 1024 MPI tasks in total. Larger simulations, using  $2048^3$ cells and $1024^3$ are presently running and require instead 32 nodes, employing 4 GPU per node and 1024 MPI tasks in total. 
For a review of the basic properties of the code we refer the reader to its method paper \citep{enzo14}, while for more specific details about the MHD version used here, we refer to our previous works \citep[][]{va17cqg,va21magcow}.

We used the following cosmological parameters: $h = 0.678$, $\Omega_{\Lambda} = 0.692$, $\Omega_{\mathrm{M}} = 0.308$ and $\Omega_{\mathrm{b}} = 0.0478$, based on the results from the \citet{2016A&A...594A..13P}. 

As a compromise  between the requirements for a reasonable resolution of halo mass growth, feedback between galaxies and their surrounding medium and shock dynamics within halos as well as their interface with filaments and cosmic voids, we resorted to a uniform resolution in our simulations. We use $1024^3$ cells and dark matter particles to cover a $42^3 ~\rm Mpc^3$ volume (additionally, to increase the statistics of object we complemented our suite also with a $2048^3$ simulation of a $85^3 \rm Mpc^3$ volume). The resulting spatial resolution ($41.5$  kpc comoving) is coarser than state-of-the-art cosmological simulations of galaxy evolution, in which typically adaptive-mesh refinement, smoothed particle hydrodynamics or moving-mesh techniques are employed \citep[e.g.][]{2020NatRP...2...42V}. However, we calibrate sub-grid models of star formation and feedback to reproduce the most important global statistics of the stellar mass assembly in galaxies, as well as the energy input of AGN from supermassive black holes during cosmic evolution.

This approach allows us to study, for the first time, how reservoirs of cosmic-ray electrons build up in the cosmic web as a function of time. Moreover, we can study how AGN produce different enrichment patterns of fossil electrons depending on the AGN feedback mechanism, while broadly reproducing the bulk properties of radio galaxies across redshift and environments (e.g., cosmic star formation history, stellar distribution function of galaxies).\\

{ It is impossible to fully explore the large number of sub-grid parameters used to parametrise star formation and feedback, AGN feedback and injection of magnetic energy and CRe associated to these processes, which are non-linearly coupled to galaxy evolution. Here we will focus on the comparison of nine resimulations of the same cosmic volume, featuring plausible variations in sub-grid prescriptions for star formation or AGN feedback, and compare them to key observable properties of galaxies (e.g. Sec.~\ref{subsec:galaxies}-\ref{radiogalaxies}).
 As we shall see in the reminder of the analysis, there is one model which best reproduces all investigated properties (B4 in the following), while the others fail in reproducing some key properties of the galaxy population. Nevertheless, even failed models are useful to understand the role of specific mechanisms in the seeding of CRe and magnetic fields on the scale of the cosmic web.

In detail, we tested a) three model variations for AGN feedback, in which we kept the star formation recipe fixed (runs A1, A2, A3); b) four model variations in which we varied the star formation recipe, and kept the AGN modelling fixed (runs B1, B2, B3, B4) and c) two model variations of one of the previous models, in which we added primordial magnetic fields (run C1) and also tested the obliquity-dependent injection of CRe by shocks (run C2).}
In all runs, for gas cooling, we assumed for simplicity a  gas equilibrium cooling model, with a primordial gas composition { and with no metal enrichment modelled or followed at run time. }
Table \ref{tab:mhd} gives the most important parameters describing the differences in our suite of simulations, while in the next Sections we will introduce in detail the adopted numerical prescriptions for magnetic fields, shock acceleration, AGN, star formation and for the advection and ageing of our CRe. 

Figure \ref{fig:map0} { and \ref{fig:mapz}} show simulated magnetic field and temperature distributions  at  { three different epochs ($z \approx 2$, $\approx 1$ and $\approx 0$), for our B4 model}. The distribution of thermal and non-thermal energy components in the simulated cosmic web is, at all epochs, a combination of the large scale dynamics of the cosmic web (halos, filaments and voids components) and of intermittent and "inside-out" impulsive release of feedback energy related to the evolution of simulated galaxies.  In the following, we will explain in more detail how the different physical implementations affects the evolution of the quantities above (and of their associated CRe populations) and in the results we will discuss how variations in these procedure may affect the measured global trends.

\subsection{Primordial magnetic fields and sub-grid dynamo amplification} 
\label{subsec:dynamo}

In most cases (run A1, A2, A3, B1, B2, B3 and B4), the magnetic field in our simulated volume is initialised with a simple  uniform volume-filling magnetic field with comoving amplitude $B_0=10^{-16} \rm G$ at the beginning of the simulation ($z=40$), i.e. at the level of lower limits from the non-detection of the inverse Compton emission from blazars \citep[][]{2010Sci...328...73N,2021Univ....7..223A,2024ApJ...963..135T}. We initialise magnetic fields with low amplitudes to make implemented magnetic feedback from astrophysical sources easily detectable in the simulation, since "magnetisation bubbles" are expected to have fields with strengths $\geq 10^{-12} \rm~G$ \citep[][]{2021MNRAS.505.5038A,2022A&A...660A..80B}.
In the C1 and C2 simulations instead we used a tangled primordial magnetic field, with fields scale dependence described by a power law spectrum: $P_B(k) = P_{B0}k^{n_B}$ ($n_B=-1.0$), whose root mean square amplitude after smoothing the fields within a scale $\lambda=1 \rm ~ Mpc$ \citep[e.g.][]{2019JCAP...11..028P} is set to $B_{\rm 1Mpc}=0.37 \rm ~nG$.  We notice that this value is about $5$ times lower than the upper limits inferred by \citet{2019JCAP...11..028P} using priors from the analysis of the Cosmic Microwave Background. This renormalisation was motivated by the latest results on extragalactic Faraday Rotation using LOFAR (\citealt[][]{2023MNRAS.518.2273C} and \citealt{Carretti24}).
{ Since our goal in this paper is to couple different CRe injection mechanisms with a plausible primordial magnetic field configuration, we focused here on the $n_B=-1.0$ spectrum, instead of using much flatter (i.e. the scale-invariant) or steeper (i.e. low-scale spectra from causal mechanisms) power spectra, which appear to be disfavoured by the comparisons of simulated and observed Faraday Rotation data (e.g., \citealt{va21}; \citealt{Carretti24},
\citealt{2024arXiv240616230M}).}

These simulations have an spatial resolution which is insufficient for resolving flows in halos with high Reynolds number and can hardly capture any small-scale amplification within halos \citep[e.g.][and references therein]{review_dynamo}. 
However, having a realistic level of magnetisation in our halos is necessary to produce realistic estimates of synchrotron radio emission from our simulated cosmic web. Following previous work  \citep[e.g.][]{va17cqg,va21}, we used a simplistic model to incorporate the expected effect of small-scale dynamo amplification, based on the run-time analysis of the solenoidal velocity field, which is the one mostly responsible for the amplification in halos \citep[e.g.][]{ry08}.  In particular, the  vorticity squared, $(\nabla \times \vec{v})^2 \equiv \epsilon_{\omega}$ (where $\vec{v}$ is the local gas velocity), is easy to measure at any timestep and it offers  a convenient way to  estimate the turbulent dissipation rate of solenoidal turbulence on the fly, as  $F_{\rm t} \simeq \eta_t \rho \epsilon_{\omega}^3/L$, where $\rho$ is the gas density in the cell, $\eta_t=0.014$ in Kolomogorov turbulence and $L$ is the stencil to compute the vorticity, as in \citep[][]{jones11}.  
We assume dynamo amplification operates below the resolved scales in the simulation, and thus we generate new 
 magnetic energy by converting, during a run time, a fraction of the kinetic energy power into magnetic field energy: $E_{\rm B,dyn} = \epsilon_{\rm dyn}(\mathcal{M}_t)F_{\rm t} \Delta t$. The conversion factor must be calibrated based on dedicated simulations of dynamo amplification, and as  tested in previous work \citep[][]{va17cqg,va21,va21magcow}, we use the analytical prescriptions derived by \citet{fed14} for $\epsilon_{\rm dyn}$, which are given depending on the local turbulent Mach number.
  
Once the additional magnetic energy is computed, we generate a 3-dimensional magnetic field $\vec{B}_{\rm turb}$, to add the existing magnetic field. We simply impose that $\vec{B}_{\rm turb}$  must be parallel to gas vorticity, so that the new generated field is also (with good approximation) solenodial by construction. The equivalent amount of kinetic energy is removed from the same cell, and momentum is removed assuming an isotropic dissipation of the small-scale velocity vectors.  This procedure is manifestly  simpler than more sophisticated subgrid models available, which measure the sub-grid turbulent energy via Favre filtering, and incorporate the electromotive force in a self-consistent way \citep{gr16}. Although our method has an advantage of producing a realistic level of magnetic fields in the simulated halos. We do not allow for this mechanism to operate at cosmic density levels lower than that of filaments, i.e., for gas densities lower than $10$ times the cosmic mean gas density, on the basis that no dynamo amplification has ever been found in this environment even with high-resolution MHD simulations; see e.g., \citealt{va14mhd}.  

The second row of Fig.~\ref{fig:mapTemps} shows the distribution of magnetic fields at $z=0.02$ for all our runs. We see that which illustrates well the much more efficient filling of cosmic volume by primordial magnetic fields (run C1 and C2), as well as slightly different scales and intensities of magnetisation by such fields than each different run (e.g., those where we used combined feedback from stars and AGN) could produce. 
 
\begin{figure}
\begin{center}
\includegraphics[width=0.49\textwidth]{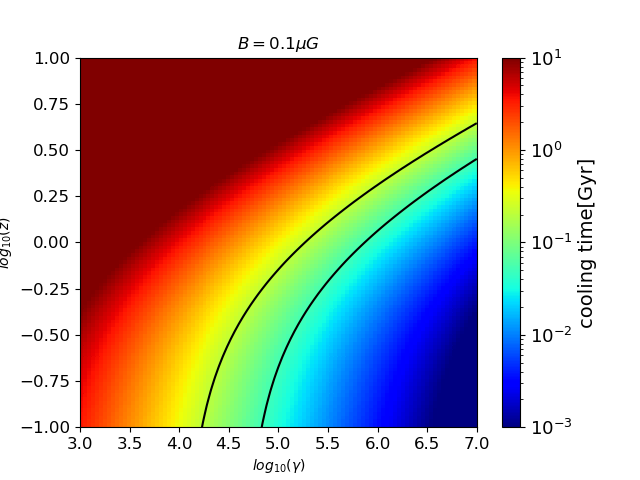}
\includegraphics[width=0.49\textwidth]{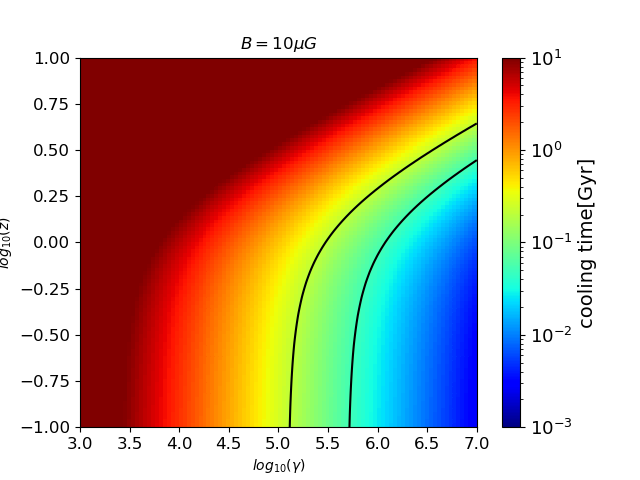}
\end{center}
\caption{Cooling time for relativistic electrons underoging synchrotron and inverse compton losses (as in Eq.\ref{eq:ic}), as function of their Lorentz factor and cosmic redshift, for two $B = 0.1 ~\rm \mu G$ and $B = 10 ~\rm \mu G$. The solid black lines mark the region corresponding to our choice of the timescale $\tau$ for the age determination of CRe injected at run-time.}
\label{fig:t_rad}
\end{figure}

\begin{figure*}
\begin{center}
\includegraphics[width=0.95\textwidth]{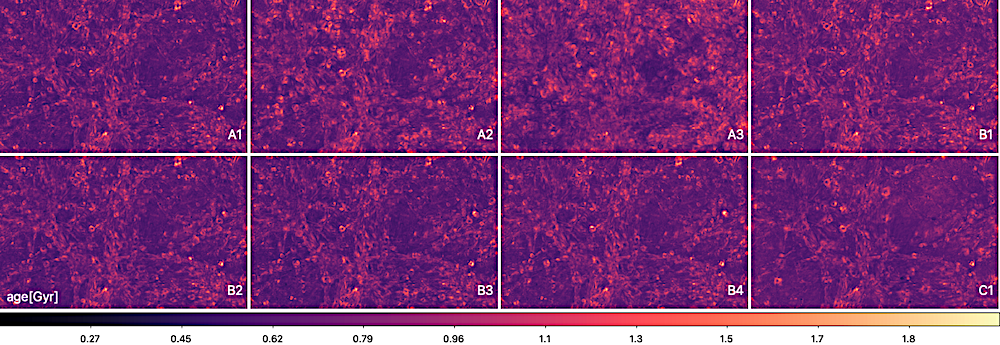}
\caption{Projected mean (CRe-weighted{ , only considering CRe from shocks}) age of CRe since their last injection by shocks (in units of [Gyr]) across the entire simulated volume for our seven runs at $z=0.02$.}
\label{fig:mapAGE1}
\end{center}
\end{figure*}

\begin{figure*}
\begin{center}
\includegraphics[width=0.95\textwidth]{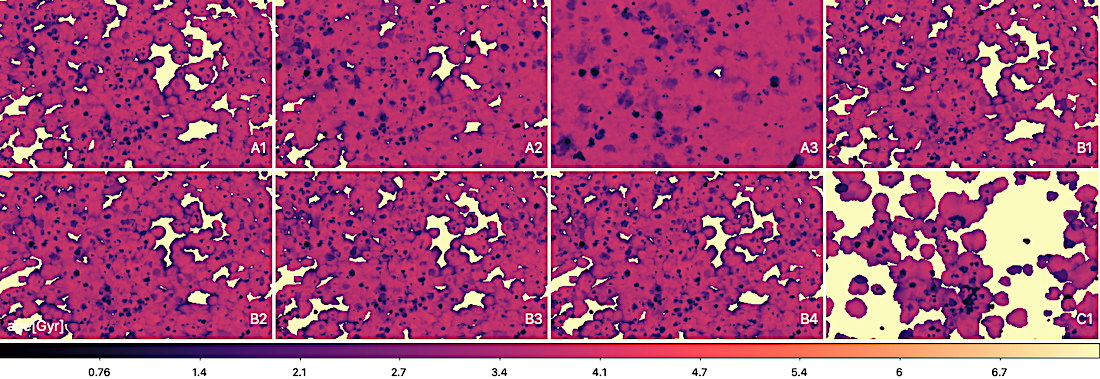}
\caption{Projected mean (CRe-weighted, { only considering CRe from AGN}) age of CRe since their last injection by AGN (in units of [Gyr]) across the entire simulated volume for our seven runs at $z=0.02$.
}
\label{fig:mapAGE2}
\end{center}
\end{figure*}

\begin{figure*}
\begin{center}
\includegraphics[width=0.95\textwidth]{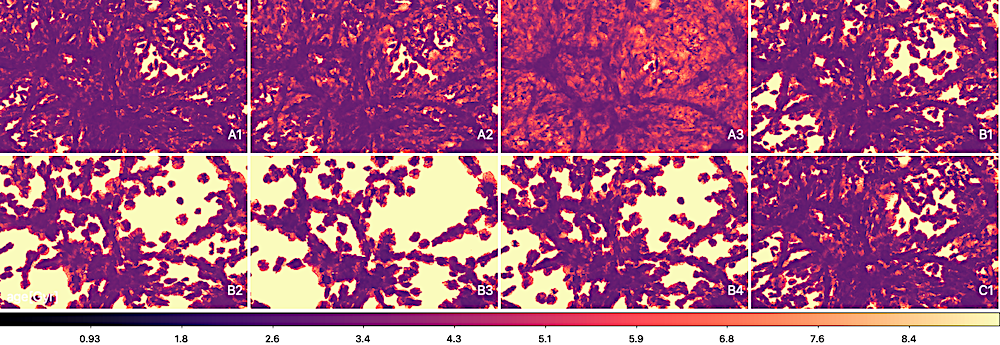}
\caption{Projected mean (CRe-weighted{ , only considering CRe from star formation}) age of CRe since their last injection by star formation (in units of [Gyr]) across the entire simulated volume for our seven runs at $z=0.02$.}
\label{fig:mapAGE3}
\end{center}
\end{figure*}  

\begin{figure}
\begin{center}
\includegraphics[width=0.49\textwidth]{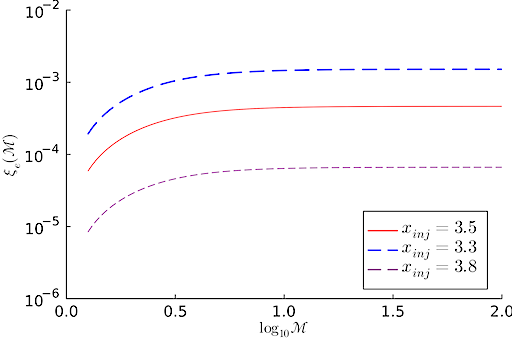}
\end{center}
\caption{Trend of injection efficiency of CRe (given in terms of the number of injected CRe with respect to the number of thermal particles) as a function of shock Mach number, for three different choices of the injection momentum, $x_{\rm inj}$ (the baseline model assumed in our simulations is the red one).}
\label{fig:xinj}
\end{figure}  

\subsection{The spatial advection and the age determination of relativistic electrons}
\label{subsec:tau}

Unlike CR protons, CR electrons can be effectively treated like a passive fluid (owing to their small pressure compared to the thermal gas and kinetic pressure, as well as to the magnetic pressure everywhere in the cosmic structures) which is frozen to the gas fluid because of the extremely small gyro-radius ({ $r_{L,e} \sim 10^{-6} \rm ~pc$ for a $\mu G$ magnetic field and $\gamma \sim 10^6$ electrons, where $\gamma$ is the Lorentz factor), which is about the highest energy which electron can reach before radiating their energy via synchrotron and Inverse Compton, in less than 1 Myr, e.g. \citealt[e.g.][]{sa99}}). 
With our modification of the  {\enzo}  code, we separately track the advection of the three kinds of CRe: the ones injected by shocks, the ones injected by AGN and the ones injected by star formation (see Sections below for details on the injection procedures). The field which is advected for each family is the {\it number density} (similarly to what {\enzo} already does for any other chemical species). 
The choice of the number density as the advected field is crucial here: this quantity is conserved even in the presence of re-acceleration processes (not included in our model). 
{ The thermalisation of CRe 
can proceed in a short time only at very high particle densities ($n \gg 0.1/\rm cm^3$), via ionisation or Coulomb losses. This means that these loss processes can affect the simulated budget of CRe in the interstellar medium within galaxies, while instead our predictions concerning the radio emission by shocks or by radio jets are unaffected by the thermalisation of CRe, because they always propagate through a much lower-density medium.}

Our model lacks a spectral ageing models for CRe (numerically very expensive), yet we want to recover the typical evolutionary timescales of our injected CRe everywhere in the simulation. We thus explore here for the first time the application of a simple, yet, powerful approach applied by  \citet{2019A&A...631A..60B} in AGN feedback simulations of single objects.  
I.e., with our {\enzo} modification we evolve, together with each of the three families of CRe, also a second "mirror" fluid, which is injected and advected exactly in the same way, but that it is additionally subject to  an artificial exponential decay with a fixed, arbitrary timescale, $\tau$, so that $n'_{CRe} \propto e^{-t/\tau}$.
Numerically, this means that all CRe decaying species are advanced in time as  
$n'_{CRe}(t+\Delta t)=n'_{CRe}(t) (1-\Delta t/\tau)$, and this  allows us, at any stage in the simulation, to retrieve the time elapsed since the last injection of CRe in all cells (separately for each CRe species) as:

\begin{equation}
t_{\rm age} = - \tau \log\big (\frac{n'_{\rm CRe}(t)}{n_{CRe}(t)} \big),
\label{eq:age}
\end{equation}
in which $n_{CRe}$ is the number density for each of the primary CRs (i.e. not subjected to the artificial decay law).
To the best of our knowledge this approach has never been applied to the simulation of CRe dynamics; it has an advantage of providing an efficient way to estimate of the typical age of CRs in $100\%$ of the simulated volume. Provided that the expected evolution of CRs is simple too, this information can be used to produce realistic predictions for the full energy spectrum of particles tracked by the total CRe density field (see Section below).

{ It should be remarked that, as long as $\tau$ is a known number, it can have any value in order to allow us deriving the effective age of the CRe in the cells, at least for the case of a single injection epoch for CRe. However, when multiple injections are involved, as in our simulations, it is important to use a physically motivated value of $\tau$, so that the  weighting of older and younger populations is realistic as well. 
Therefore, in our models we fixed $\tau=0.1 ~\rm Gyr$, which is the representative value of the cooling time of relativistic electrons in the $\gamma \sim 10^4$ range, which dominate synchrotron radio emission for magnetic field intensities in the $B \sim 0.1-10 \rm \mu G$. 

In detail, the cooling time is given by \citep[e.g.][]{sa99,bj14}: 

\begin{equation}
    \tau_{\rm cool} =\frac
    {0.77 \rm ~Gyr} {(\gamma/{300})\left[\left(\frac{B}{3.25 \rm \mu G}\right)^2 + (1+z)^4\right]}.
    \label{eq:ic}
  \end{equation}

The range of $z$ and $\gamma$ for which the cooling time (considering synchrotron as well as Inverse Compton losses) is of the order of $\tau$ is shown in Fig.~\ref{fig:t_rad}, for $B=0.1 \rm \mu G$ and $B=10 \rm \mu G$.
Our choice of $\tau=0.1 \rm ~Gyr$ thus ensures that, with good approximation, whenever we compute the radio emission from our distribution of CRe (Sec.~\ref{subsec:sync}) we are correctly weighting the contribution from the CRe in the cell which will actually dominate the radio emission in the real case - especially for long evolutionary timescales ($\geq 0.1 \rm Gyr$). On the other hand, in the case of multiple AGN bursts on very short timescales, and in presence of significant losses in dense environments, the presence of multiple complex component may make our approach too simplistic. However, this limitation is not too severe for the large-scale global analysis performed in this work. 
Additional tests given in the Appendix  (Sec.~\ref{sec:A1}) show that the exact choice of this parameter (at least in the $0.01-1 \rm~ Gyr$ range is not crucial in the age determination of all CRe species analysed in this work.}

The different visual patterns of the elapsed time since their last injection for the CRe associated to different mechanisms is shown in Figures \ref{fig:mapAGE1}-\ref{fig:mapAGE3}, for our runs at $z=0.02$. Here we produced maps of the age of CRe within each cell, projected along the LOS by weighting for the CRe density. The quantitative information is partially lost due to the smearing of the age distribution along the LOS, yet the maps give a qualitative view of the differences in ages related to the various mechanisms. 

Compared to the other mechanisms, shocks inject on average the "freshest" electrons at all epochs, given the ubiquitous process of shock formation. This yields to a relatively young ($\leq 0.1-0.4 \rm Gyr$ in the maps) population of CRe  at the periphery of halos and filaments, and occasionally associated to the launching of shocks within halos (Fig.~\ref{fig:mapAGE1}). Most of the projected volume has been filled, at some time in the past, with the injection of small quantity of CRe, since the process of shock formation begins early and can sweep a large fraction of the cosmic web. 
It should be noticed that the distribution of shock age and morphologies is not the same in all runs, due to galaxy feedback processes. Indeed, beside directly injecting new CRe in jets or winds (see next Sections), feedback also triggers new shocks, which in turn inject additional CRe as they expand into the intergalactic medium, { an effect which has been already identified by previous} cosmological simulations \citep[e.g.][]{ka07,2016MNRAS.461.4441S}. 

Figure \ref{fig:mapAGE2} illustrates the typical bubble structure of the regions enriched with CRe by AGN feedback instead. In all our models, AGN feedback becomes significant only for $z \leq 2$, which explains why there are projected regions with no enrichment of CRe at all, even in this very overdense selection of the simulation. As it will be discussed later in Sec.~\ref{subsec:pdf}, the 3-dimensional filling factor of AGN and stellar feedback bubbles is typically $\leq 10-35\%$ in our runs. While the first expanding bubbles of CRe have long injection timescales ($\geq 2-3 \rm ~Gyr$ in the maps), the recurrent repeated activity of AGN is marked by the smaller and darker blobs at the centre of bubbles, which are associated with very recent ($\leq 0.1 \rm ~Gyr$ in the maps) feedback episodes.

In the case of the age distribution of CRe injected by stellar feedback (Fig. \ref{fig:mapAGE3}) the bubbles are confined to a smaller volume, and they visually show a larger correlation with the underlying cosmic web distribution. Their average age is also much larger compared to the other two mechanisms, following from the fact that the peak of star formation was at $z \sim 2$ in all runs. More recent episodes of injection of CRe by stellar feedback tend to be hard to be seen in projection, as they are covered by the information of older populations along the LOS.  

This approach to monitor the injection and circulation of relativistic electrons has a strong predictive power, despite its great simplicity, due to aforementioned fact that the number of CRe does not evolve even in the presence of processes which are not included (yet) in our model, like Fermi reacceleration or loss processes of CRe.  Of course, the actual radio emission from CRe crucially depends on their spectral energy distribution, and in this case we can only approximately guess it, as it will be discussed in  Sec.~\ref{subsec:sync}. 

This fluid approach to study CRe is also complementary to previous work by our group \citep[][]{wi17, va21jets,2023MNRAS.526.4234S,va23}, in which we employed Lagrangian particles to track relativistic electrons, injected in post-processing in our snapshots.  Here the application of a Eulerian fluid approach allows us to monitor the evolution of fluid in $100 \%$ of the simulated domain, unlike what can be done through the application of a finite number of tracers - for which only the central regions of single galaxy clusters could be well sampled. Moreover, this approach allows us to  evolve electrons at run-time and use the parallelization as well as the GPU acceleration available in the {\enzo} code, and apply this approach to very large simulations. This means that we can study the distribution of electrons in a $O(1)$ fraction of the $\sim 1-8 \cdot 10^9$ cells of our simulations at all timesteps, i.e. 4-5 orders of magnitude more data that we can do with Lagrangian tracers.

Finally, we shall notice that our approach neglects the spatial diffusion of cosmic rays, which is mostly negligible on the scales of interest here (this is addressed in more detail in 
 in Sec.~\ref{sec:discussion}).

\begin{figure*}
\includegraphics[width=0.99\textwidth]{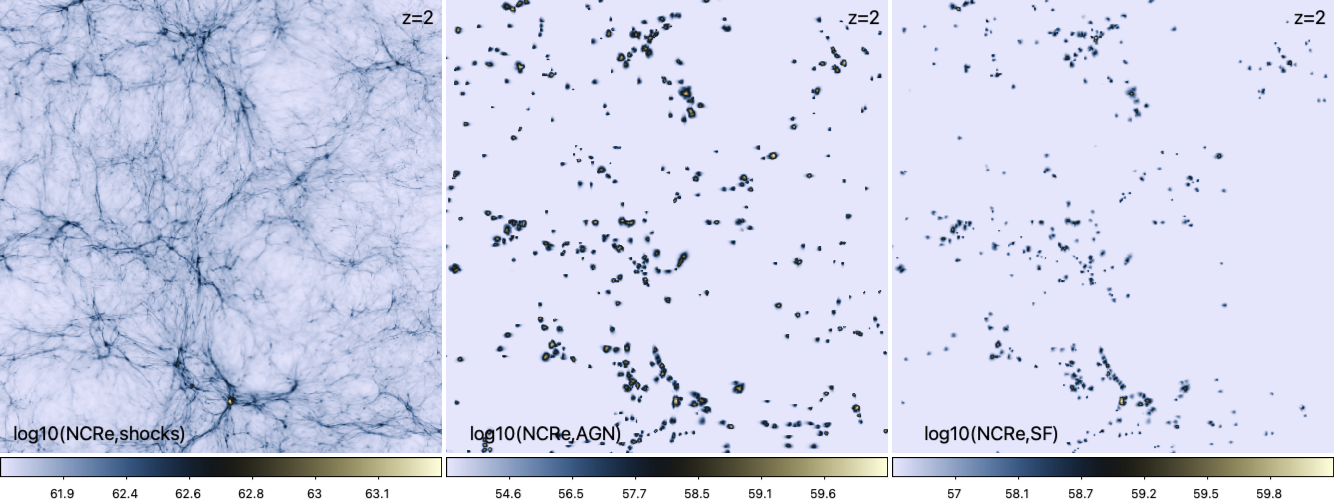}
\includegraphics[width=0.99\textwidth]{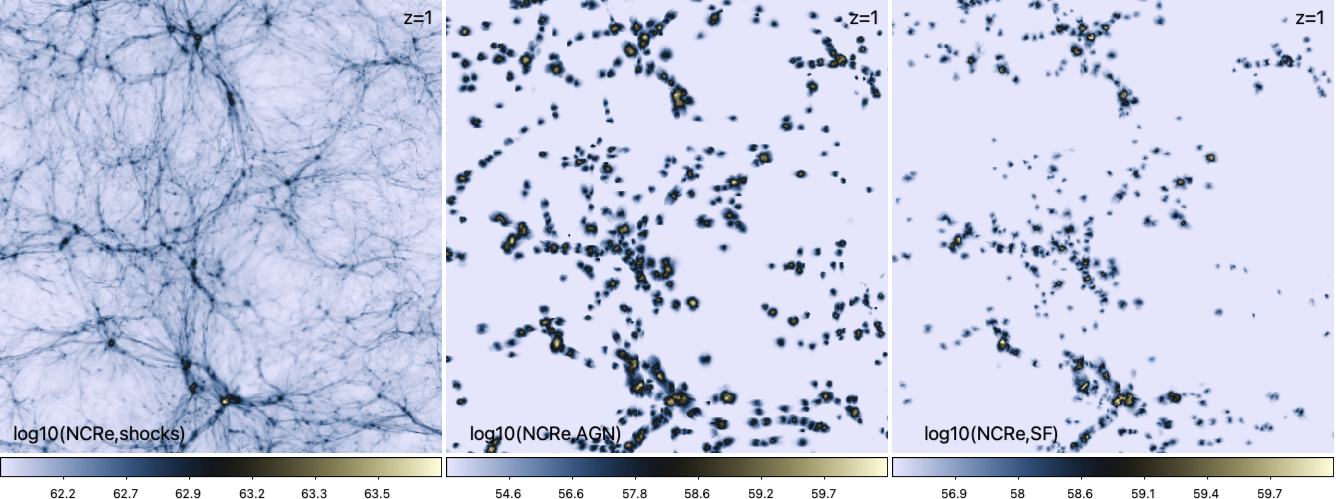}
\includegraphics[width=0.99\textwidth]{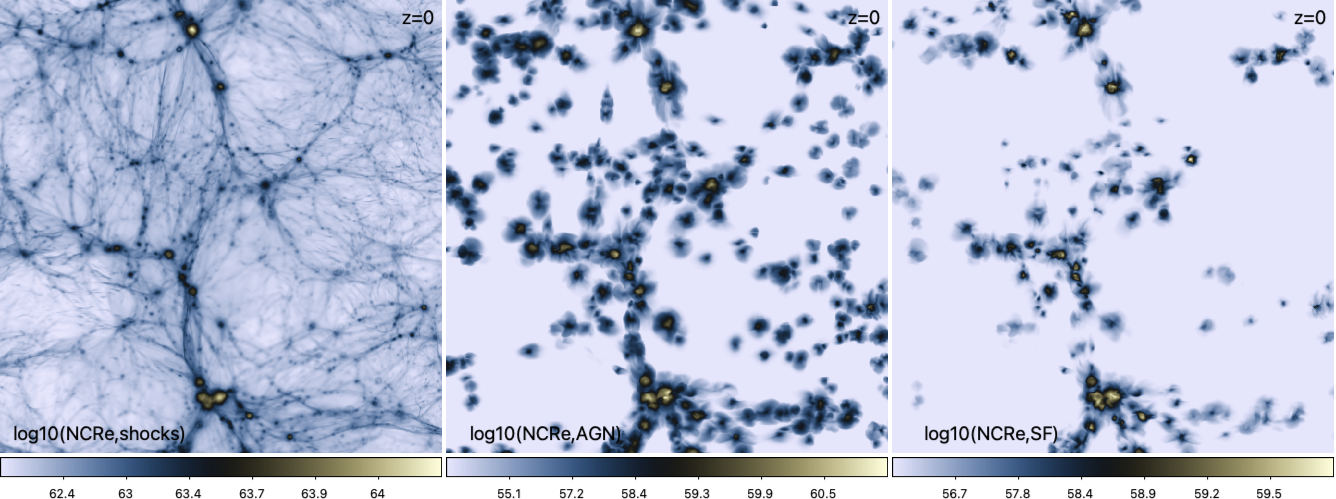}
\caption{Projected number of CRe  across the full simulated $42.5 \rm  Mpc^3$ volume in our { B4 run (considering only $5 \rm ~Mpc$ along the line of sight to better isolate single structures), for the $z \approx 2$, $z \approx 1$ and $z \approx 0$ epochs (from top to bottom). The first column shows CRe injected by shocks, the second by AGN feedback and the third by star formation}. The full movie of these fields can be found at \url{https://youtu.be/xMTSK_rxZZU?si=hj4Yckl3hBuhYF-v}.}
\label{fig:map1}
\end{figure*}

\begin{figure*}
\begin{center}
\includegraphics[width=0.33\textwidth]{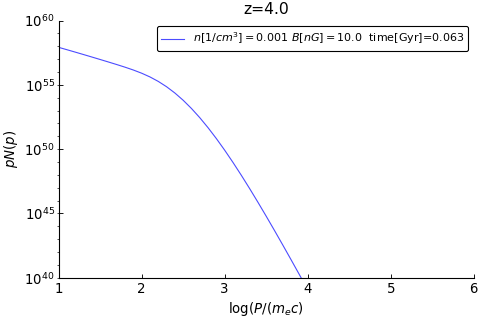}
\includegraphics[width=0.33\textwidth]{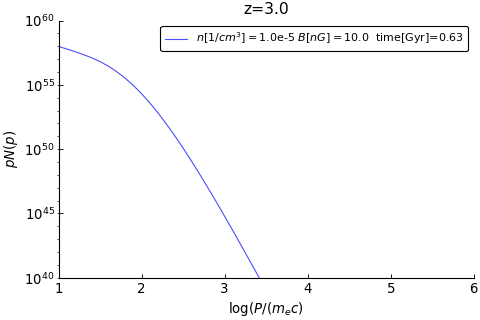}
\includegraphics[width=0.33\textwidth]{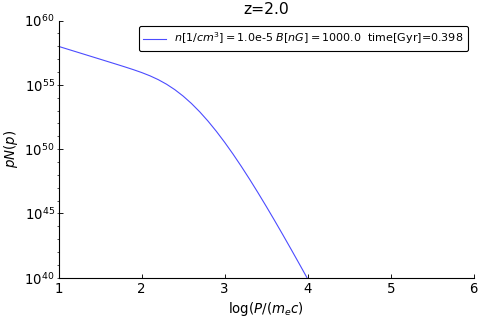}
\includegraphics[width=0.33\textwidth]{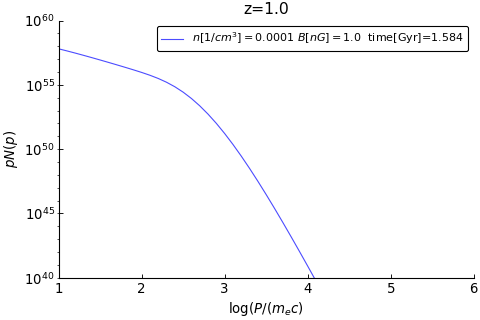}
\includegraphics[width=0.33\textwidth]{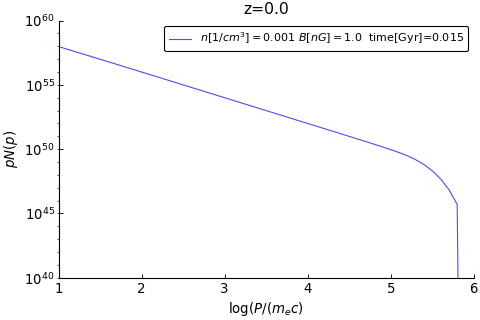}
\includegraphics[width=0.33\textwidth]{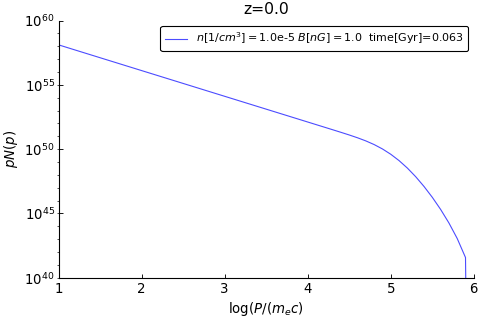}

\end{center}
\caption{Example of six different CR electrons spectra computed by our post-processing model, for different (random) variations of the combinations between gas density, magnetic field strength, redshift and { time elapsed since the injection} of the initial power-law distribution of CRs. Spectra like this are extracted from our tabulated list of spectra generated with the ROGER code, and applied to compute in post-processing the synchrotron emission from every cell in our simulated volumes.} 
\label{fig:spec}
\end{figure*}

\subsection{The injection of relativistic electrons by shocks}
\label{subsec:dsa}
To inject at run-time a fluid of fresh CRe accelerated by shocks based on Diffusive Shock Acceleration (DSA), we use an
 on-the-fly shock finder, which we coded in {\enzo} and used already in previous works \citep[][]{scienzo}, which modifies the original algorithm by  \citet{ry03}. 
In essence: a) we select candidate shocked cells based on their negative velocity divergence; b)  we perform 1-dimensional scans along the three spatial axes and check whether the local changes in gas temperature and entropy are equiverse \citep[e.g.][]{ry03}; c) we use the local gradient of temperature sets the shock propagation verse, and d) we compute the local Mach number  from the pressure jump using ideal Rankin- Hugoniot jump conditions.

{ For CR electrons to be injected via DSA, their Larmor radius must be of the same order of that of CR protons, implying a dependence on the local plasma conditions (i.e. downstream gas temperature, magnetic field strength and topology). Determining the exact injection momentum there (and, if there is a clean cut injection momentum at all) is still a theoretical challenge, which particle in cell (PIC) simulations have only since recently started to tackled with dedicated simulations \citep[e.g.][]{2007ApJ...661..190A,Arbutina2021,2023ApJ...958L..32G,Gupta24}. A general finding is that, while the injection of CR protons from the thermal distribution is favoured in quasi-parallel shocks  \citep[e.g.][]{2018ApJ...864..105H,ryu19}, the situation for CR electrons is more complex, as their injection can proceed both in quasi-parallel or quasi-perpendicular shock geometries, through a combination of different acceleration mechanisms.
\citep[e.g.][]{2020ApJ...897L..41X,2022ApJ...932...86S,Gupta24}.  However, the lack of fully self-consistent 3-dimensional PIC simulations capable of scanning the entire range of parameters relevant for collisionless non-relativistic structure formation shocks makes it difficult to generalise from the above work. 

Given the present uncertainties, here we rely on a simpler semi-analytical recipe, in which the injection of CR electrons is inferred based on the one of CR protons \citep[e.g.][]{2024arXiv240918425K}. In this view}
both, protons and electrons need to be pre-accelerated to suprathermal momenta greater than the so-called injection momentum, $p_{\rm inj}= x_{\rm inj}  p_{\rm th,p}$ in order to diffuse across the shock transition layer and fully participate in the diffusive shock acceleration process, where $p_{\rm th,p} = \sqrt{2m_pk_B T_p}$ is the thermal momentum of protons in the shock downstream  and $x_{\rm inj}$ the injection parameter, which is not yet constrained by the DSA theory.
In a shock, a fraction of the post-shock thermal particle density ($n_p$) becomes a number density of accelerated CR electrons ($n_{\rm CRe}$), { as:
\begin{equation}
n_{\rm CRe}=\xi_e(\mathcal{M})~ n_p = n_p ~\frac{4 x_{\rm inj}^3}{\pi^{3/2} (\alpha_{\rm inj}-3)} \exp{(-x_{\rm inj}^2)}, \label{eq:shocks}. 
\end{equation}
\citep[e.g.][]{2018JKAS...51..185K,2020JKAS...53...59K,kang21,2024arXiv240918425K}. 
In all our simulations we assumed $x_{\rm inj}=3.5$, which is also in line with the latest indications from 1-dimensional kinetic PIC simulations \citep[e.g.][]{Gupta24}. This means that for typical $\sim 10^6-10^8 \rm K$ post-shock gas temperatures, CR electrons are injected with a Lorentz factor $\gamma \sim 3-30$. Therefore it is worth stressing that, in the reminder of the paper we shall loosely use the term "Cosmic Ray electrons" (CRe) to refer to all electrons accelerated by DSA in this way, from electrons with $\gamma \sim $ a few, up to the super-relativistic regime of electrons emitting synchrotron radiation in the radio band ($\gamma \sim 10^4$), or beyond.}  The injection spectral index is computed from the DSA theory:  $\alpha_{\rm inj}=4 \mathcal{M}^2/(\mathcal{M}^2-1)$, which is appropriate for plane shocks in the test particle regime \citep[e.g.][and references therein]{2011JKAS...44...49K}. The trend of the $\xi_e(\mathcal{M})$ used in this paper, and the comparison with small different variations of $x_{\rm inj}$ are shown in Fig.~\ref{fig:xinj}. 

{ In line with \citet{2024arXiv240918425K} we also considered a low Mach number limiter for the injection,   $\mathcal{M_{\rm thr}}=2.3$ because, since several PIC simulations haver suggested a suppression of CR proton acceleration for 
 low Mach number shocks in high $\beta$ plasmas below ${\mathcal M}\sim 2.25$ \citep[e.g.][]{2018ApJ...864..105H,ryu19}. In choosing this low Mach number limiter, we are also motivated by the fact that very weak shocks are typically difficult to measure in cosmological simulations, owing to the amount of transonic and supersonic velocity fluctuations typically found in the same environment \citep[e.g.][]{ry03,va11comparison,2016MNRAS.461.4441S}.}
 
The shock obliquity (i.e. the angle  between the upstream magnetic field and the shock velocity vector) is a key parameter ruling the acceleration of cosmic rays. Several recent works, mostly using particle-in-cell simulations, have explored the relations between shock obliquity, Mach number and the efficiency in the injection of both relativistic protons \citep[e.g.][]{2014ApJ...783...91C,Park2015,2018ApJ...864..105H,2019ApJ...883...60R,Bohdan2020a} and electrons \citep[][]{guo14a,guo14b,Matsumoto2017,Bohdan2017,Bohdan2019b,Arbutina2020,2020ApJ...897L..41X,Arbutina2021}. 
In the case of electron, a robust finding is that for quasi-perpendicular shocks ($\theta_B \geq 45^\circ$)
and strong enough Mach numbers, electron  are pre-accelerated by the motional electric field by drifting along the shock front, via "shock drift acceleration", and a significant fraction of these  electrons eventually takes part to diffusive shock acceleration, and it becomes highly relativistic.

Therefore, in one of our runs (C2) we tested the role of shock obliquity by allowing the injection of the CRe fluid only for quasi perpendicular shocks ($\theta_B \geq 45^\circ$), by measuring at run-time the angle between the pre-shock magnetic field vector and the shock velocity vector, inferred at run-time by our shock finder. The result of this model are compared with the more standard implementation, in which the role of obliquity is not considered and all shocks equally contribute solely as a function of their Mach number (C1).
Both C1 and C2 runs were run assuming the same primordial stochastic model with $P_B(k)=A_B  k^{-1}$, in order to work with a more realistic distribution of angles between shocks and magnetic fields, compared to the more simple setup with a uniform primordial magnetic field used in all other runs. 

The { first} column in Fig.~\ref{fig:map1} gives the evolving spatial distribution of CRe injected by shocks, for the example of our { B4} model, at three different epochs. As already discussed above, shocks inject overall CRe in a very large fraction of the cosmic volume. DSA converts a nearly constant fraction of the gas matter into CRe ($\sim 10^{-4}-10^{-3}$, see Fig.~\ref{fig:xinj}), and it starts almost together with the structure formation processes, hence the large-scale distribution of CRe looks very similar to the standard gas cosmic web, although on small-scales some differences appear, due to the injection of recent and powerful shocks, triggered either by mergers or by AGN.

\subsection{The injection of relativistic electrons  and magnetic fields  by active galactic nuclei}
\label{subsec:bh}

In most cosmological simulations, SMBH are modelled with Lagrangian "sink" particles, which accrete mass from the surrounding cells and need to be injected in the simulation at arbitrary times and positions \citep[e.g.][for the case of {\enzo}]{2011ApJ...738...54K}. The feedback from SMBHs is added either by introducing isotropic thermal heating, or kinetic energy through directional jets along arbitrary directions \citep[e.g.][]{2015ApJ...811...73L}. 

This approach comes with a number of numerical challenges:  from the need of stopping and restarting simulations to inject SMBH seeds (which kills performances in large simulations), to the problems connected in SMBH drifting away from the shallow potential of their host galaxy if simulated with insufficient resolution \citep[see e.g.,][]{2024arXiv240312600D},  to the ad-hoc prescriptions needed to produce realistic SMBH masses \citep[e.g.][]{2022MNRAS.511.3751H}. Considering that our main purpose here is to produce realistic forecasts for the distribution of CRe injected by AGN, without necessarily resolving the puzzle of how SMBH co-evolve with their host environment, we explored a simplified approach which bypasses all aforementioned challenges.

 We avoided the use of sink particles entirely, assuming instead that each gas density peak in the simulation harbours an SMBH at every timestep, with a fiducial mass assigned based on observational scaling relations.. In detail, we measure at run time the total baryonic mass around density peaks in the simulation (only considering cells at least with a total matter density $100$ larger than the cosmic mean density at the same epoch), within a fixed comoving radius of $\leq 83 \rm kpc$ (equal to 2 cells), and we 
 use a scaling relation to compute the average SMBH mass which would be hosted by a real galaxy at that location. 
 Among the many possible best-fit relations derived by \citet{2019ApJ...884..169G} for several observable proxies of SMBH, we use the relation between the enclosed gas mass within the galactic potential, as in their Fig.6; that is $M_{BH} \approx 2 \cdot 10^7 M_{\odot } (M_g/(10^6 M_{\odot})^{0.57}$, which is easy to measure at run-time.
{ To avoid overcrowding of SMBH in the same galaxies, we implemented a local exclusion criterion, which prevents to consider more than one single SMBH cell within the same cube of $7^3$ cells (i.e. closer than a distance of $3$ cells, or $124.5 \rm ~kpc$, from the central gas density peak). This choice somewhat limits our capability of properly computing the SMBH masses of halos undergoing close mergers, since only one of the two gas density peaks (in case they are separated by less than $3$ cells) is assigned with a SMBH mass. In the Appendix (Sec.~\ref{sec:A2}) we present a model variation of this scheme, showing that varying the minimum allowed separation of gas density peaks does not have a significant impact on our simulations.}

 Of course, given the coarse resolution of our simulation, a very detailed measurement of the gas mass may be prone to numerical uncertainties, and other quantities can be used as well to guess the SMBH mass (i.e. total stellar mass, central gas temperature, central gas density etc). We defer a more detailed study of the potential use of different scaling relations to future work, where a higher spatial resolution will be employed."  We impose $M_{BH} \geq 10^6 M_{\odot}$ as an additional requirement to trigger SMBH feedback (because such low mass halos are dominated anyway by stellar feedback). 

For every identified  SMBH location, during run-time we compute the instantaneous mass accretion rate with the standard Bondi–Hoyle formalism: $\dot{M}_{\rm BH}=4 \pi \alpha_B G^2 M_{\rm BH}^2\rho/(v_{\rm rel}^2+c_s^2)^{3/2}$, where $c_s$ is the sound speed of the gas at the SMBH's localtion, $(v_{\rm rel}$ is the relative velocity between the accreted gas and the SMBH (which we assume $=0$ here),  $\rho$ is the local gas density and   
$\alpha_B$ is an ad-hoc parameter meant to compensate for the lack of resolution around the Bondi radius \citep[e.g.][]{2009MNRAS.398...53B,gaspari12,2016Natur.534..218T}.  The bolometric luminosity of the black hole is defined as  $L_{\rm BH} = \epsilon_r \dot{M}_{\rm BH} c^2$, where $\epsilon_r$ is the radiative efficiency of the SMBH, and $P_j=\epsilon_{\rm BH}~L_{\rm BH}=\epsilon_{\rm BH} \epsilon_r \dot{M}_{\rm BH} c^2$ is the feedback power, in which  $\epsilon_{\rm BH}$ is  the factor that converts the bolometric luminosity of the SMBH into the thermal feedback energy (we use  $\epsilon_{\mathrm{BH}}=0.05$ in line with the literature),  and  $\epsilon_r=0.1$ is the radiative efficiency of the SMBH. 

We followed the approach of other cosmological simulations and adopted two distinct feedback modalities depending on the temperature of the accreted matter \citep[e.g.][]{2018MNRAS.481..341S}:
\begin{itemize}
\item "cold gas accretion" when $T \leq 5 \cdot 10^{5} \rm K$, in which case we use a low $\alpha_B$ and distribute most of the feedback energy in the form of dipolar dumps of thermal energy, with velocity vectors pointing at the opposite sides of the SMBH (roughly mimicking "quasar feedback" in halos accreting cold gas via mergers), or

\item "hot gas accretion"  when $T > 5 \cdot 10^{5} \rm K$, in which case we use a large $\alpha_B$ and assume that most of the feedback energy is in the form of kinetic jets (thus mimicking "radio feedback" in halos mostly accreting hot gas).

\end{itemize}

The actual values of $\alpha_{B,cold}$ and $\alpha_{B,hot }$ used in all runs is given in Table \ref{tab:mhd}. 
To estimate the energy ratios going to each energy form, we first consider in all cases that $\epsilon_{\rm B,AGN}=10\%$  is the fraction of energy released in magnetic energy,  through pairs of magnetised loops wrapped around the direction of kinetic jets. Every time jets are activated, their  direction is randomly drawn along any coordinate axis of the simulation.   In all cases we injected new CRe within the jets, with a number density equal to a fixed fraction of the number density of the thermal gas in jets, $n_{\rm CRe,AGN}=\xi_{\rm AGN} n_{\rm jet}$, with $\xi_{\rm AGN} = 10^{-3}$ in our runs.  This specific value of $\xi_{\rm AGN}$ has been calibrated a-posteriori, based on the comparison between the simulated radio luminosity function of our AGN with the observed one (Sec.~\ref{radiogalaxies}), yet it is close to typical choices in the literature \citep[e.g.][]{2012ApJ...750..166M,nolting19a}. 

For the thermal and kinetic energies, we assumed that a) in "cold gas  accretion" feedback, the remaining $72 \%$ of energy is thermal and $18\%$ is kinetic, while in b) the "hot gas accretion" feedback, $81\%$ of the remaining energy is kinetic and the rest is thermal. Similar numbers are quoted in the recent literature \citep[e.g.][]{2018MNRAS.481..341S,2019MNRAS.486.2827D} and even though they are susceptible to changes when working at a different resolution, they still represent parameters giving the best results in our runs (see Sec\ref{sec::res}).

The second column in Fig.~\ref{fig:map1} gives the evolving spatial distribution of CRe injected by AGN, for the example of our { B4 model}, at three different epochs. The distribution of CRe injected by AGN is, as expected, much less volume filling than the one of CRe injected by shocks, but it reaches large values in the proximity of very active AGN.

\begin{figure}
\includegraphics[width=0.46\textwidth]{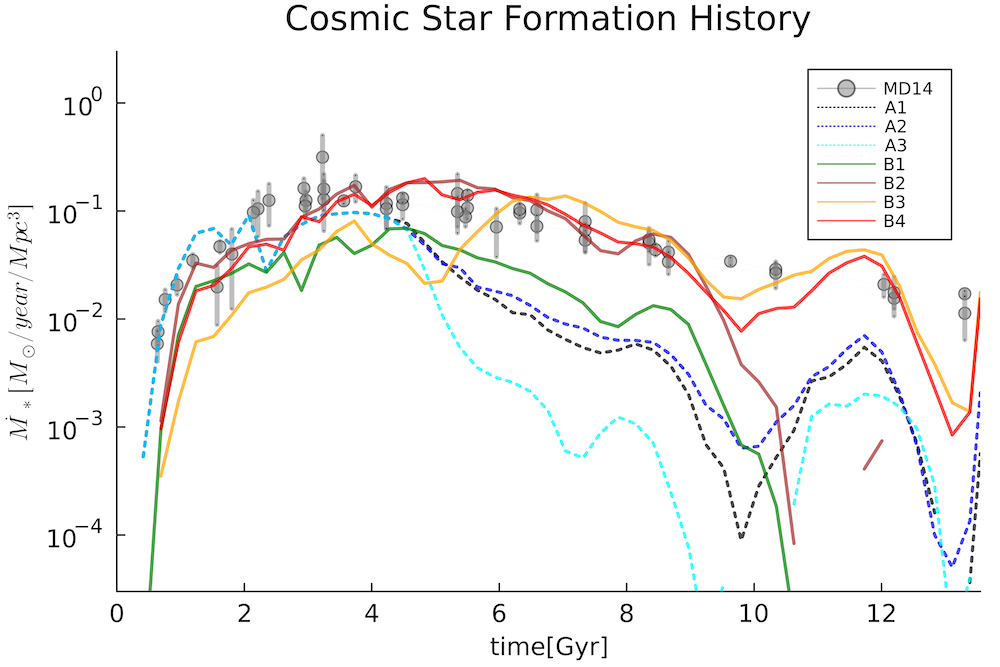}
\caption{History of the simulated cosmic star formation density (i.e. normalised for $\rm Mpc^3$ comoving)  in our suite of simulations, as a function of cosmic time. 
The grey points with error bars show the observed cosmic star formation derived in \citet{2014ARA&A..52..415M}.}
\label{fig:SFR}
\end{figure}

\begin{figure*}
\includegraphics[width=0.45\textwidth]{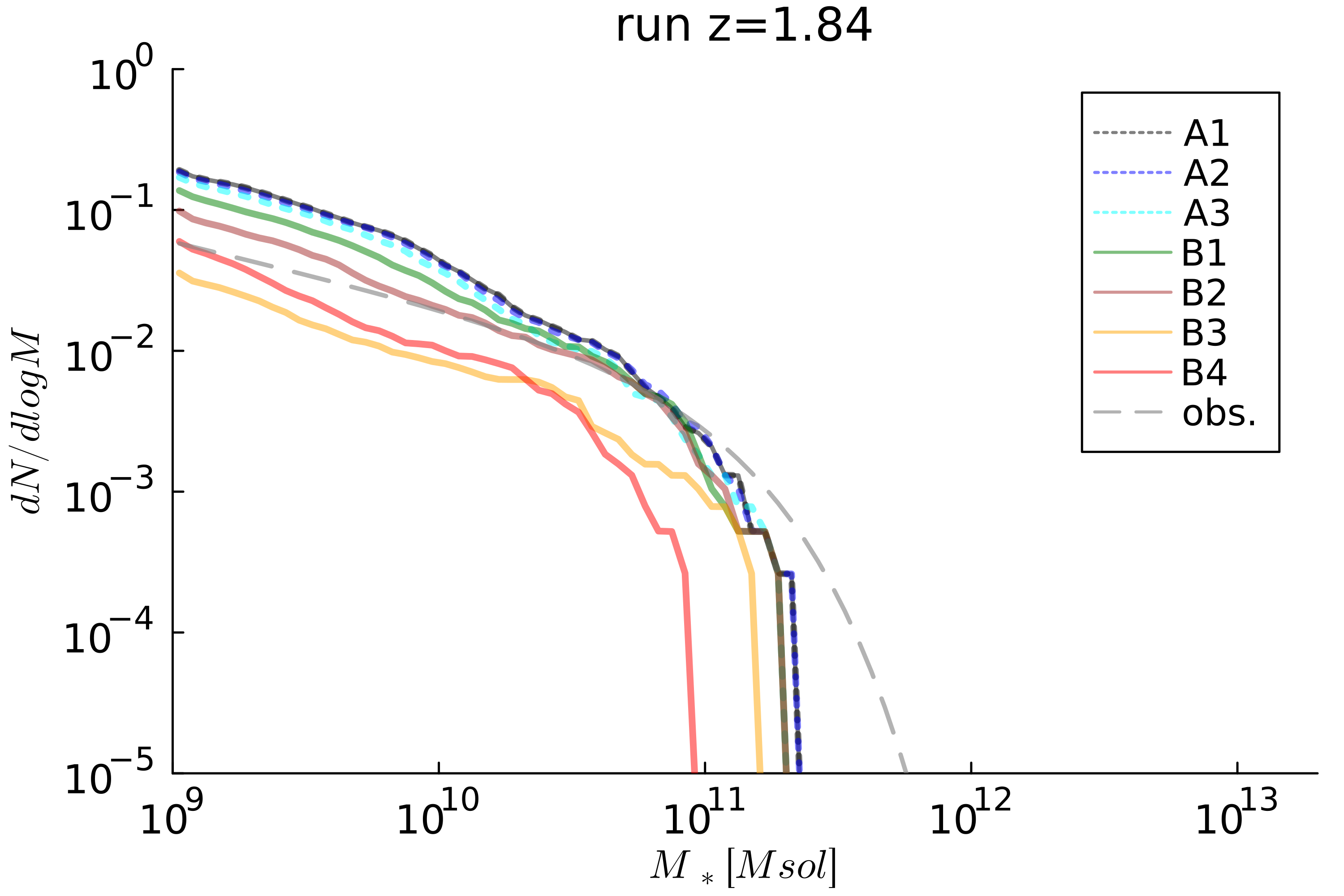}
\includegraphics[width=0.45\textwidth]{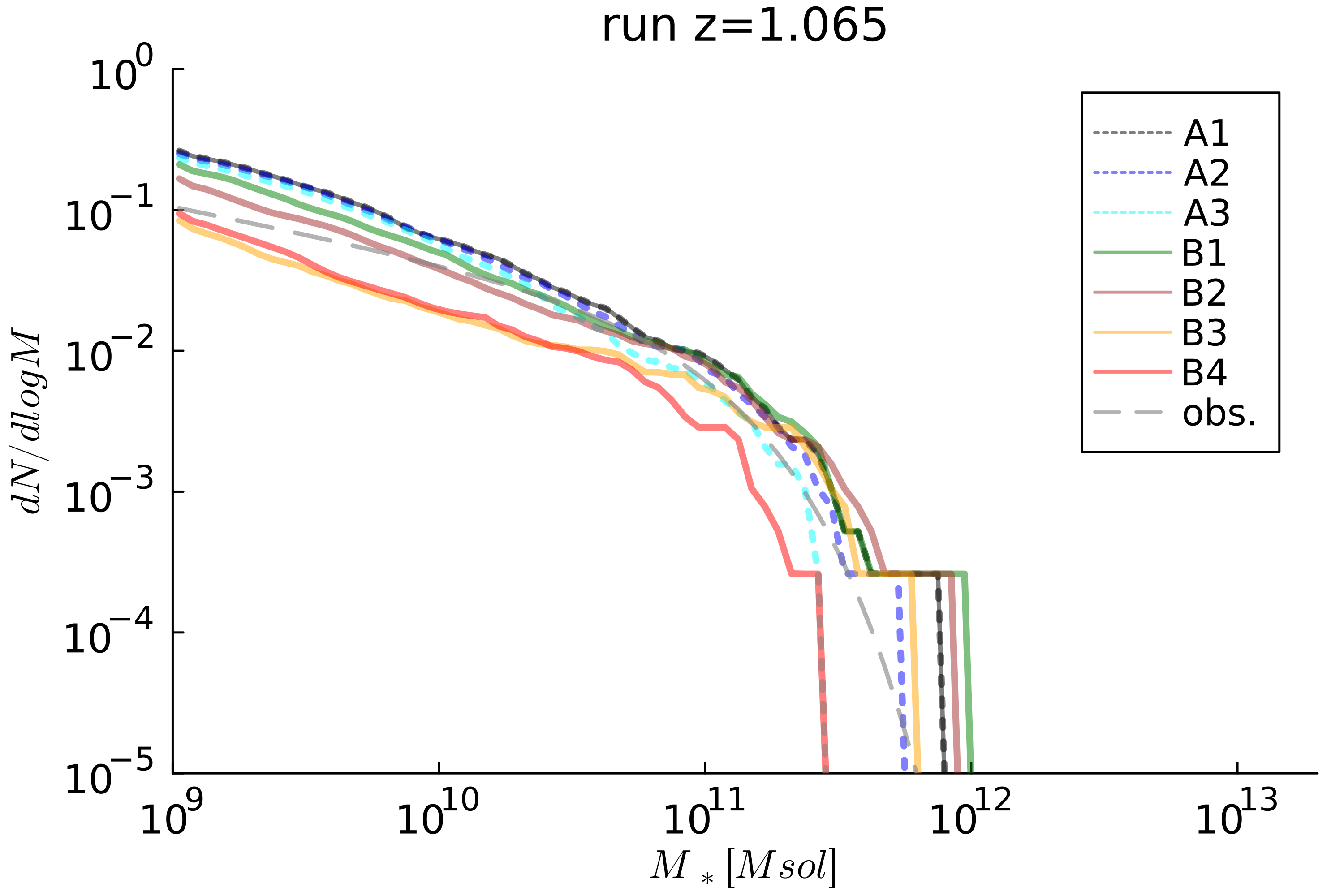}
\includegraphics[width=0.45\textwidth]{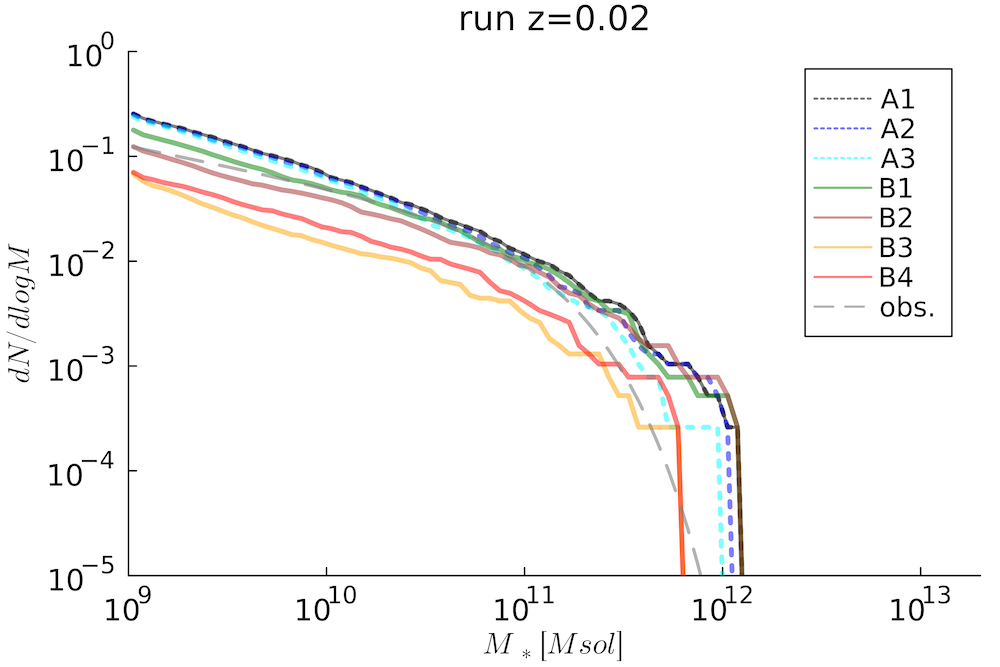} 
\includegraphics[width=0.45\textwidth]{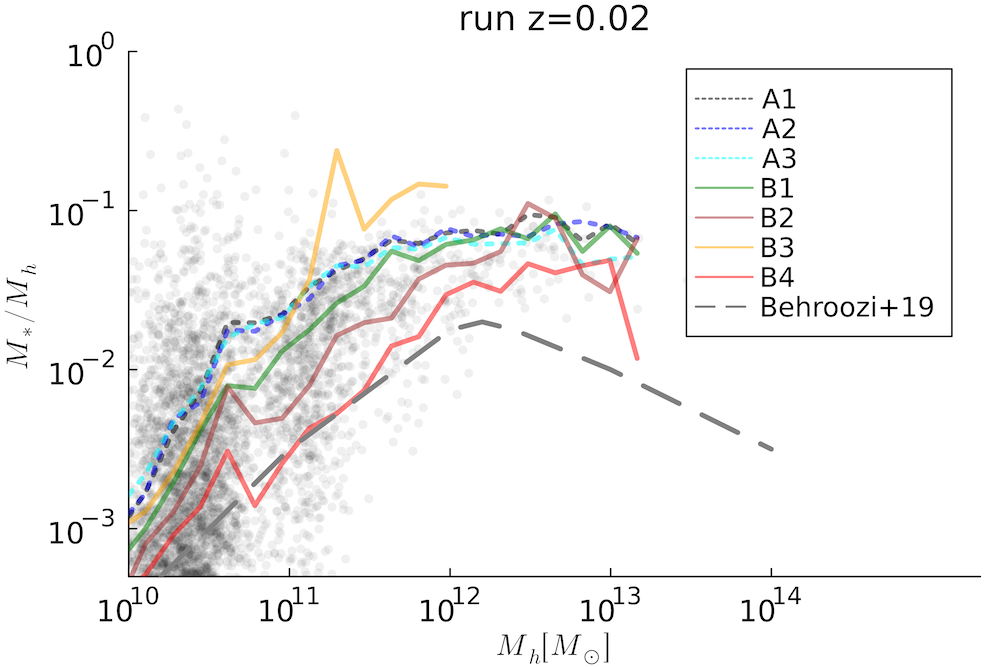}
\caption{Stellar mass luminosity function for our simulated galaxies in all models and for three different epochs, compared with the best-fit derived by \citet{2021MNRAS.503.4413M} for observations at approximately equal redshift bins. The last panel shows the distribution of the stellar mass fraction as a function of the host halo mass at $z=0.02$, compared to the best-fit relation derived from observations by \citet{2019MNRAS.488.3143B}.}
\label{fig:stellarMF}
\end{figure*}

\subsection{The injection of relativistic electrons and magnetic fields by stellar feedback} 
\label{subsec:sfr}
To couple the injection of CRe to star formation, we use the  \citet[][]{2003ApJ...590L...1K} model, implemented in  {\enzo}. Its key parameters are the minimum gas density threshold to form stars, $n_{*}$, the dynamical timescale of the star formation process, $t_{*}$, and the minimum mass, $m_{*}$, for newly formed stars. Whenever $n \geq n_{*}$, the gas is contracting ($\nabla \cdot \vec{v} < 0$), the local cooling time is smaller than the dynamical timescale ($t_{\rm cool} \leq t_*$) and the baryonic mass in the cell is larger than $m_{*}$, a new "star particle" (actually representing an entire stellar mass distribution) is formed with a mass $m_* =m_b \Delta t/t_{\rm *}$. This recipe has been proposed to reproduce the observed Kennicutt's law \citep[][]{1998ApJ...498..541K} and in our simulations can well match the observed cosmic star formation history \citep[e.g.][]{va17cqg}. Table \ref{tab:mhd} gives the parameters that overall  best reproduce the cosmic star formation history in our runs, and are meant just to give an effective sub-grid model prescription of the star formation process, averaged over the large $41.5^3 \rm ~kpc^3$ volume resolution of our runs. 
 
 The feedback from star formation depends on the assumed fractions of energy/momentum/mass ejected per each formed star particles, $E_{SN}= \epsilon_{SF} m_* c^2$, with $\epsilon_{SF}=10^{-8}-5 \cdot 10^{-8}$ as fiducial parameter, as in previous work \citep[][]{va17cqg}.
$90\%$ of the feedback energy is released in the thermal form (i.e. hot supernovae-driven winds), distributed among the 27 nearest cells around the star particle, while $10\%$ in the form of magnetic energy, assigned to dipoles during each feedback episode. 

CRe are injected into the same cells, using a fixed fraction of the local gas density, $\xi_{\rm SF} = n_{\rm CRe}/n_g$. This fraction depends on several processes:   a) the direct injection of cosmic rays by shocks driven supernova remnants and pulsar wind nebulae and b) the continuous injection of secondary CRe from the hadronic collisions between thermal protons and CR protons; c) the thermalisation of CRe due to collisional and ionisation losses in the dense interstellar medium \citep[e.g.][]{2017ApJ...847L..13P}.
{ With preliminary tests with smaller volumes at the same spatial resolution (e.g. as in the Appendix), we calibrate a reasonable value for such "macroscopic" injection efficiency to $\xi_{\rm SF}=10^{-5}$ as fiducial value, as this ensures that the synchrotron emission from star forming galaxies in the high luminosity end of our distribution is in line with observations  \citep[e.g.][]{2023MNRAS.523.6082C,2023A&A...669A...8H}. However, as we shall discuss in Sec.\ref{radiogalaxies}, our simplistic treatment of CRe from star formation cannot produce a match of the entire observed power distribution of star forming radio galaxies at different redshifts (see also Sec.\ref{sec:discussion} for further discussion).}

The { last} column in Fig.~\ref{fig:map1} gives the evolving spatial distribution of CRe injected by star formation for our { B4 model}, at three different epochs. The distribution of CRe injected by this mechanism has a low filling factor, and has overall a more patchy distribution compared to the CRe injected by the other two mechanisms. This is because the sources of the stellar feedback are more than sources of CRe from active AGN, considering that stellar feedback is prominent in the majority of low-mass galaxies, which dominate the galaxy distribution.

\begin{figure*}
\includegraphics[width=0.33\textwidth]{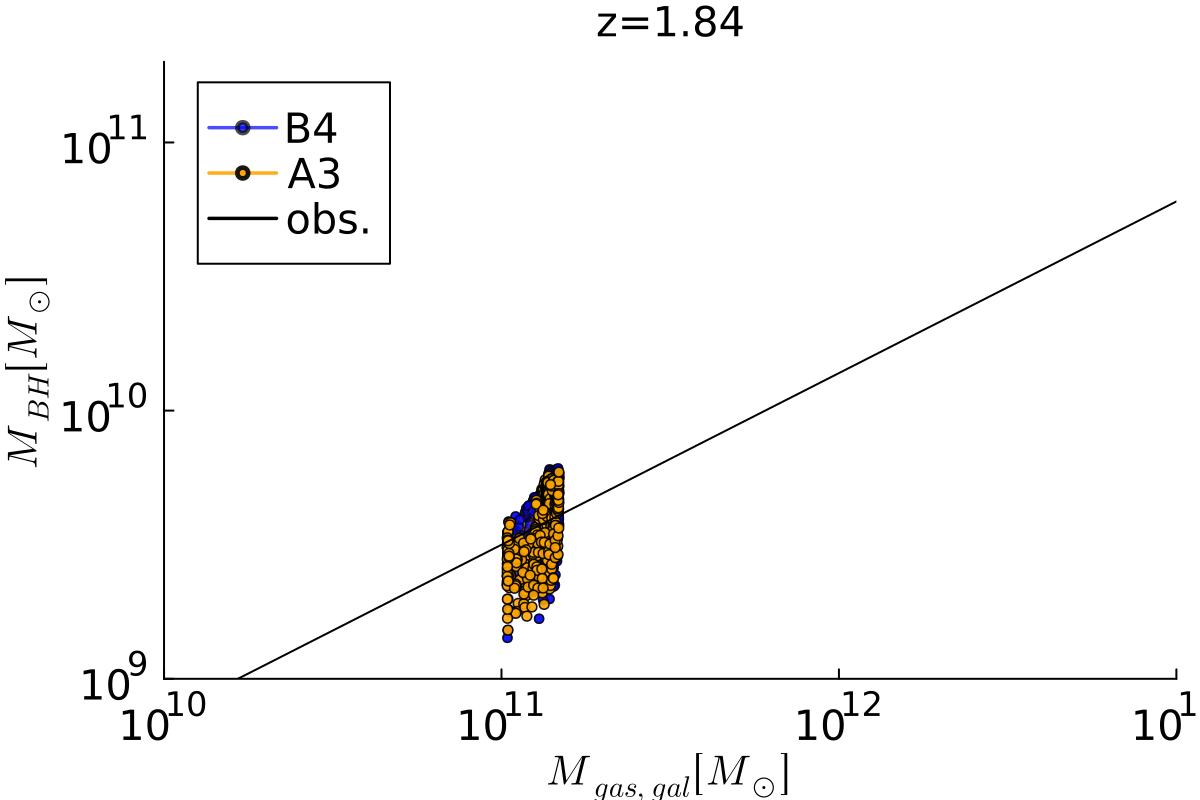}
\includegraphics[width=0.33\textwidth]{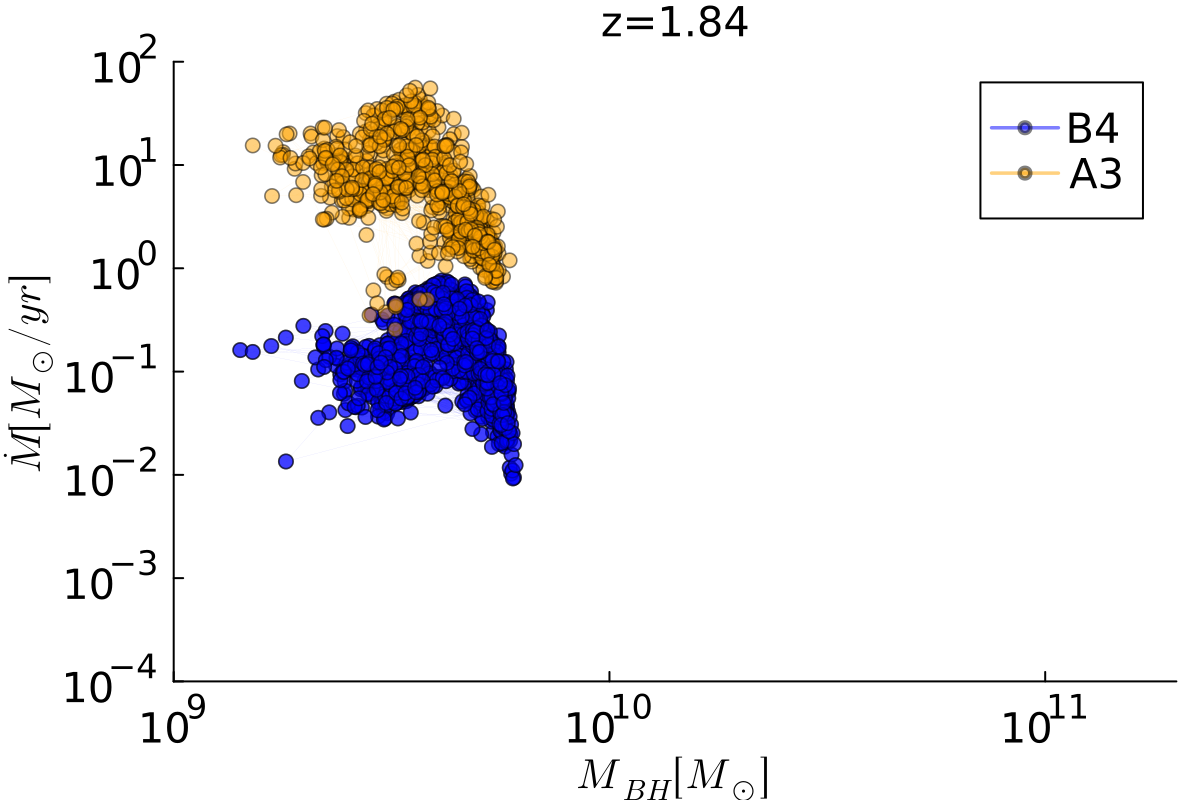}
\includegraphics[width=0.33\textwidth]{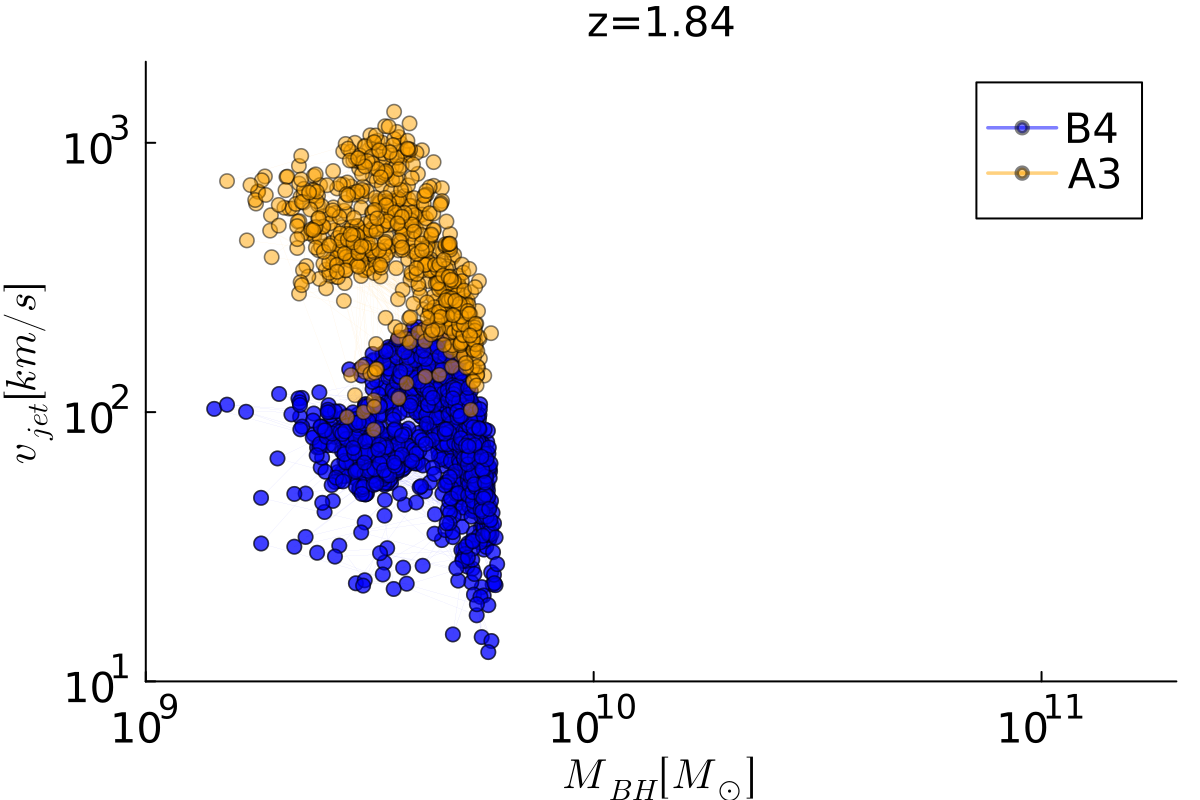}
\includegraphics[width=0.33\textwidth]{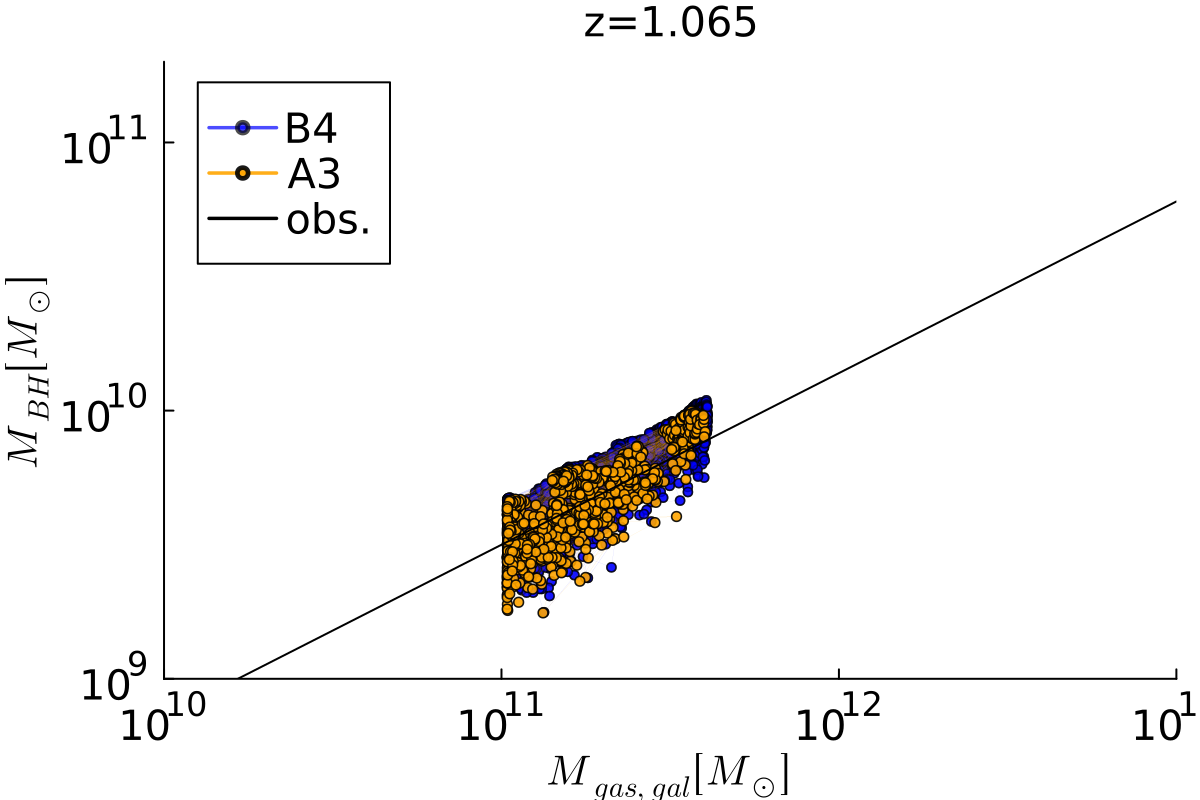}
\includegraphics[width=0.33\textwidth]{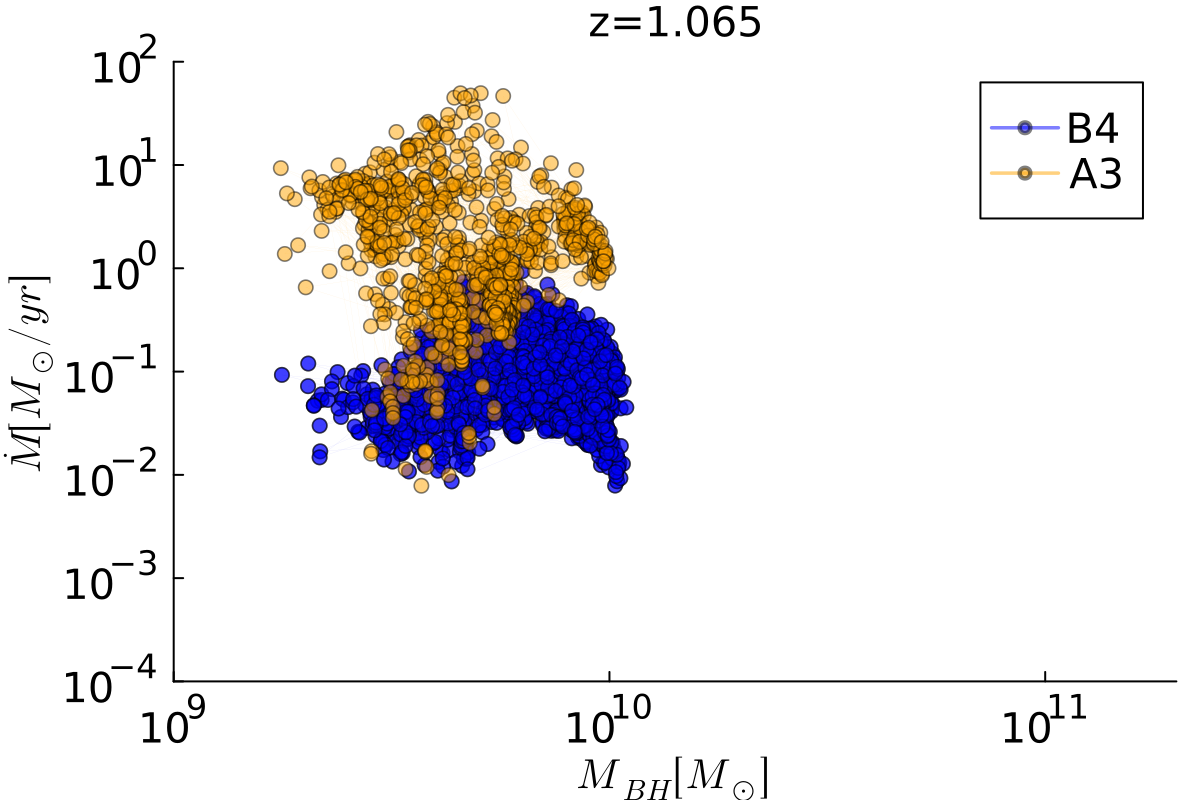}
\includegraphics[width=0.33\textwidth]{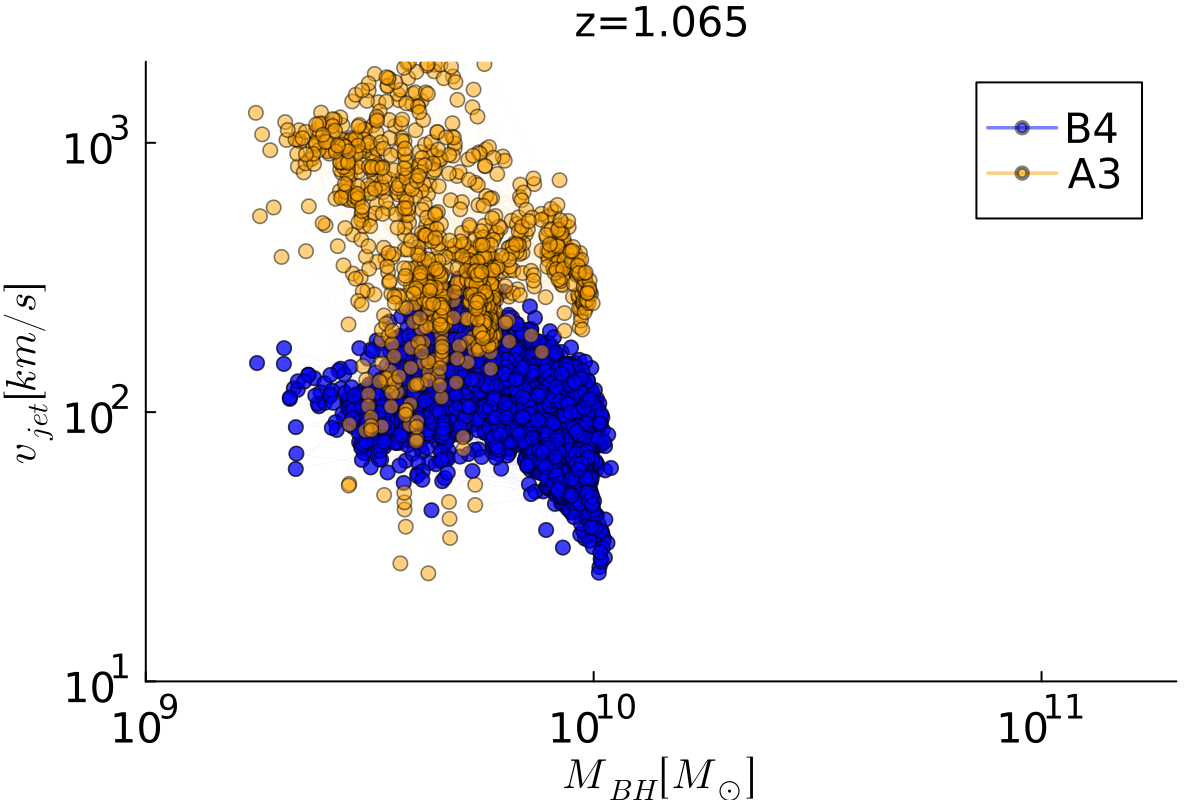}
\includegraphics[width=0.33\textwidth]{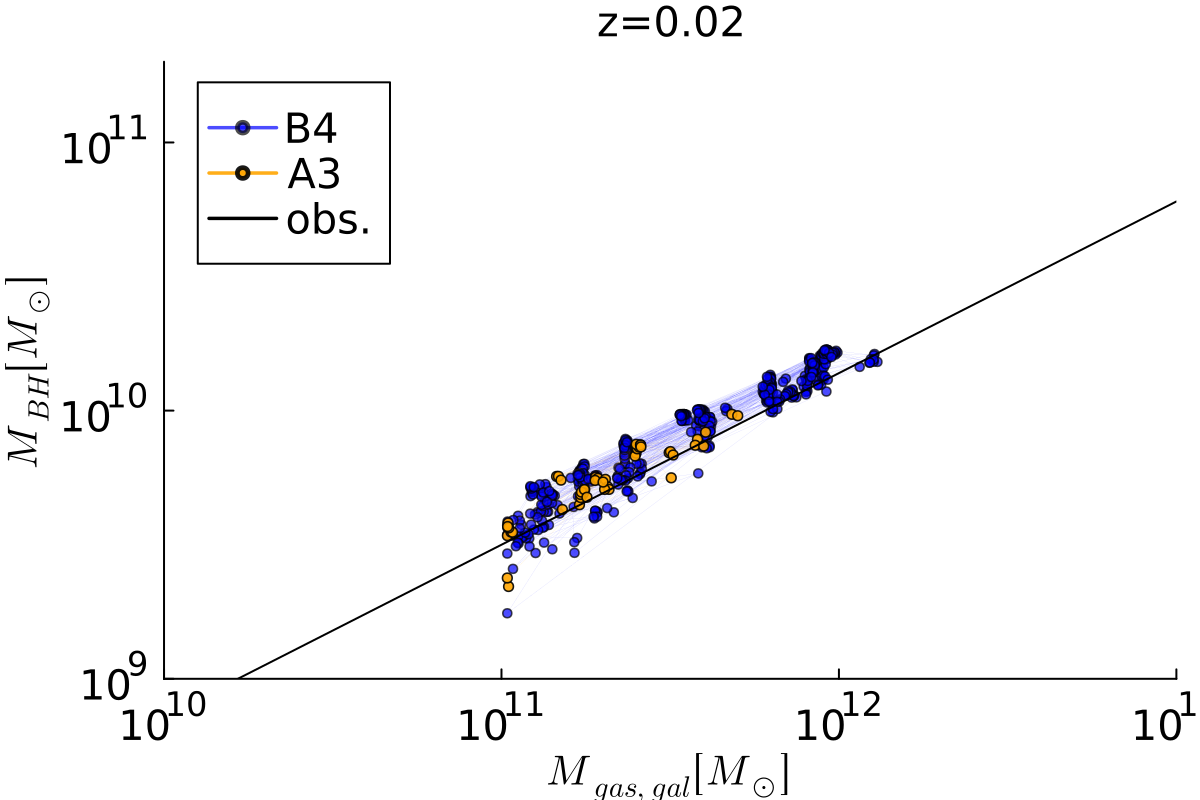}
\includegraphics[width=0.33\textwidth]{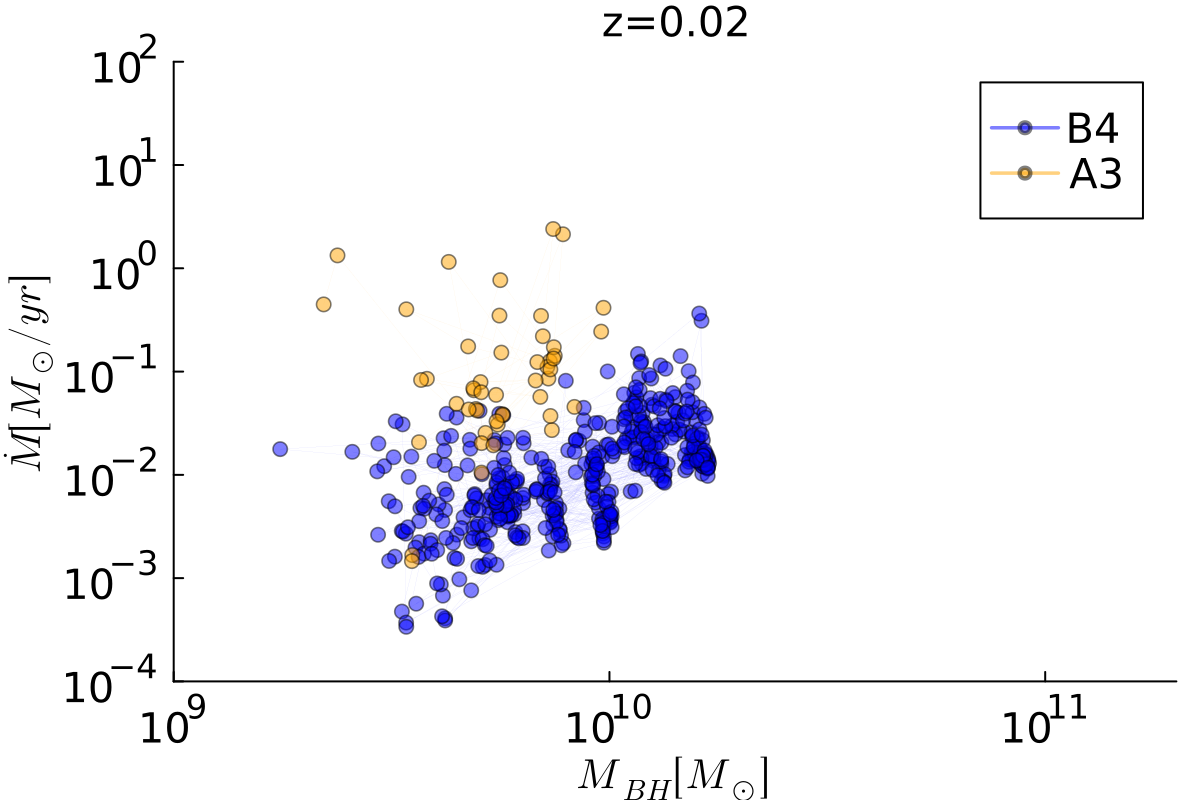}
\includegraphics[width=0.33\textwidth]{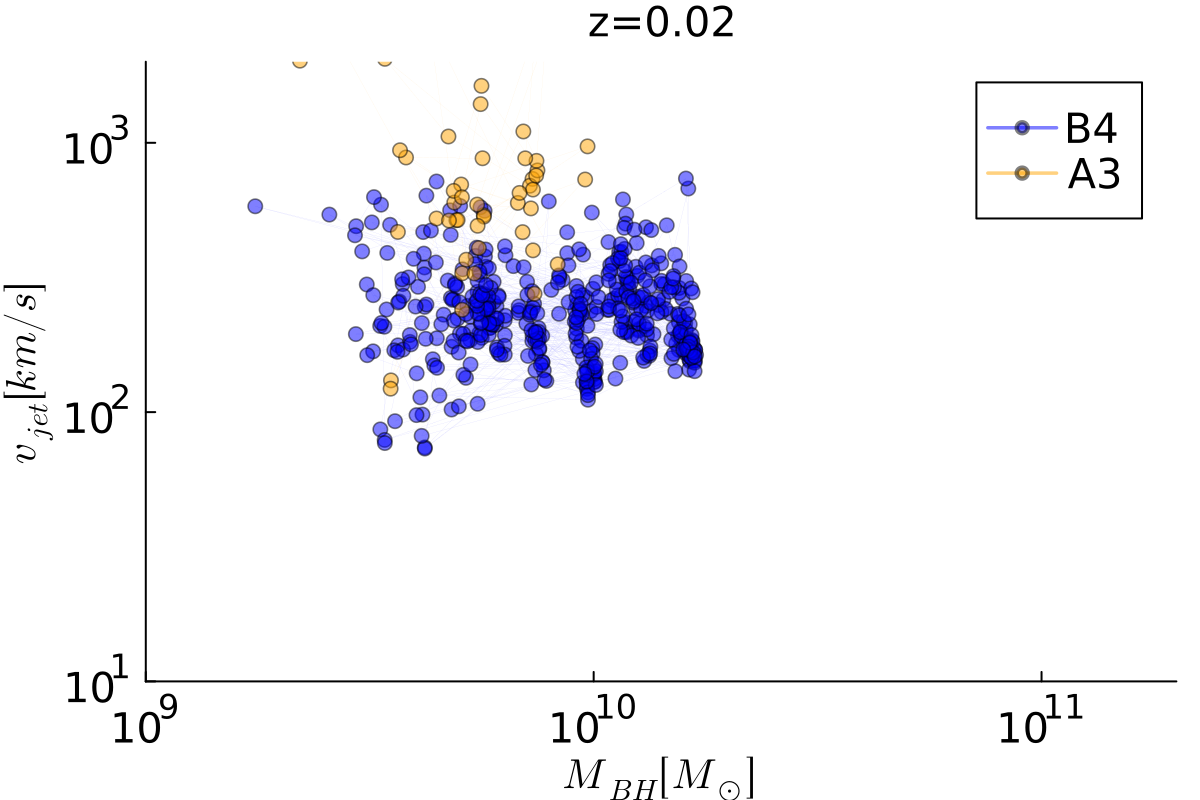}
\caption{Simulated scaling relations for the supermassive black holes modelled at run-time in our A3 and B4 runs, at three different epochs. The first column show the relation between the gas mass of the host galaxy and the SMBH mass, 
compared with the scaling relation inferred from observations by \citet{2019ApJ...884..169G}. The second column shows the relation between the SMBH mass and the mass accretion rate onto the SMBH; the third column gives the relation between the SMBH mass and the jet velocity associated with feedback events.}
\label{fig:mBH}
\end{figure*}

\begin{figure}
\includegraphics[width=0.45\textwidth]{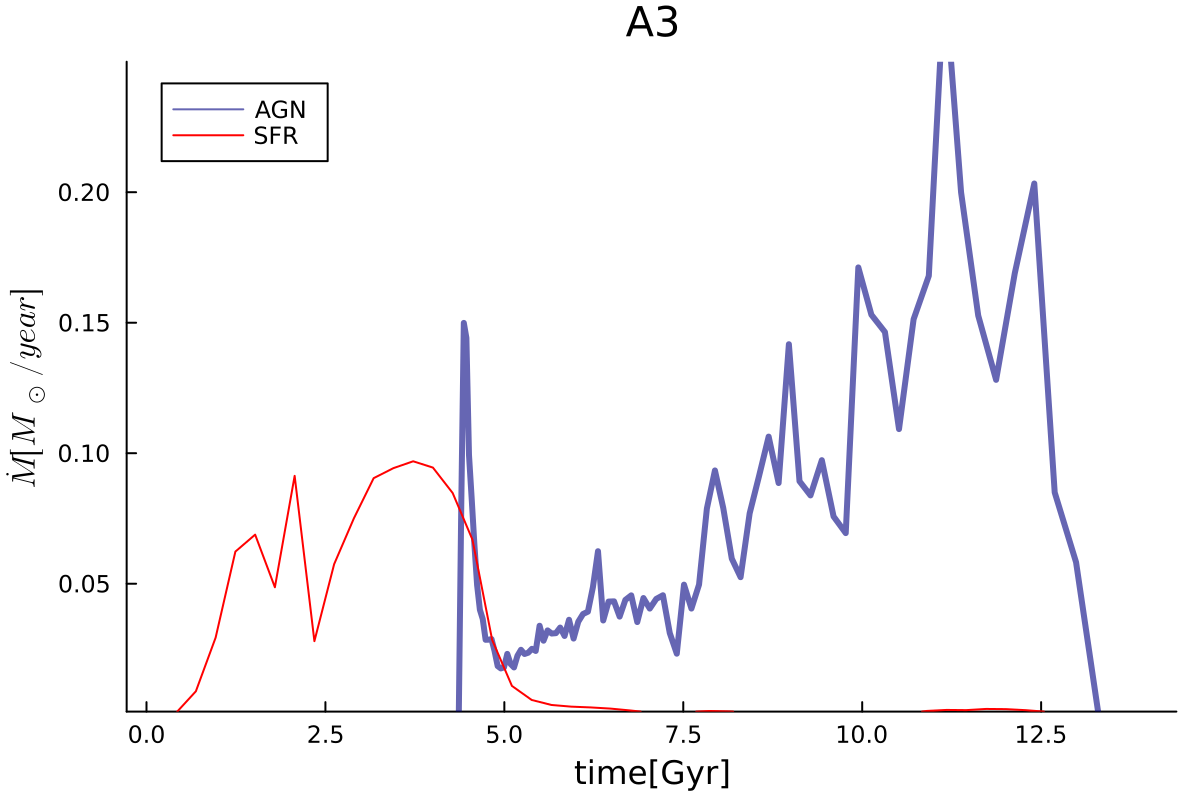}
\includegraphics[width=0.45\textwidth]{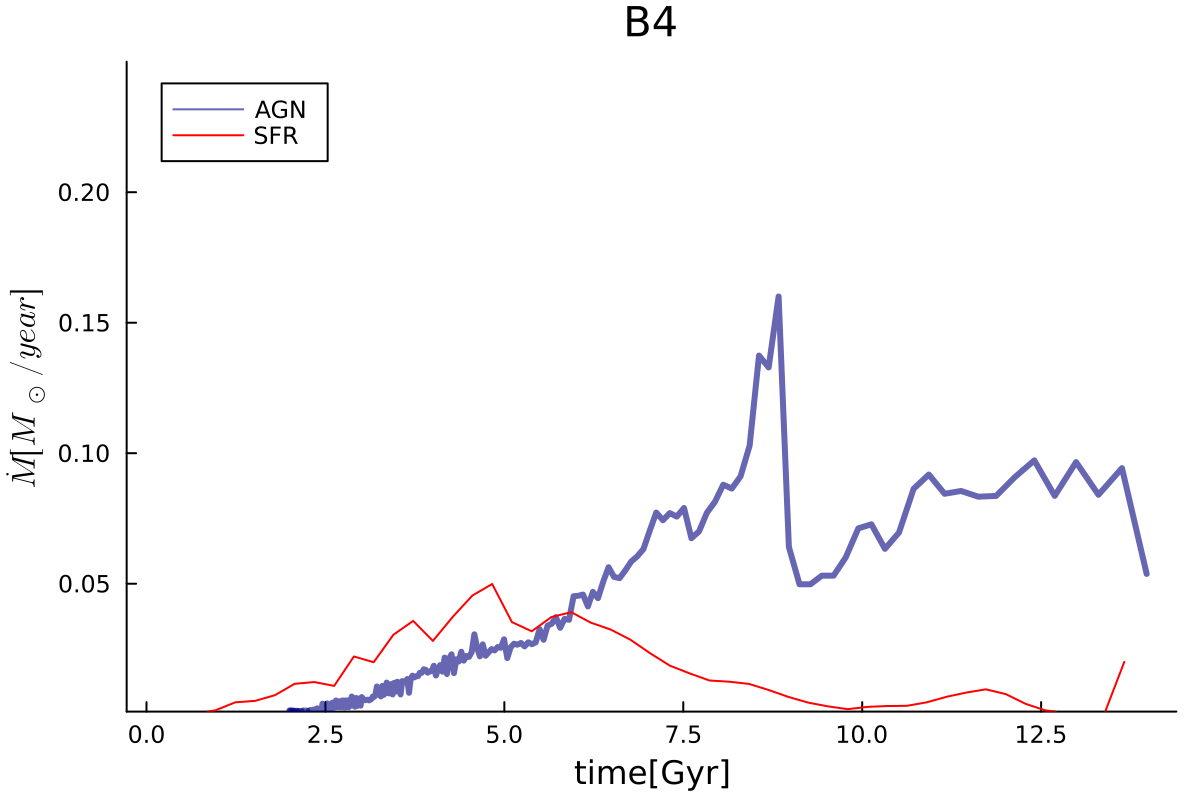}
\caption{Evolution with cosmic time of the matter accretion rate onto star forming particles (red lines) and of the matter accretion rate onto AGN (blue lines) for our A3 and B4 models.}
\label{fig:AGN_SF}
\end{figure}

\subsection{Synchrotron emission}
\label{subsec:sync}
While the radio emission from cosmic shocks can be computed based on the prompt acceleration of relativistic particles  \citep[e.g.][]{hb07}, here the presence of multiple injections from different processes require a more sophisticated treatment. 

Ideally one should continuously update the CRe spectra under the effect of cooling and acceleration processes \citep[e.g.][]{sa99}; however, this approach is presently only doable for a few objects at a time, using Lagrangian tracers injected in Eulerian simulations \citep[e.g.][]{2013ApJ...762...78Z,wi17,va21jets} or limited to the same number of gas resolution elements, like in Smoother Particle Hydrodynamics \citep[e.g.][]{boss,2023ApJ...957L..16B}.

Here we explored a different approach,  which has the advantage of  allowing us to keep track of the time elapsed since the last injection of CRs in each simulated cell. In a nutshell, we applied pre-tabulated CRe spectra to predict the momentum distribution of CRe in all cells, based on their local value of density, magnetic fields, redshift and time since last injection of CRe, and use these spectra to predict the synchrotron radio emission. 
For the tabulated spectra,  we used  the ROGER Fokker-Planck solver \footnote{\url{https://github.com/FrancoVazza/JULIA/tree/master/ROGER}} developed by \citet{va23} to produce a grid of template electron spectra and of their synchrotron emission spectra, in which we scanned 
a large range of discrete possible values of the magnetic field ($10^{-9} \rm G \leq |B| \leq 10^{-5} \rm G$), gas density ($10^{-5} \leq n \leq 10^{-2} \rm part/cm^3$), redshift ($0 \leq z \leq 6$) and age ($0.01 \rm Gyr \leq t_{\rm age} \leq 10 \rm Gyr$) since the injection. 
A total of  $276,480$ possible combinations of the above parameters, using $10^2$ momentum bins equally spaced in momentum bins and with up to $100$ subcycling timestep, was used to produce a large library of 
momentum spectra for CRe, and for their synchrotron emission spectra (in the $50 \rm ~MHz-5 \rm ~GHz$ range).
For every snapshot in which radio emission is needed, we selected for every cell in the simulation the  momentum and synchrotron spectra relative to the closest bin in density, magnetic field, age and redshift, and we normalise the CRe spectra and the synchrotron emission based on the actual CRe number density in the {\enzo} simulation.  The same procedure is repeated separately for the three 
CRe fluid fields, which allow us to  compute, with reasonable approximation, the radio emission produced by all CRe in the simulation.

Some caveat on this procedure are worth noticing.
First, it is reasonable to compute the emission from our CRe at a given epoch as long as the physical condition in the given cell did not change much since the first injection of CRe. In the presence of fast advection of CRe and/or of CRe injected since a long time, our solution for the synchrotron emission is likely to overestimate the true emission, as strong adiabatic losses, or the effect of stronger magnetic fields in the past, are underestimated. Our approach also neglects the role of shock and turbulent re-acceleration processes, which on the other hand may further stretch the high energy part of the CRe energy distribution on long timescales \citep[e.g.][]{beduzzi23}. 

In any case, CRe undergoing cooling losses converge onto the same evolving distribution for a large range of initial different parameters: after a first rapid ($\leq 0.1 \rm ~Gyr$) cooling stage { which mostly affects} the highest energy tail of their distribution (usually $\gamma \sim 10^4$), the rest of the distribution settles to a slow
evolution on long ($\sim 1-10 \rm ~Gyr$) timescales, since collisional or adiabatic losses are negligible for the majority of the gas density values occupied by CRe \citep[e.g.][]{sa99}.   Therefore, our approach gives an accurate answer only as long as the evolution of CRe is either dominated by recent injections (e.g. prompt shock emission or radio jets), or else for very long timescales during which loss processes dominate over re-acceleration terms.

\section{Results}
\label{sec::res}

We divide our results in two main parts: first, a validation part where we show how our modelling compares with the basic known properties of galaxies; and second part, where we highlight the important aspects of magnetic fields and cosmic ray electrons evolution, which can be studied for the first time with our new simulations.

\subsection{Properties of galaxies}
\label{subsec:galaxies}
The evolution of the cosmic star formation rate density (SFR), i.e. the rate of star formation per unit time and volume, is given in Fig. \ref{fig:SFR}. This is the first global statistics to assess our subgrid implementation of star formation physics.
While we show here only the best results from our implementations and discard all failed attempts, we notice that even small parameter variations (i.e. the minimum mass of the formed star particles, the strength of the stellar feedback and the combination with different choices for the AGN feedback model) are found to significantly alter the total star formation across time. The simulated SFR is compared with the collection of observational data by \citet[][]{2014ARA&A..52..415M}. 

As a noticeable difference, the first three models (A1,A2,A3) show a drop of the cosmic SFR at $\sim 5 \rm ~Gyr$ since the start of the simulation ($z \sim 1.3$), while the other { four (B1, B2, B3 and B4) show a more prolonged star formation rate; in particular, models B3 and B4 roughly match also the star formation rate in the local Universe.
We deem model B4 to be the best performing here, because compared to B3 it can also best reproduce the peak of the cosmic star formation history at $t \sim 3 \rm ~Gyr$ ($z \sim 2$).}  
{ The above differences can be understood because of the largest threshold mass used for star formation in the last four models, and the longer assumed dynamical time for star formation in the latest two in particular (B3, B4). As a result, less cold gas is consumed at high redshift, yielding a more realistic balance between star formation and the circulation of cold and hot baryons until the end of the run).}

The last two models (C1 and C2) are not showed here for clarity, as in all galaxy properties they can just overlapped to model B4, as they employed exactly the same galaxy formation physics. 
It should be noted that our volume ($42.5^3 \rm Mpc^3$) is still too small to be entirely free from cosmic variance effects, which can be dominated by the rare tail of high mass galaxies. 
In any case, the inefficient star formation in the low-z part of our simulated evolution is a deficiency in our models, which is likely to underestimate the amount of CRs injection from star formation in the local Universe. On the other hand, considering that the peak of the cosmic star formation rate is well captured in all these runs, and that the overall injection of CRs from galaxies is dominated by AGN feedback during most of the simulated evolution, we consider the latter an acceptable failure of our model, to be addressed in future work. 

Next, we compute the distribution of the stellar mass and of the stellar mass fraction in all halos identified in the simulation, which is a more stringent test for the relative effects of star formation and stellar/AGN feedback across the full mass range of our galaxy distribution.  The first three panels of Fig.~\ref{fig:stellarMF} show the galaxy stellar mass distribution at three different epochs ($z=1.84$, $z=1.065$ and $z=0.02$), which are compared to the corresponding best-fit relation obtained by analysing the Hubble Space Telescope observations in similar redshift bins by \citet{2021MNRAS.503.4413M}.
 The more prolonged star formation in B- runs implies an increased amount of halos with a large stellar mass compared to the first three models, and in general all models at $z=0.02$ show a slight excess of massive galaxies compared to the best fit of observations \citep[e.g.][]{2019MNRAS.488.3143B}.
This is better shown by the last panel, which gives the final distribution of the stellar mass fraction as a function of the halo mass at $z=0.02$ for our galaxies. All models have the tendency of overproducing the stellar mass fraction, although { our "favoured" model B4} has this problem only at the high mass end of the distribution.  
We notice that similar trends with halo masses are found also by recent simulations adopting more sophisticated prescriptions for galaxy physics, e.g. by \citet{2024arXiv240407252S} in the SIMBA simulation. 

 Far from being an attempt to self-consistently model the process of star formation in the Universe, our approach seeks an optimal combination of numerical parameters which can reasonably reproduce the observable properties of star formation on large scales, using the physical information available at the spatial and temporal resolution of our simulations. Based on these tests, we can conclude that with the right tuning of sub-grid parameters, despite its coarse spatial resolution our model can reasonably well capture the time evolution of the global SFR, as well as distribute the stellar component across a large range of halo masses, reasonably close to reality (with the exception of very large masses). 
{ Of all investigated models, the combination of  parameters in run B4 can produce a realistic stellar mass distribution, a realistic history of cosmic star formation as well as a realistic correlation between black hole mass and halo baryonic mass. Therefore, this is the model in which our predictions for the  total magnetic energy and the total amount of CRe injected by galaxy feedback can be considered as best anchored to real galaxies.  Additional tests on the role played by our subgrid parameters for star formation and AGN physics on the simulated cosmic star formation history are discussed in the Appendix (Sec~\ref{fig:appendix2}).}

\subsection{Properties of the simulated black hole population} 

Next, we computed the scaling relations between the properties of simulated SMBH, to test the result of our implementation of SMBH evolution and growth (Sec.~\ref{subsec:bh}). 
As we will do later also with other comparisons, we focus here on two representative examples of our models, i.e. model A3 which is characterised by an intense AGN feedback activity in the past, and model { B4} which presents a much more moderate one (runs C1 and C2 presenting only negligible differences in galaxy properties compared to run B4).  

The first column in Figure \ref{fig:mBH} gives the relation between the simulated black hole mass and the gas mass of the host galaxy for three sample epochs, in which the  solid line gives the best fit of the observational trend, as reconstructed by \citet{2019ApJ...884..169G} for low-z objects, $M_{\rm BH} \propto M_g^{0.57}$. As expected given our procedure, we can recover at all time steps the theoretical relation between the two quantities, meaning that the mass associated with our simulated SMBH is realistic at all epochs. { It should be noticed that our points do not always stay exactly on top of the theoretical relation. The reason is that, in this plot, the gas mass is measured in post-processing, after the code has generated the output of one evolutionary timestep since the creation of SMBHs.
Therefore, the gas mass is measured at that location after a small time delay (typically a few Myr) during which evolutionary processes of further growth and feedback have started affecting the surrounding gas mass. }
The second column of Fig.~\ref{fig:mBH} gives the relation between the mass accretion rate onto SMBHs, and the third column shows the injected jet velocity from the same objects. 

The two simulated models clearly differ in their simulated AGN activity over time, with the A3 model being much more active for $z \geq 1$, and the { B4} producing more active SMBH (albeit with a low matter accretion rate, $\leq 1 ~M_{\odot}/\rm yr$) down to $z=0$. Such differences are entirely ascribed to the largest minimum SMBH mass  assumed in run { B4}, which resulted into an overall lower power activity by SMBH, but more prolonged in time, since a moderate amount of gas was allowed to remain closer to a typical SMBH until the end of the simulation. The impulsive vs more self-regulated AGN feedback of these two models is well captured by the distribution of $\dot{M}$; it is systematically lower at all times in the { B4 run, while in run A3  correspondingly higher jet velocities are reached, which clearly had the effect of expelling a larger fraction of gas away from halos already at high redshift.}

{ Figure \ref{fig:AGN_SF} gives the evolution of the mass accreted onto star forming regions or on AGN, for the A3 and B4 cases. In the A3 model, there is an initial very efficient stage of stellar mass accretion, followed by a sharp drop due to the effect of stellar feedback, as well as by bursty cycles of AGN feedback. In the last Gyrs of evolution, star formation is negligible while the AGN accretion is poorly self-regulated and characterized by episodic bursts. In the B4 model instead, the stellar mass accretion is more prolonged in time, and the balance between stellar and AGN feedback produces a more regular pattern of matter accretion onto AGN, as well as significant star formation also at later times.}

\begin{figure*}
\begin{center}
\includegraphics[width=0.33\textwidth]{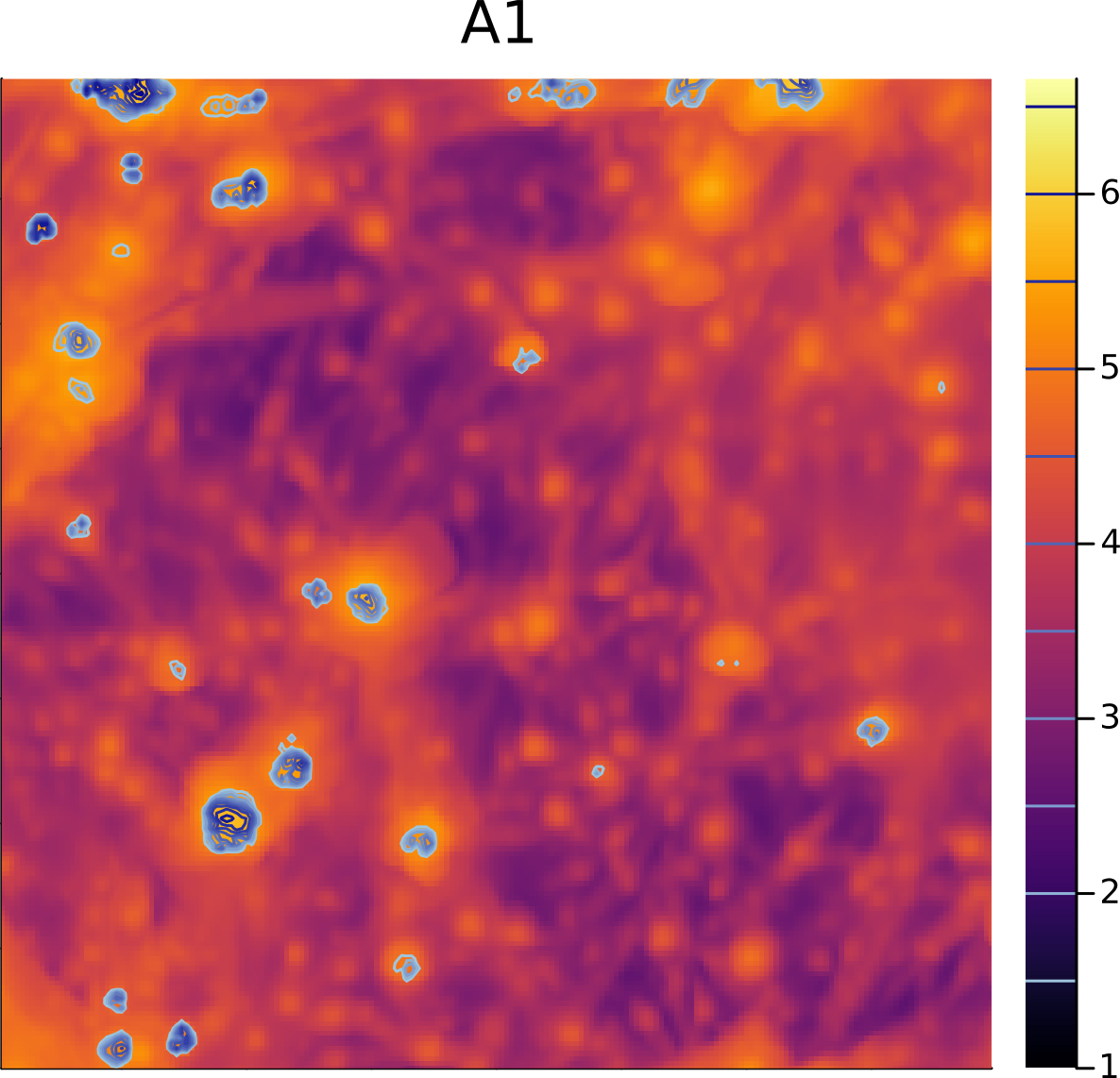}
\includegraphics[width=0.33\textwidth]{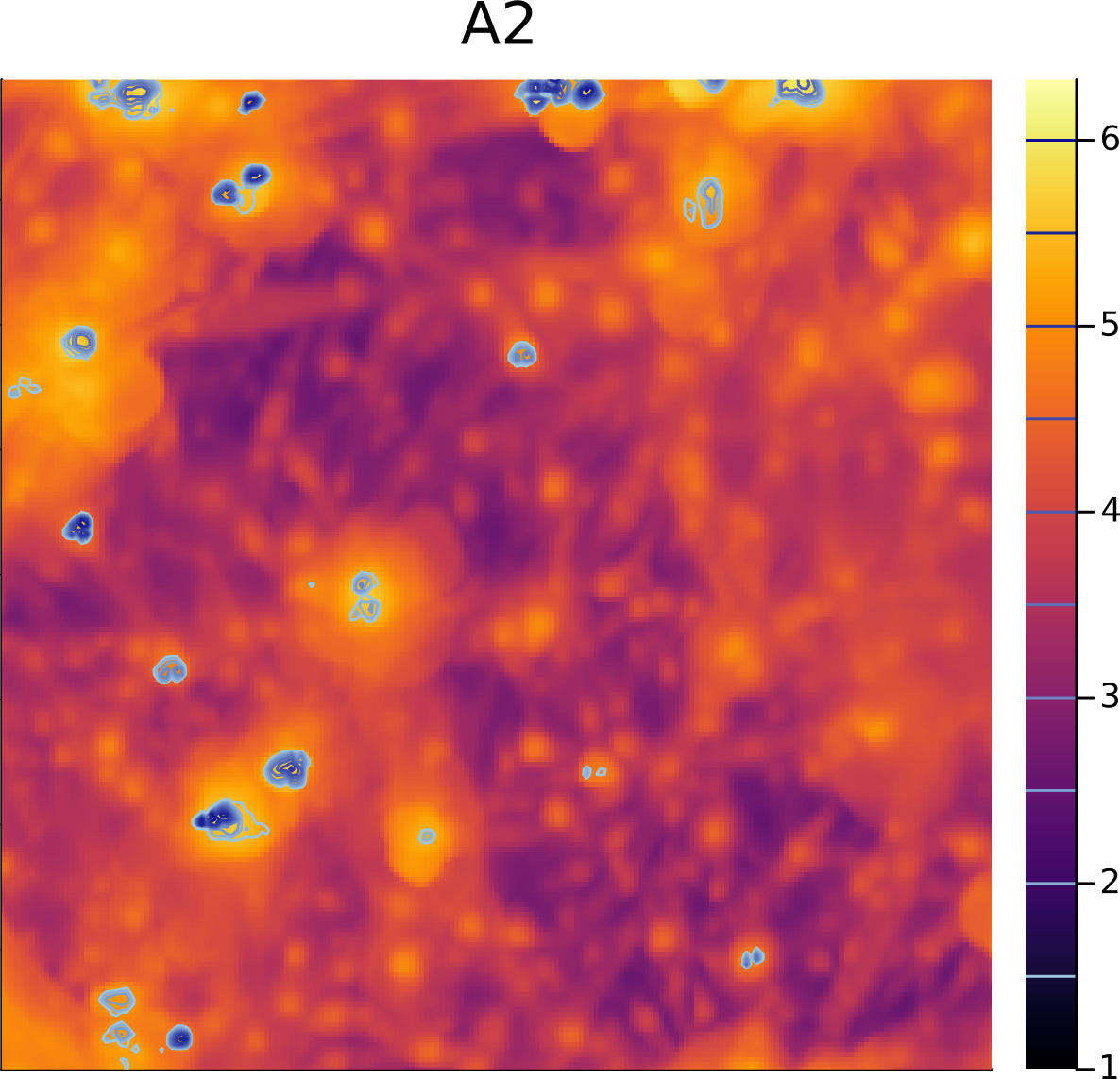}
\includegraphics[width=0.33\textwidth]{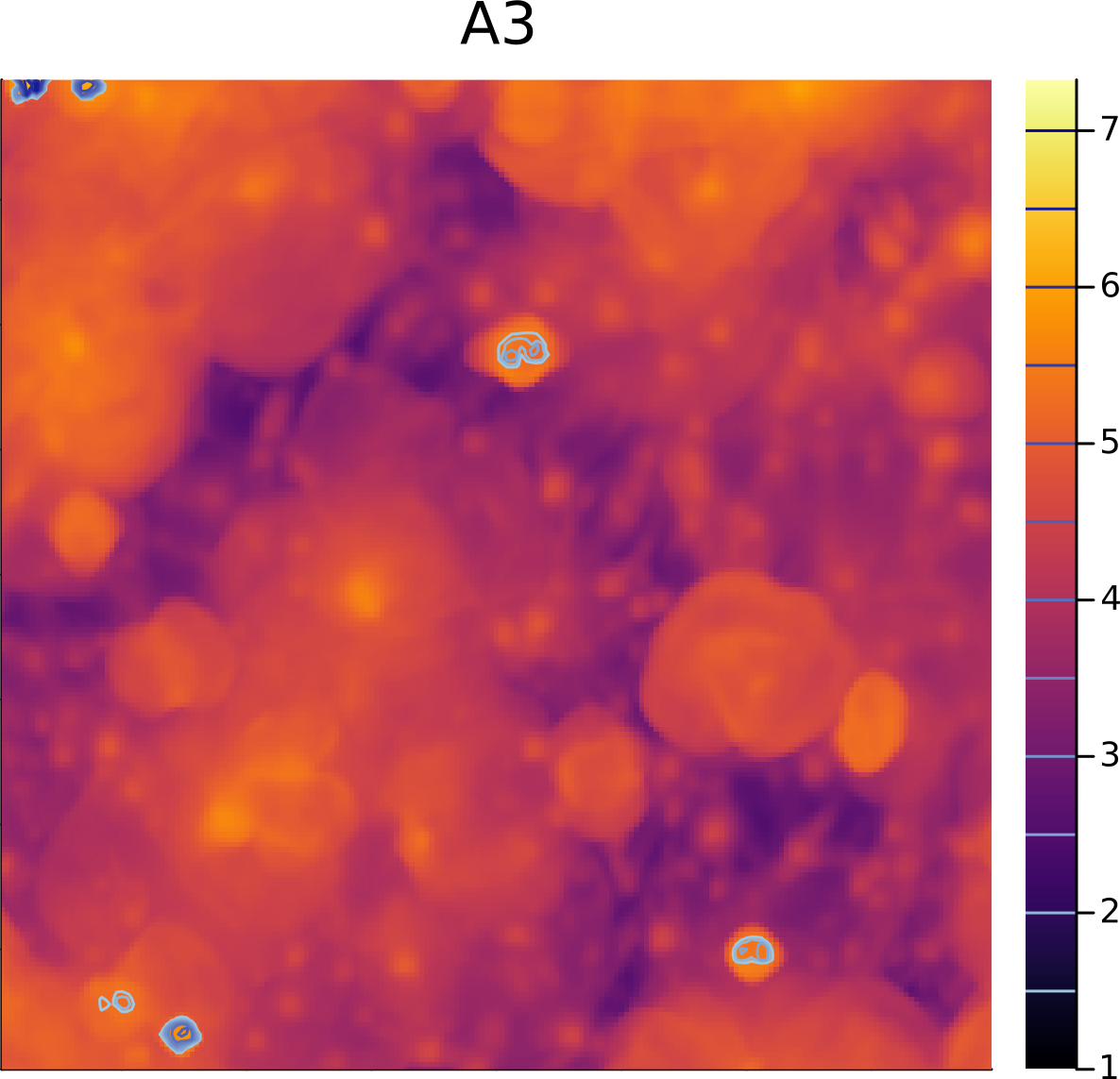}
\includegraphics[width=0.33\textwidth]{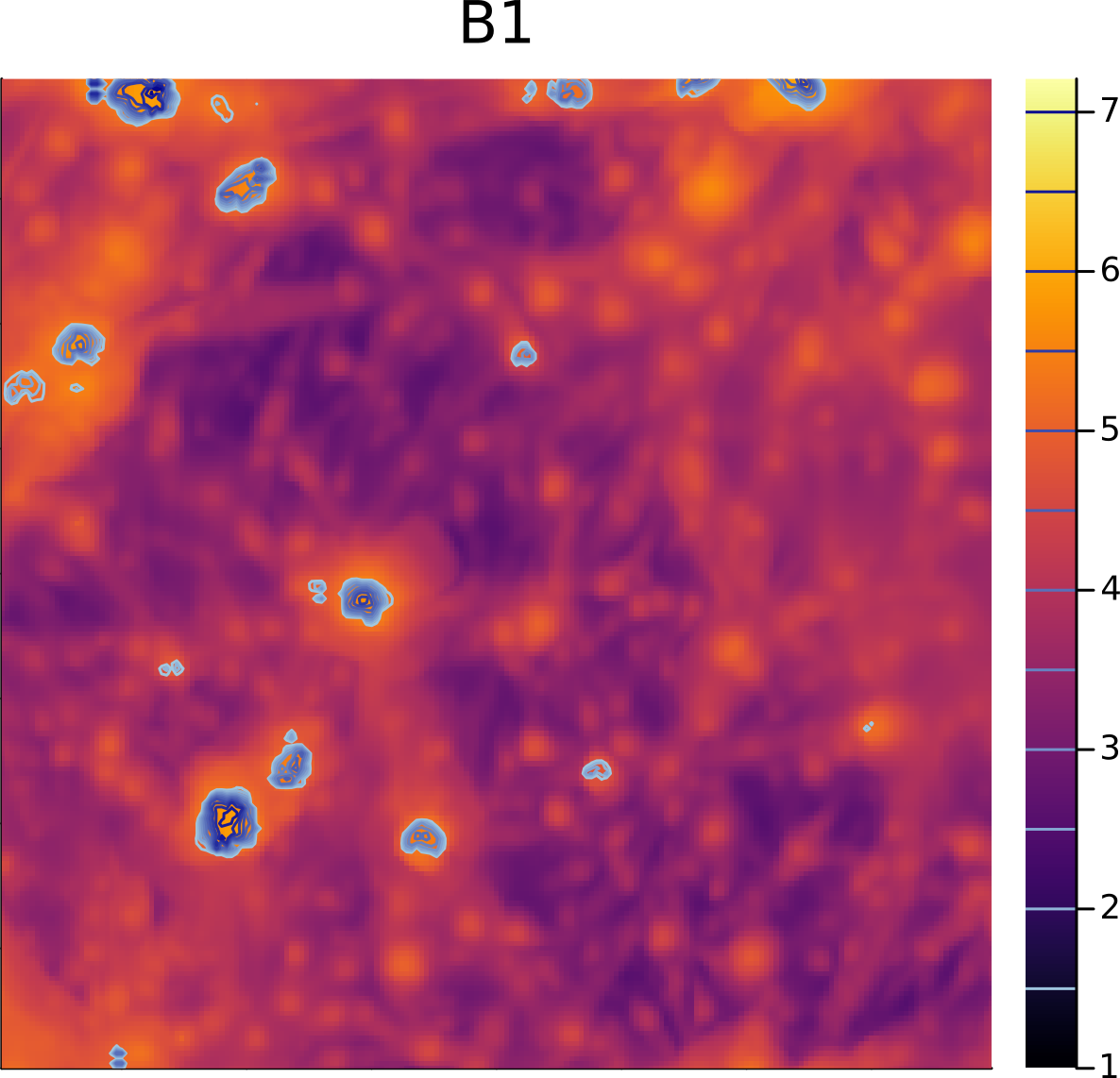}
\includegraphics[width=0.33\textwidth]{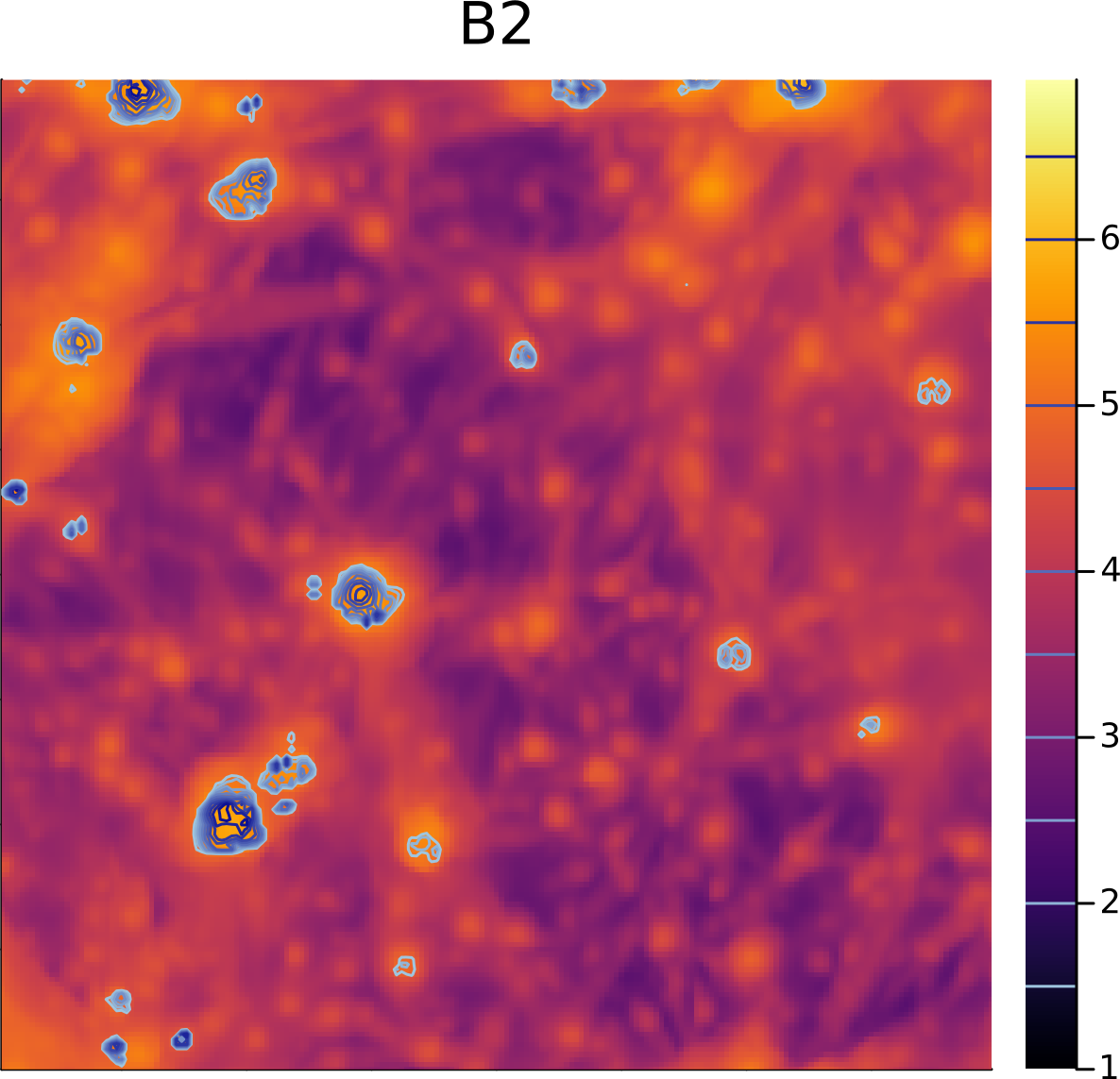}
\includegraphics[width=0.33\textwidth]{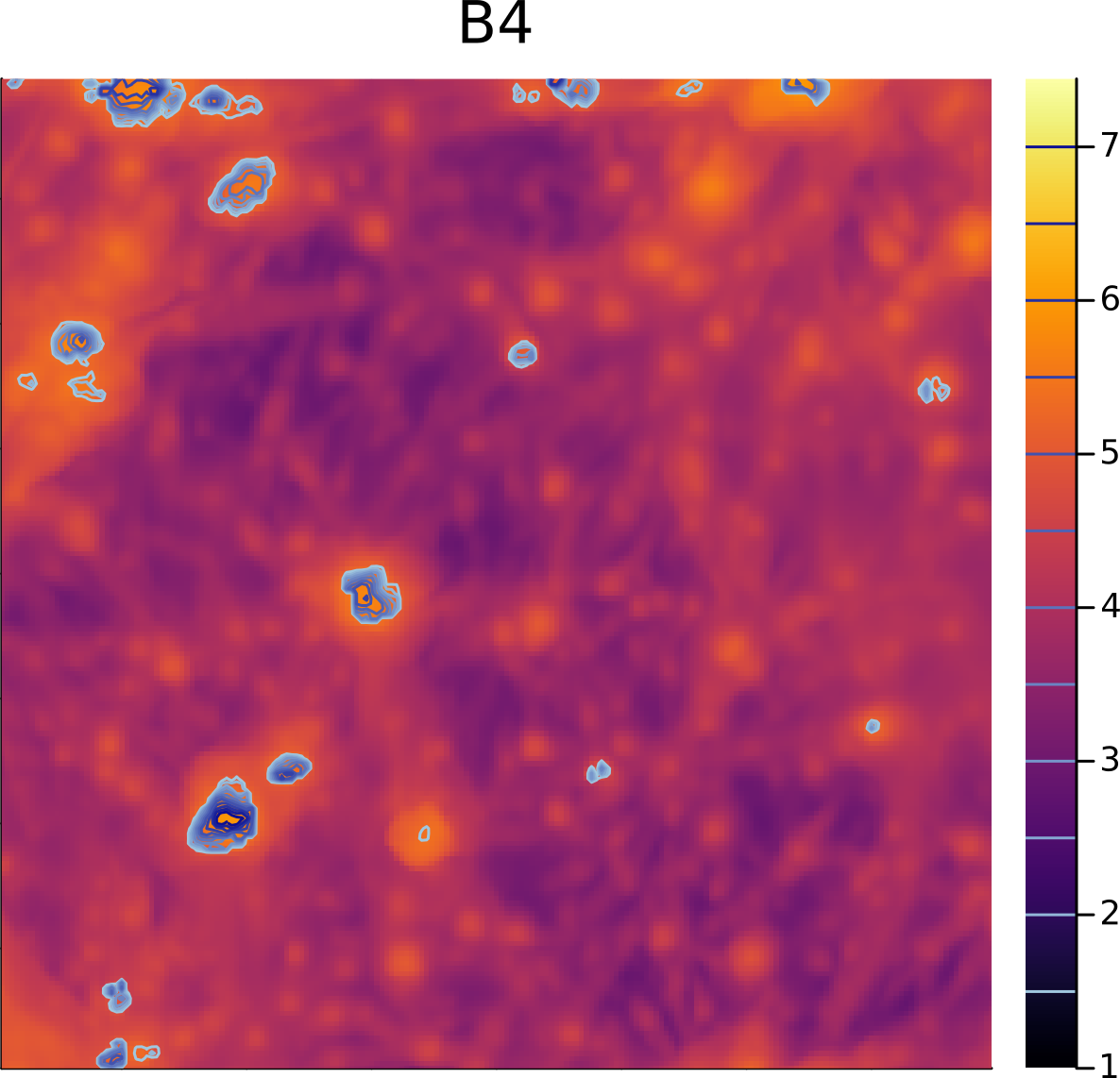}
\caption{Projected mass weighted gas temperature (colors, in units of $\rm log_{\rm 10}[K]$) and over imposed contours of detectable synchrotron radio emission at 150MHz from CRe injected by AGN 
in a subselection of our volume (about $10 \rm ~Mpc$ across) in six runs at  $z \approx 1.0$.}
\label{fig:map_radiogals}
\end{center}
\end{figure*}

\subsection{Radio emission from galaxies} 
\label{radiogalaxies}

We extracted the distribution of radio power emitted from our simulated radio sources, by generating their integrated emission at different epochs, based on the normalisation of the CRe density and of the local magnetic field strength, as well as on the average age since the injection of CRe in each cells (Sec.~\ref{subsec:sync}), and then by computing the emission within the projected area covered by the $R_{500}$  \footnote{$R_{500}$ is the radius enclosing a total matter overdensity $500$ times larger than the critical cosmic one, amd $M_{500}$ is the corresponding total enclosed mass.} of each galaxy identified in the simulation. Since CRe injected by AGN or by star formation are differently tracked, we computed the emission produced by the two mechanisms separately, which further allows us to roughly compare with the distribution of emission from star forming and radio galaxies.

Figure \ref{fig:map_radiogals} gives the example of a  crowded fields in our simulations at $z \approx 1$, in runs A1,A2,A3,B1,B2 { and B4 (B3 being extremely similar to B4)}. The contours mark the radio detectable emission from the CRe injected by AGN at 
 150MHz, assuming detection being larger than $3 \sigma_{\rm noise}$ where $\sigma_{\rm noise}=1.05 \mu Jy/\rm arcsec^2$ is of the order of the sensitivity of LOFAR for the LOTSS survey for a $\theta=25"$ beam. The colors give instead the { mass-weighted gas temperature for the same volume.} 
Several double lobe objects are visible in the image, often formed at the same location in all runs but with morphological differences (both in power and jets orientation) due to the stochasticity of the gas accretion events and launching jet directions from the simulated SMBH. 
{ By knowing the typical time elapsed since the injection by AGN of the CRe emitting in each cell, as allowed by our method (see e.g. Fig.\ref{fig:mapAGE1}), we observe that all our detectable radio sources are "young", i.e. $\leq 0.1-0.2 \rm Gyr$. The latter thus represents the typical duty cycle of our simulated radio sources at low redshift, which is in good agreement with the typical duty cycle inferred from recent observations of radio galaxies in galaxy clusters and groups \citep[e.g.][]{2019A&A...622A..17S,2021biava,brienza22}.}

The basic feature of a double lobed radio emission, loosely resembling Fanaroff-Riley type II radio galaxies in the cosmic web, is correctly captured by our implementation. However,  the small-scale details and morphology are obviously poorly captured by the coarse spatial resolution of our runs. 

Limited to the radio emission from CRe injected by star formation (only relevant in the faint luminosity end of the distribution)  we introduced an ad-hoc fix to our procedure, to account for the fact that large fraction (i.e. from $60$ to $80\%$) of the radio emission from galaxy halos is (likely) dominated by "fresh" secondary electrons from the hadronic collision process \citep[e.g.][]{2021MNRAS.508.4072W}. In reality, the continuous injection of secondary electrons well accounts for the observed flat distribution of radio spectral indices ($\alpha_R \leq 1.0$ where $I(\nu) \propto \nu^{-\alpha_R}$ is the emitted radio power).  In our case instead, the absence of the continuous injection of secondary CRe makes the population of CRe injected by star formation events dominated by 
$z \geq 1$ events, which results into long average age for the CRe in the cells, and consequently too steep spectra associated with this population in particular.
As a simple fix, limited to this population we impose an upper value of $10 \rm ~Myr$ to their age, in order to compute their emission spectrum. 

\begin{figure*}

\includegraphics[width=0.33\textwidth]{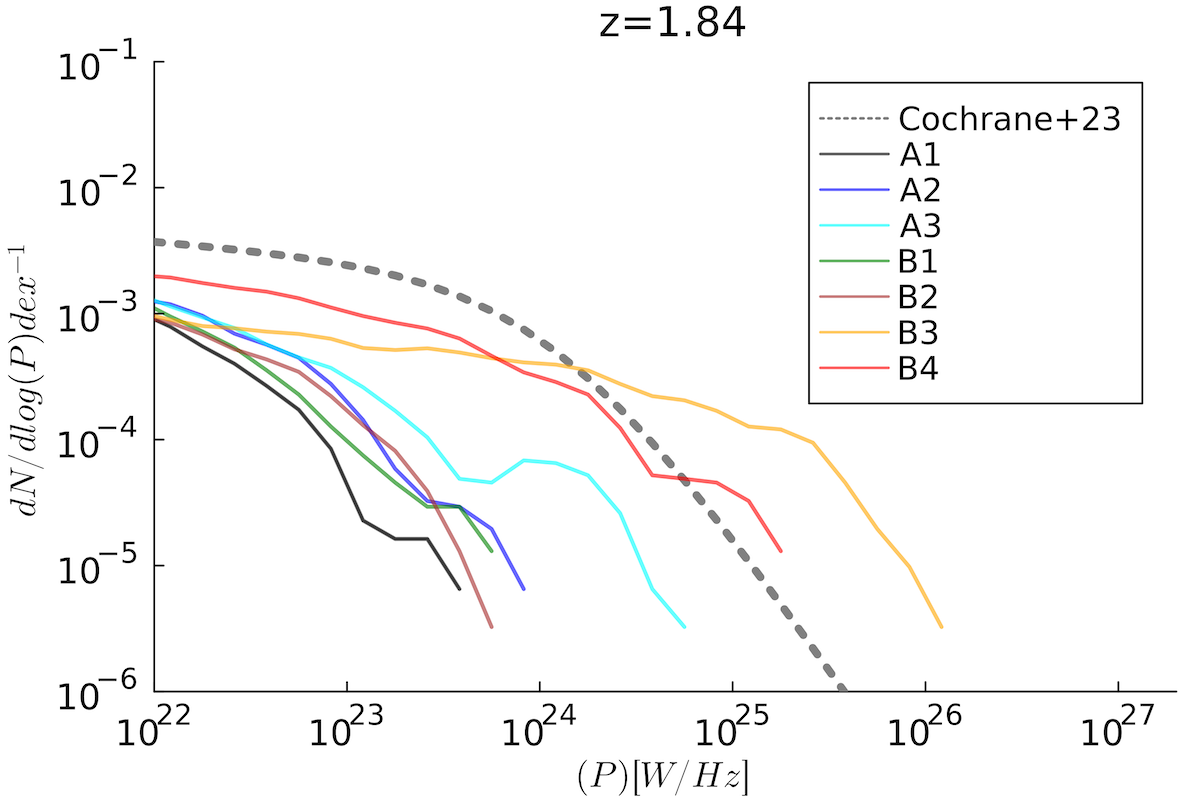}
\includegraphics[width=0.33\textwidth]{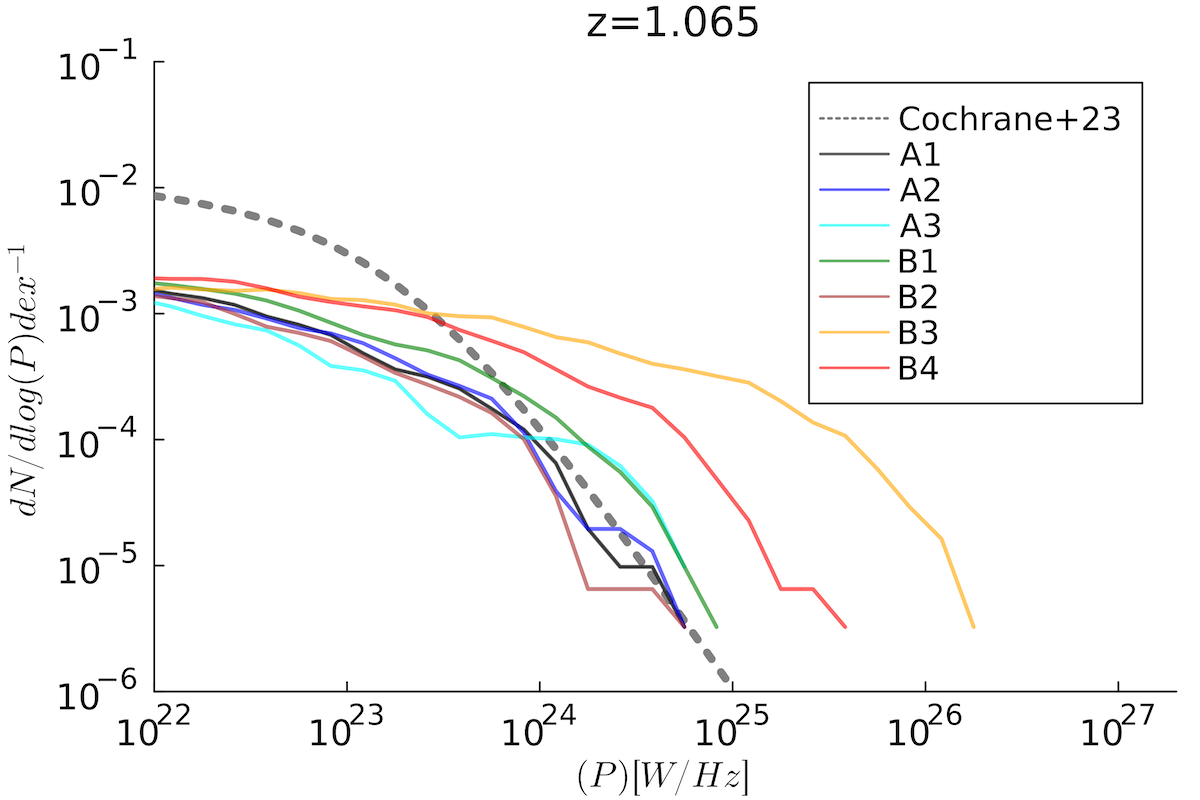}
\includegraphics[width=0.33\textwidth]{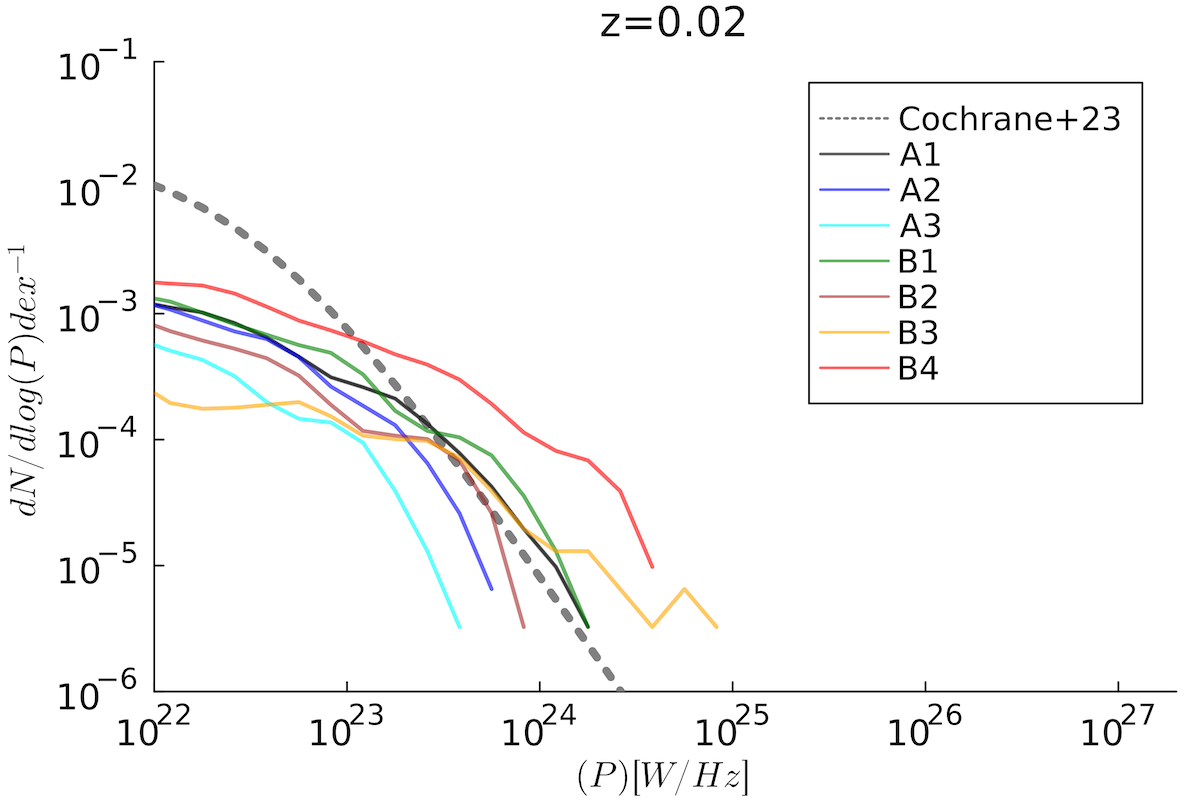}
\includegraphics[width=0.33\textwidth]{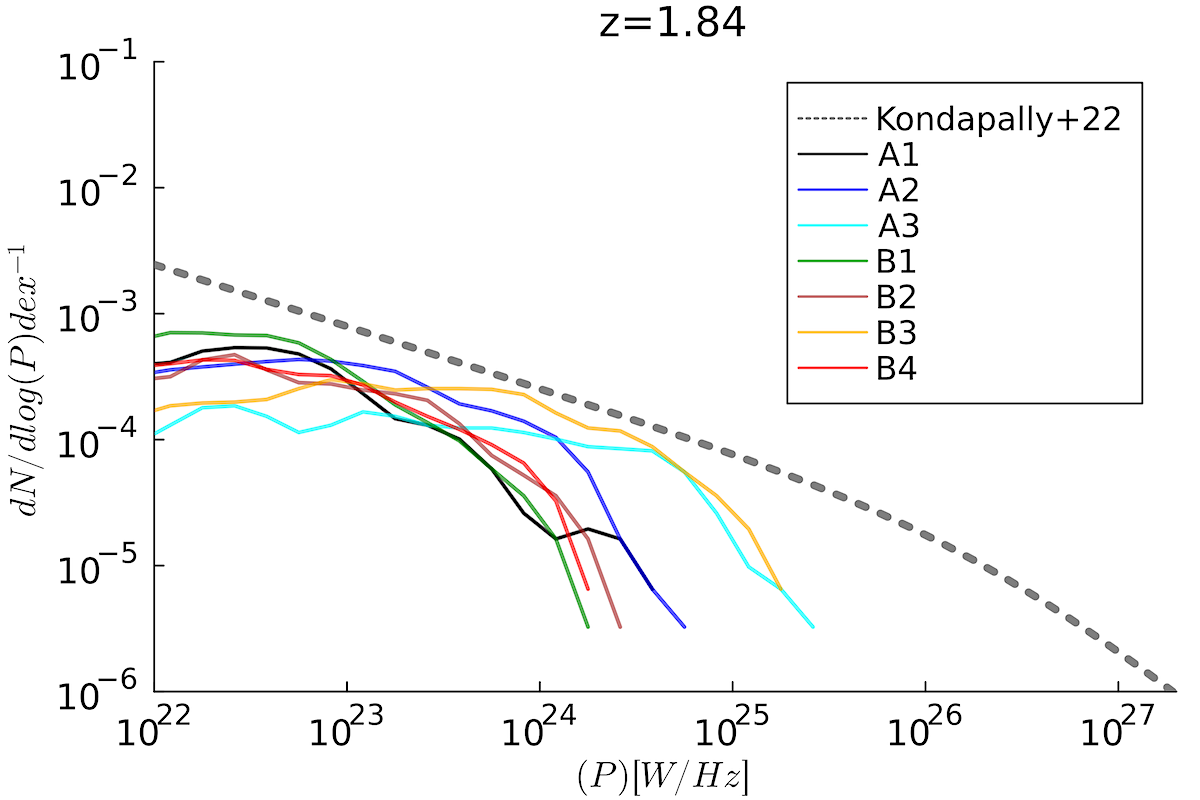}
\includegraphics[width=0.33\textwidth]{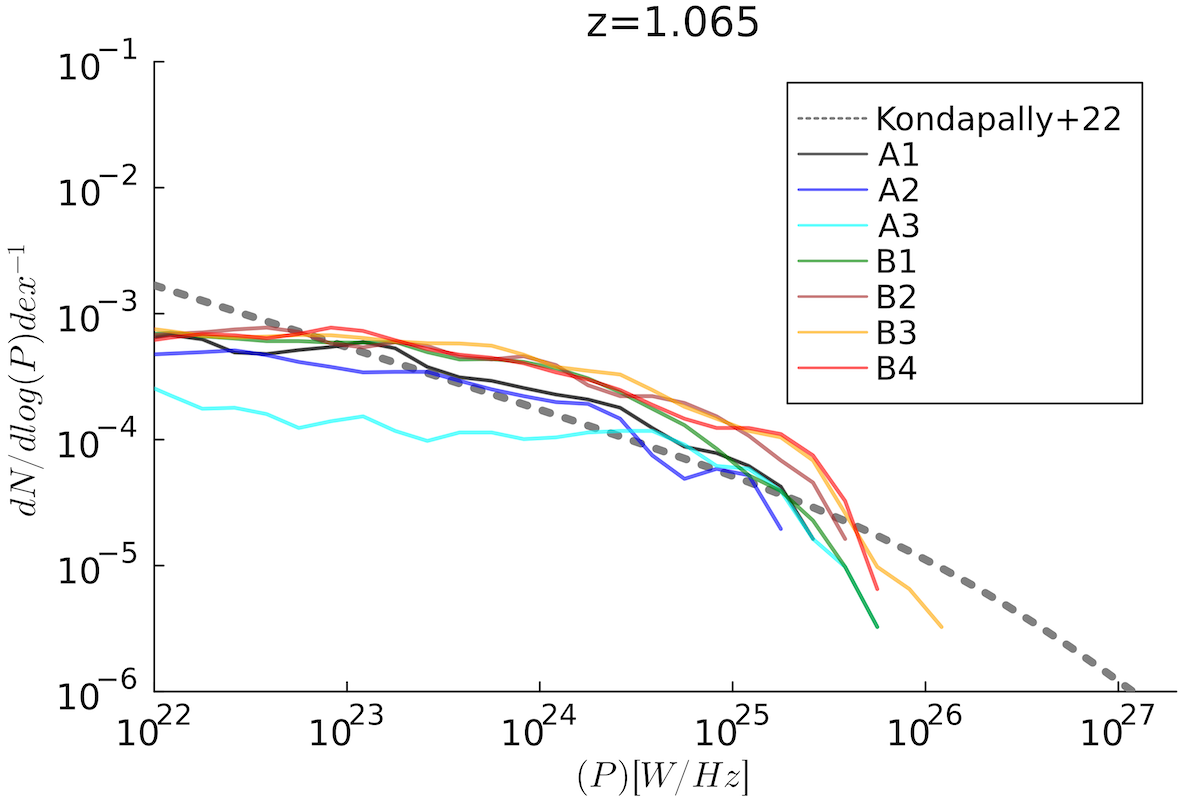}
\includegraphics[width=0.33\textwidth]{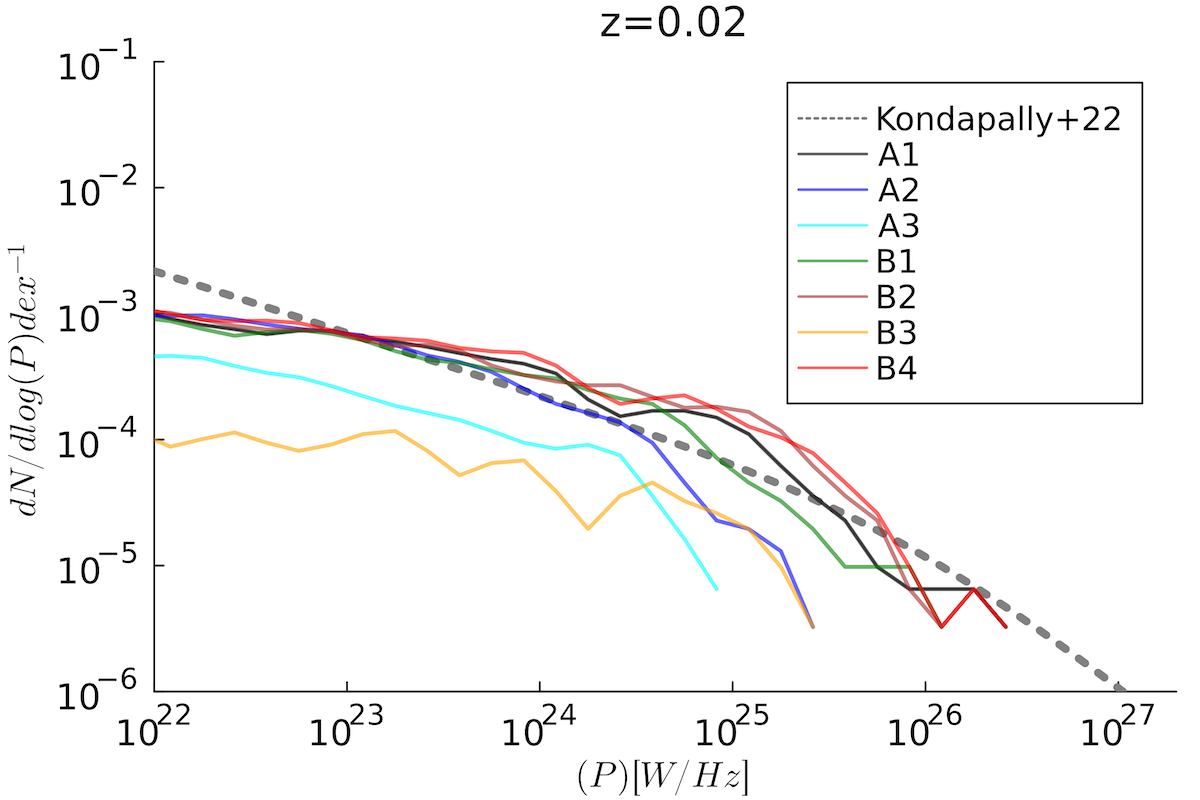}

\caption{Top row: evolving distribution function of radio power at $150$ MHz from our simulated galaxies, in which only the radio emission from CRe injected by star formation is considered.
Bottom row: evolving distribution function of radio power at $150$ MHz from our simulated galaxies, in which only the radio emission from CRe injected by AGN is considered. The dotted grey lines in the top rows 
give the best-fit relation of observed data with LOFAR, in the same redshift bins,  from \citet{2023MNRAS.523.6082C} (top) and \citet{2022MNRAS.513.3742K} (bottom).}
\label{fig:RG_function}
\end{figure*}

Figure \ref{fig:RG_function} shows the evolving distribution function of the radio power of our simulated galaxies at $150$ MHz, dividing the contribution from CRe injected by star formation { (top panels) or by AGN (bottom panels). 
In all cases, we compare with the recent results of LOFAR observations, derived by  \citet{2023MNRAS.523.6082C} and by \citet{2022MNRAS.513.3742K} for star forming galaxies or AGN, respectively. Within the scatter related to the small simulated volume, the B4 model appears as the best in reproducing the real radio luminosity function of AGN, across nearly four orders of magnitude in power both at $z \approx 0$ and $z \approx 1$, while all models underpredict the real AGN radio luminosity function at $z \approx 2$. The comparison between the radio luminosity function of star forming galaxies and of our star formation-related CRe is less satisfactory, with an underprediction of low-power galaxies, as well as with an overprediction of high luminosity objects. The definition of radio luminosity functions based on physical processes is not straightforward, leading to biases in observed samples \citep[e.g.][]{morabito25}. However, the mismatch of our simulated radio luminosity function of star forming galaxies is larger than any realistic observational bias. Hence, it likely implies that our modelling of the injection of CRe by star formation processes is too crude (also considering the fact that we do not follow the production of secondary CRe, nor follow their diffusion and streaming through the multi-phase interstellar medium in halos and discs; see e.g., \citealt{2021ApJ...922...11A,hopkins22}).}

\begin{figure}
\begin{center}
\includegraphics[width=0.495\textwidth]{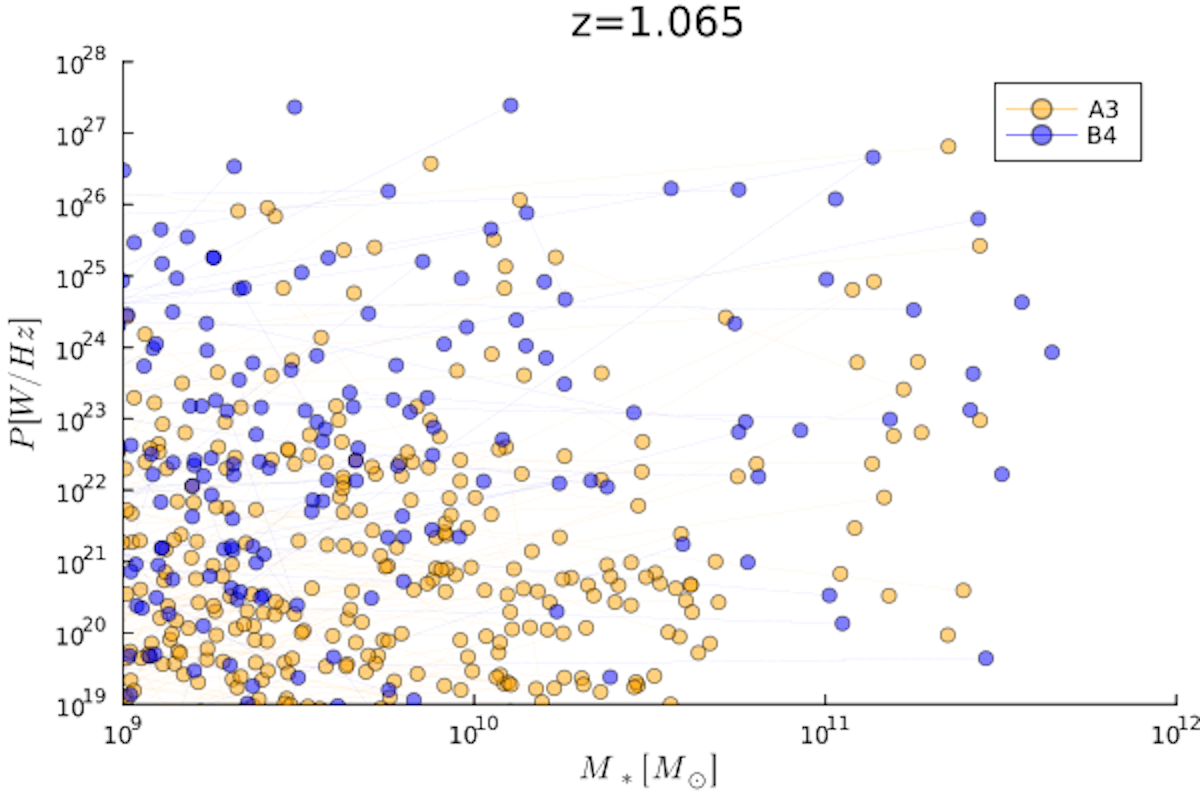}
\includegraphics[width=0.495\textwidth]{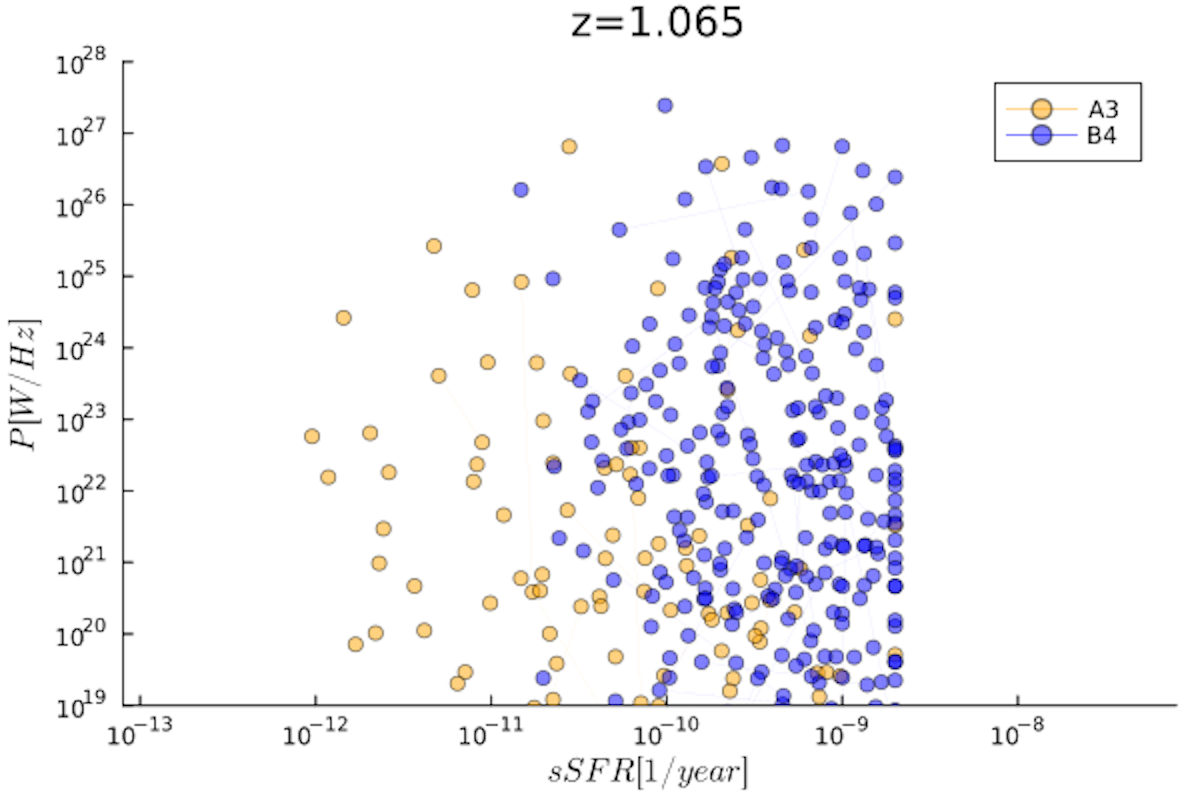}
\caption{Relation between the emitted radio power at 150 MHz and the stellar mass (top panel) or the specific star formation rate (bottom panel) for our A3 and B4 run at $z=1.065$.}
\label{fig:radiogals_vs_star}
\end{center}
\end{figure}

A remarkable recent result from recent radio surveys (e.g. with LOFAR) is the hint of a relation between the radio galaxy morphological class (e.g. Fanaroff-Riley class I and II) and the specific star formation rate (i.e. star formation rate divided by the galaxy stellar mass) of host galaxies, while links between morphological class and stellar mass, environment, or else, are more elusive to detect \citep[][]{mingo22}. Inspired by this work, we measured in Fig.~\ref{fig:radiogals_vs_star} the relation between the radio emission from our galaxies and their stellar mass (top panel) or their specific star formation rate (bottom panel) again in run A3 and { B4}. We used here the $z=1.065$ epoch,  which where most of the objects  analysed by \citet{mingo22} come from. As in real observations, in both these models there is no clear relation between radio power and stellar mass \citep[e.g. Fig. 16 left of ][]{mingo22}. However, our simulated catalogues also do not show any clear correlated trend between radio power and specific star formation rate, { in disagreement with observation. While in principle the cosmic volume sampled by our simulations is rather small and the high power radio galaxies studied in \cite{mingo22} are not present here, this lack of correlation may suggest that additional variations of our jet feedback models, combined with a higher resolution and larger volumes, are necessary to reproduce reality.}

While very valuable attempts have instead been produced using post-processing on the simulated distribution of galaxies \citep[e.g.][]{2021MNRAS.503.3492T},  our simulation (to the best of our knowledge) is the first in the real radio luminosity function of radio galaxies is reasonably well matched  by the result of direct cosmological simulations, lunching magnetised jets and CRe at every feedback event.

  \begin{table*}
      \centering
      \begin{tabular}{cccccc}
 network&  number of nodes & average node degree& average clustering& average neighbour degree& average triangles\\ \hline 
           A3 AGN&  2280&  14.75&  0.82&  15.62&  159.96\\
           A3 SF&  6588&  13.30&  0.78&  14.24&  125.59\\
           B4 AGN&  2668&  12.17&  0.84&  12.65&  109.92\\
           B4 SF&  5938&  13.81&  0.82&  14.64&  136.39\\ 
      \end{tabular}
      \caption{Average value of the network parameters (see Sec.~\ref{network} for more explanations) measured on the distribution of halos illuminated by star formation or by radio galaxies in run A3 and run { B4} at $z=0.02$.}
      \label{tab:network}
  \end{table*}

\subsection{Analysis of the network of galaxies}
\label{network}

The schematisation of the continuous matter distribution of the  cosmic web as a network, through its reduced representation by means of nodes connected by edges, has 
been recently proliferating in cosmological simulations and has led to powerful new probes of cosmological parameters and galaxy clustering  \citep[e.g.][]{2018MNRAS.477.4738D,2020MNRAS.495.1311T,2020FrP.....8..491V,2021JCAP...04..061B,tsizh23,2023MNRAS.523.5738O,2024MNRAS.529.4325B}. 

Here we explore the application of network analysis to highlight  potentially detectable differences between the distributions of radio galaxies and star forming galaxies at $z=0.02$, comparing again model A3 and { B4} as before.  Can network analysis tell these two models apart, based on either the distribution of radio emission from star forming galaxies or AGN? 

We used the 3-dimensional distributions of identified matter halos for both runs, and with a standard procedure  \citep[see e.g.][for discussions]{2018MNRAS.477.4738D,2020MNRAS.495.1311T} we built a connected network by assuming that all closer than a $r=1 ~\rm Mpc$ distance are connected by an edge, or are disconnected otherwise.
In order to test whether different radio emission components to the total radio luminosity functions are characterized by the same network, for each model we further doubled the catalogue of halos, and we alternatively removed the halos without detectable radio emission (using $P \geq 10^{20} \rm W/Hz$ at 150 MHz as a limiter for simplicity) from CRe injected by star formation or by AGN. 

We used the same algorithm by \citet{2020MNRAS.495.1311T} (to which we refer the curious reader for the full mathematical prescription of all parameters) to compute the average values of all most important network parameters, namely: a) the average node degree, which is the average number of edges connected to a node; b) the average clustering coefficient,  which gives the fraction of connections between neighbours, compared  to all possible connection
between them, therefore quantifying the existence of structure in the local vicinity of nodes; c) the average neighbour degree, i.e. the  average degree of the neighbours of a given node, normalized by number of neighbours; d) the average number of triangles, which gives the  number of triangles, formed by edges and having nodes at the vertices, that include a given node. The list of the average parameters for the two models and their further subdivision into star forming (SF) galaxies and AGN radio galaxies is given in Tab.\ref{tab:network}.

A first striking difference to notice, is that regardless of the model, the networks traced by SF galaxies or by AGN are very different in all parameters, with the second showing a smaller number of nodes and a higher degree of clustering and density of nodes. This fully reflects the approximate mass segregation of the two effects (with star formation, and its related radio emission, being more relevant in the low mass end of the galaxy mass distribution) and it echoes the topological mass bias \citep[e.g.][]{2024MNRAS.529.4325B} according to which halos with different mass ranges of halos samples give a biased view of the underlying cosmic network. 

Furthermore,  the analysis  shows that the distribution of radio AGN in model A3 is significantly denser than the distribution of radio AGN in the { B4} model (as marked by the much higher average node degree, neighbouring degree and triangles), despite the number of nodes is significantly larger (by { $17 \%$}) in model { B4}.  On the other hand, the network traced by the radio emission by the SF galaxies  has $\sim 3$ more nodes than the network of AGN galaxies in both models, but is significantly less dense than the latter, with average parameters varying by less than $\sim 10\%$ in all cases. 

 This first pilot exploration of the simulated network of radio emitting galaxies already shows the potential of our suite of simulations to, both, assess the biases associated with the use of specific samples of  radio emitting galaxies  (considering that this will be one of the main aims of the Square Kilometre Array, e.g. \citealt{2015aska.confE..67P}) , and it also highlights the impact of different prescriptions for galaxy evolution on the reconstructed network parameters. 

\begin{figure}
\begin{center}
\includegraphics[width=0.48\textwidth]{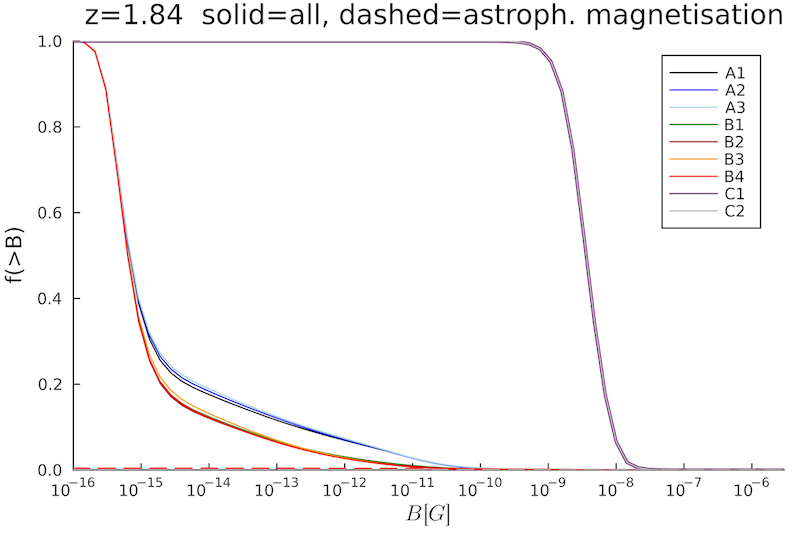}
\includegraphics[width=0.48\textwidth]{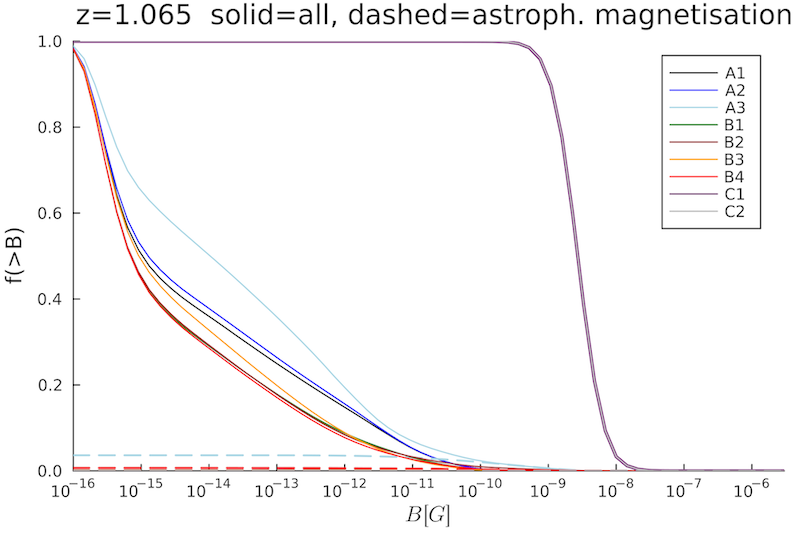}
\includegraphics[width=0.48\textwidth]{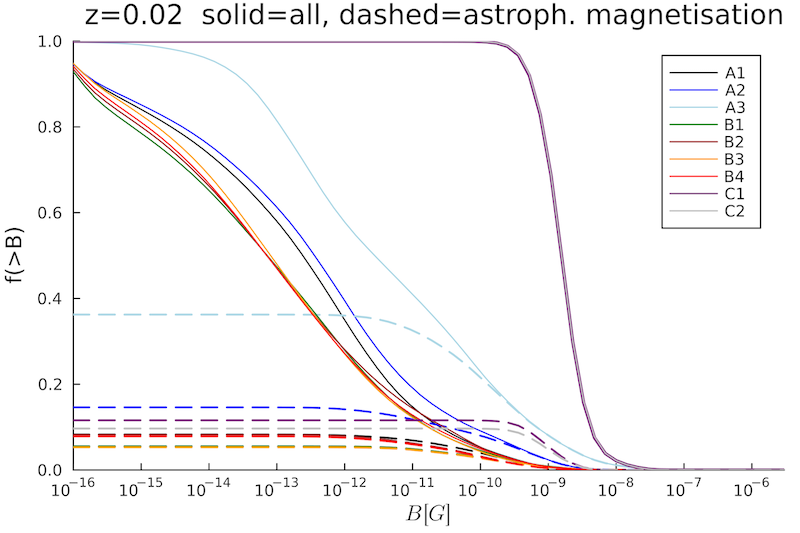}
\caption{Volume filling factors of comoving magnetic field strength in our runs at three different epochs. The solid lines give the cumulative distribution for all cells in the simulation, while the dashed lines show the contribution from magnetic fields only injected by "astrophysical" processes (i.e. AGN feedback and star formation).}
\label{fig:pdfB}
\end{center}
\end{figure}

\begin{figure}
\begin{center}
\includegraphics[width=0.48\textwidth]{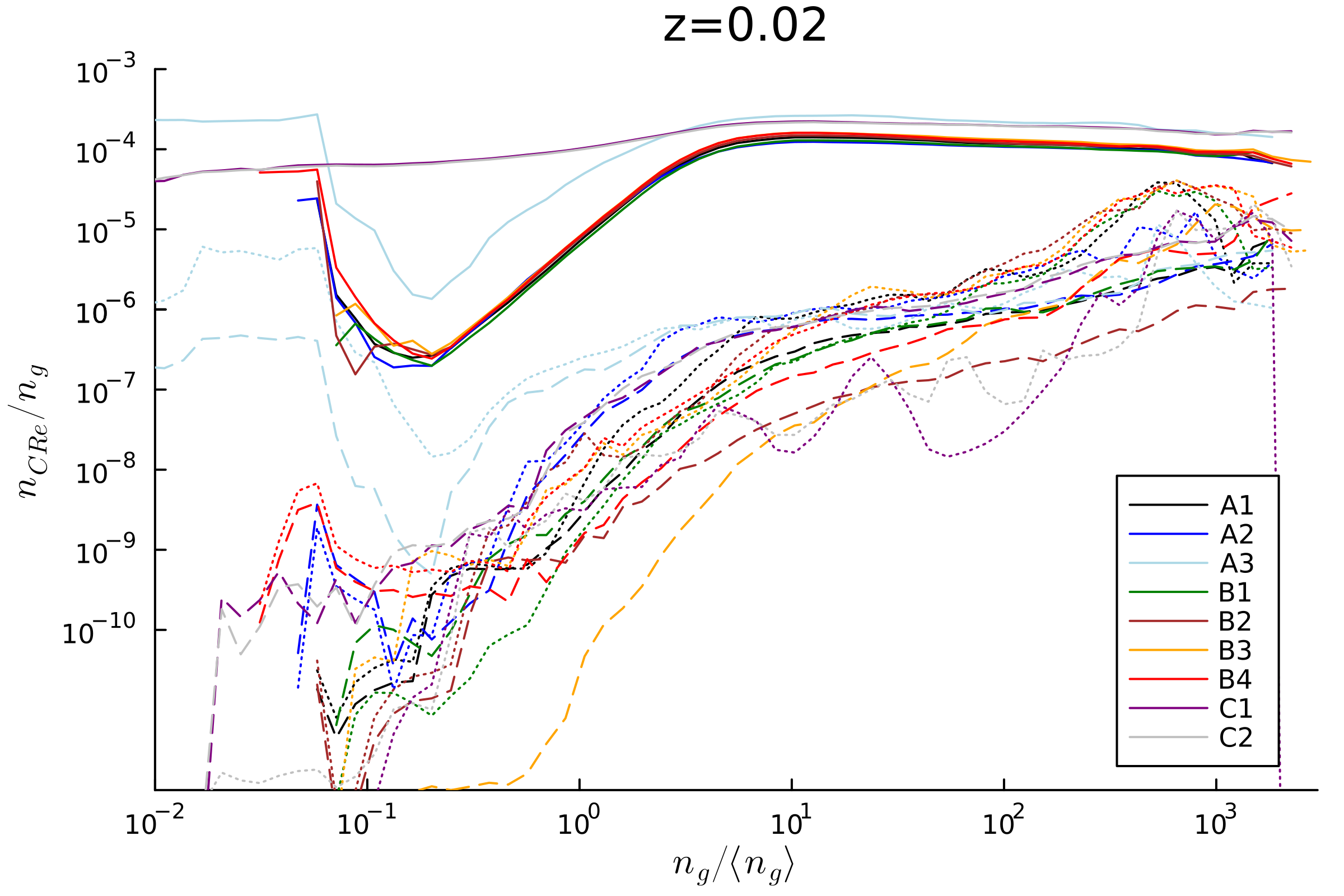}
\includegraphics[width=0.48\textwidth]{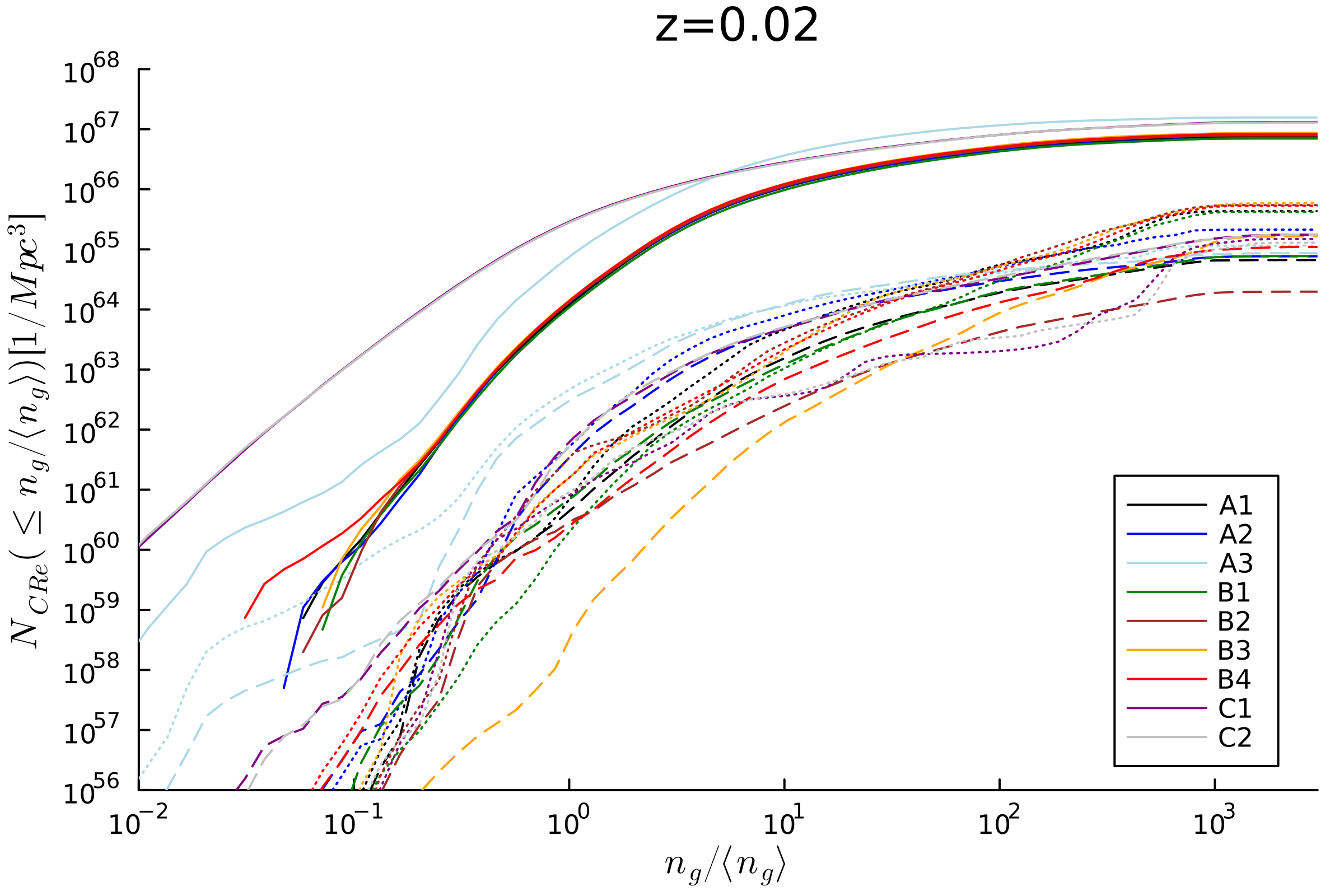}
\includegraphics[width=0.48\textwidth]{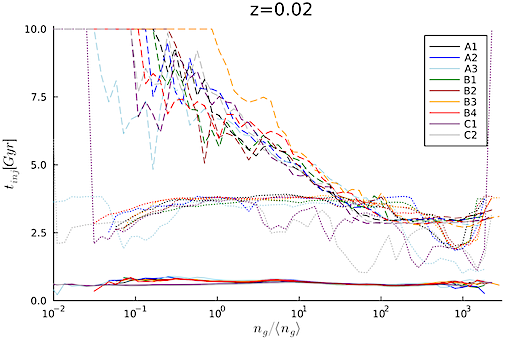}
\caption{Top panel: average ratio between CRe density and thermal gas proton density as a function of the gas over-density in the simulation by the end of our runs. Middle panel: cumulative absolute number of CRe normalised for a comoving $\rm Mpc^3$ volume and averaging over an increasing gas overdensity. Bottom panel: average time elapsed since the last injection of CRe as a function of gas overdensity. The solid lines give the trend for CRe injected by shocks, the dotted lines for CRe injected by AGN and the dashed lines for CRe injected by star formation.}
\label{fig:CRe_dens}
\end{center}
\end{figure}

\begin{figure}
\includegraphics[width=0.45\textwidth]{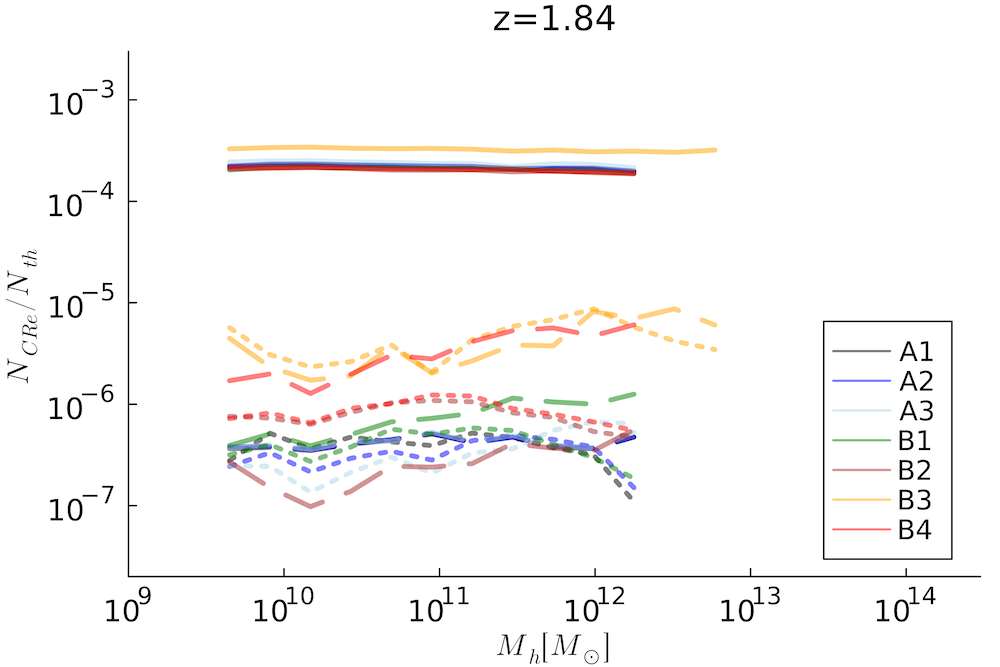}
\includegraphics[width=0.45\textwidth]{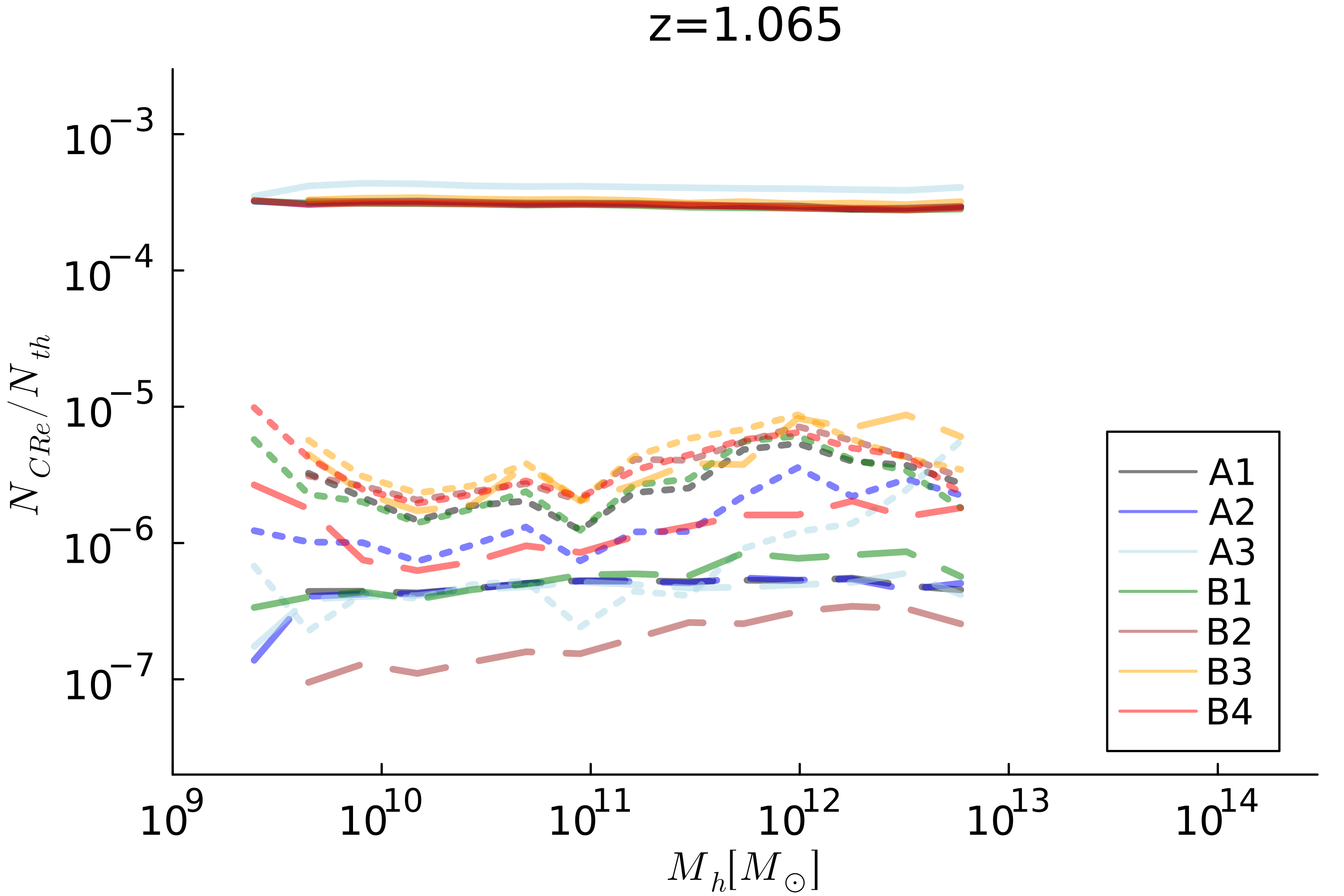}
\includegraphics[width=0.45\textwidth]{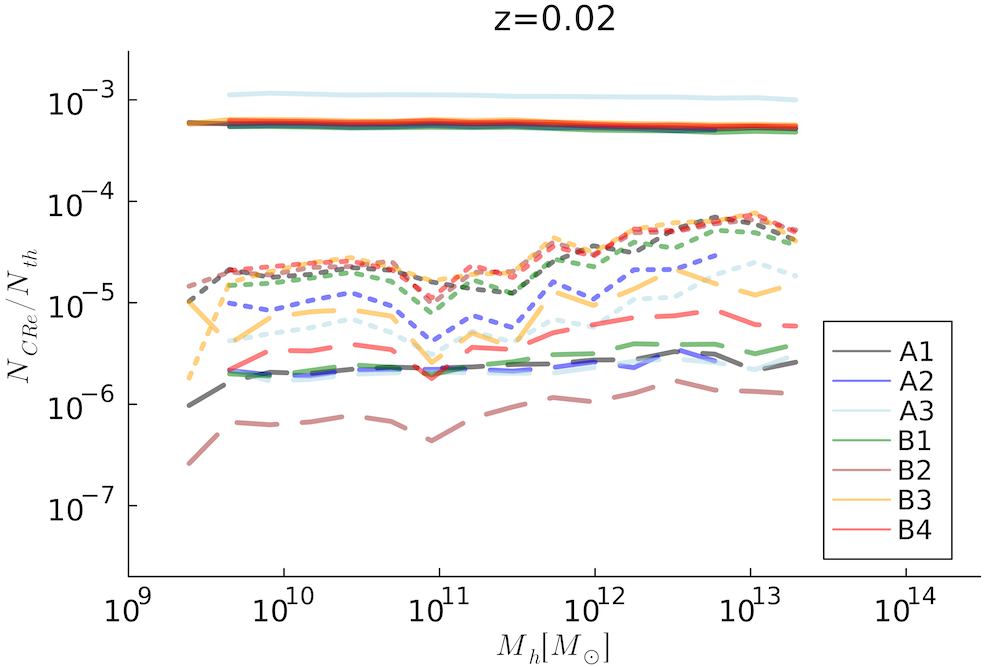}
\caption{Distribution of the ratio between CRe density and baryon density within $R_{500}$ of our halos at three different epochs and for all runs. The solid lines show the distribution of CRe injected by shocks, the dotted lines by AGN and the dashed lines by star formation.}
\label{fig:halosCRe}
\end{figure}

\subsection{Volume filling properties of magnetic fields and of cosmic ray electrons}
\label{subsec:pdf}
The interpretation of several key observational probes of primordial magnetic fields crucially depends on the accurate removal of the possible magnetic field contamination by  astrophysical processes \citep[e.g.][for recent discussions and tests]{va17cqg,2021Univ....7..223A,va21magcow}. A simple way to assess the possible contribution of primordial and astrophysical mechanisms to the magnetisation of the Universe is to compute the volume filling fraction of magnetic fields produced by a given process. 
For example, \citet{2024ApJ...963..135T} have recently estimated that a volume filling factor $f \geq 0.67$ with a magnetic fields strength $B\geq 10^{-15} \rm G$  is required to explain the observed lack of Inverse Compton Cascade emission from blazars at $\geq \rm ~GeV$ energies \citep[e.g.][]{2010Sci...328...73N}.

The volume filling factors of magnetic fields for our runs at at the usual three epochs are shown in Fig.~\ref{fig:pdfB}. 
The solid lines refer to all cells in the simulation, and show the very clear effect of primordial magnetic fields (runs C1 and C2) compared to all other models: strong enough primordial magnetic fields ($3.7 \cdot 10^{-10} \rm G$ in this case)  can clearly dominate the magnetisation of virtually $\sim 100\%$ of the cosmic volume (with the exception of the tiny volume fraction occupied by galaxies).
In runs with negligible primordial field ($=10^{-16} \rm G$), on the other hand, astrophysical sources can dominate the magnetisation of a significant fraction of the cosmic volume. To quantify how much, and since there still is a minimal amount of primordial magnetisation in all our runs, we use our tracking of CRe fields to compute exactly which fraction of the cosmic volume has been magnetised by galaxies in all runs. Of course a fraction of the magnetic fields in these cells will also be of primordial origin, but by construction these regions were mostly contaminated by feedback processes.

Therefore, in Figure\ref{fig:pdfB} we also mark the volume filling factor only of cells affected by feedback in our runs  with dashed lines. 
At the reference value of $10^{-15} \rm G$, all models without a strong primordial magnetic field have a filling factor  $f < 20 \%$, in line with \citet{2024ApJ...963..135T}, with the exception of the A3 model which has $f \sim 35 \%$ due to its enhanced AGN activity.  These filling factors are reached only for $z \leq 1$, as the magnetisation bubbles emerge from their host galaxies and start percolating only at evolved cosmic epochs At $10^{-12} \rm G$, the filling factor is $f \leq 15 \%$ (again with the exception of run A3 ($f \sim  37\%$), which is in line with similar statistics computed by  \citet{2021MNRAS.505.5038A} for Illustris TNG simulations ($f \sim 11-15\%$).  
In all cases, such low values of $f$ supports the idea that blazar observations can be a powerful probe of primordial magnetism \citep[e.g.][]{2010Sci...328...73N}, because no plausible feedback scenario could magnetise a large enough fraction of the Universe's volume to compete with primordial magnetic fields \citep[e.g.][]{2022A&A...660A..80B,2024ApJ...963..135T}. 

A similar volumetric analysis of the distribution of CRe with respect to thermal gas particles is useful to assess how many fossil electrons are available across the cosmic web, as a result of all simulated processes.  The top panel of Figure \ref{fig:CRe_dens} gives average ratio between CRe density (of all species of injected CRe) and the thermal gas proton density as a function of the gas overdensity in the simulation by the end of our runs. 

A very general finding is that, regardless of model-to-model variations, at large overdensities the energy density of CRe is { always dominated by the time-integrated injection by shocks (solid lines)}. The relative abundance of models is similar to the trend of the luminosity functions (Sec.~\ref{radiogalaxies}), in which the smallest amount of CRe injected by AGN is found for model A3, and the highest for model { B4} Interestingly however, the abundance of CRe injected by AGN in the A3 model is the highest of all, but only when gas overdensities $\leq 1$ are included: this means that the stronger feedback activity found in the A3 model at $z \geq 1$ has expelled a lot of gas enriched with CRe in the extreme periphery of halos, or beyond. This trend is also consistent with the similarly high trend of the  volume filling factor of magnetic fields, found above. 

Shocks are instead found to dominate the injection of CRe { at all overdensities, with increasing importance with decreasing density}. While all runs without strong primordial magnetic fields display a drop in the ratio of CRe density to thermal gas density, owing to the fact that shocks are almost never formed in the low density range typical of voids, in our C1 and C2 run with primordial stochastic magnetic fields there is a non negligible injection of CRe also at very low density, due to a population of shocks driven by the dynamics of magnetic field fluctuations in voids, as first reported in our previous work \citep[][]{va21}.

Star formation is never the dominant injection mechanism of CRe, even if its ratio compared to the AGN injection depends on the details of the implemented galaxy formation prescriptions, and can be more important than shocks for a large, $\geq 10^2$ overdensity.
{ Considering the limited resolution of our runs and the simplistic adopted star formation model, this result may need further confirmation with more sophisticated galaxy formation models in the future.}

The middle panel of Fig.~\ref{fig:CRe_dens} gives the cumulative absolute number of CRe, injected by the different mechanisms, normalised for a comoving $\rm Mpc^3$ volume and averaging over an increasing gas overdensity. 
This shows that through most of cosmic volume the budget of CRe is dominated by shock acceleration. While a magnetic obliquity dependant acceleration efficiency (see also Sec.~\ref{subsec:obliq}) does not change this picture, the total number of CRe injected by shocks, especially at low densities, is affected by the additional shock dynamics induced by galaxy formation-driven winds and outflows \citep[e.g.][]{ka07,2016MNRAS.461.4441S}.

The ratio between the number of CRe injected by various mechanisms within the $R_{500}$ of halos in the volume,  and the number of thermal gas protons within the same volume is given 
 in Fig.~\ref{fig:halosCRe} for the three epochs considered already elsewhere. 
By the end of the simulation, the number density of CRe is dominated by the contribution from { shocks}.  On the other hand, the CRe injected by star formation is small and only competes with the other two at $z=1.84$ , i.e. close to the peak of the cosmic star formation history in all models. 
The trend with halo mass is approximately constant for shocks injected CRe, while we report a clear increase with mass of the abundance of CRe injected by AGN, consistent with the dominance of this feedback mechanisms in high mass halos. { In summary, in all simulated halos for $z \leq 2$ the number density of CRe is $\sim 0.02-0.1 \%$ of the total number density of thermal protons,  and it is dominated by the production at shocks. Once more it is worth stressing that total number of CRe includes very low energy electrons, with momenta a few times above the thermal momentum of protons ($\sim 1-10 ~ \rm MeV$), as well as super-relativistic electrons typically emitting synchrotron radiation in the radio band ($\sim 1-10 \rm GeV$), or beyond, with energy distribution depending on the local spectral evolution.}

Finally, the bottom panel of Fig.~\ref{fig:CRe_dens} gives the the average time elapsed since the last injection of CRe, as a function of environment, separately for the three families of CRe and for all runs. This shows in a more systematic way the trends we have highlighted already:  even at $z=0.02$ the injection by shocks is a very active process,  and the average time since the last shock injection is $\leq 0.5 ~\rm Gyr$ at all densities. CRe injected by AGN have instead average ages in the $\sim 1-3 \rm~ Gyr$ range, with variations depending on the assumed routine for triggering AGN activity.  Instead in all our models, the average age of CRe injected by star formation is much higher, i.e. $\sim 2 ~\rm Gyr$ within halos and increasing to $\sim 10 \rm ~Gyr$ in voids and filaments, which were contaminated only in the very first star formation episodes, which could drive some amount of CRe via outflows, before the full formation of cosmic structures.

\begin{figure*}
\begin{center}
\includegraphics[width=0.99\textwidth]{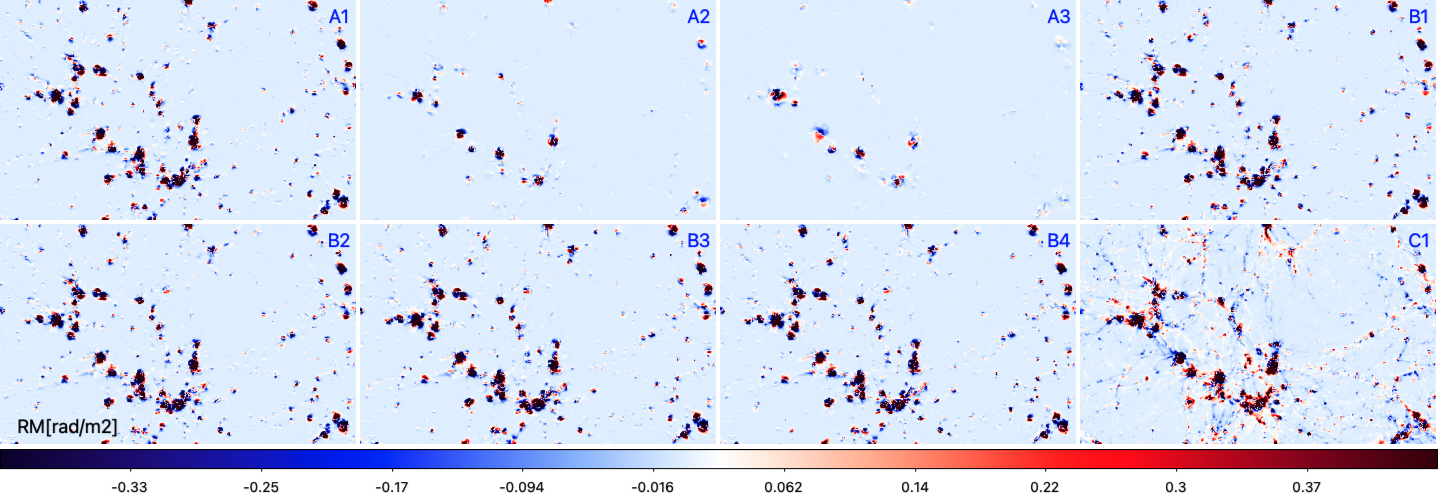}
\caption{Projected Rotation Measure maps (in units of [$\rm rad/m^2$] across the entire simulated volume for our seven runs at $z=0.02$. Each panel is approximately $21.25 \times 42.5  ~\rm Mpc^2$ large.} 
\label{fig:mapRM}
\end{center}
\end{figure*}

\begin{figure}
\includegraphics[width=0.49\textwidth]{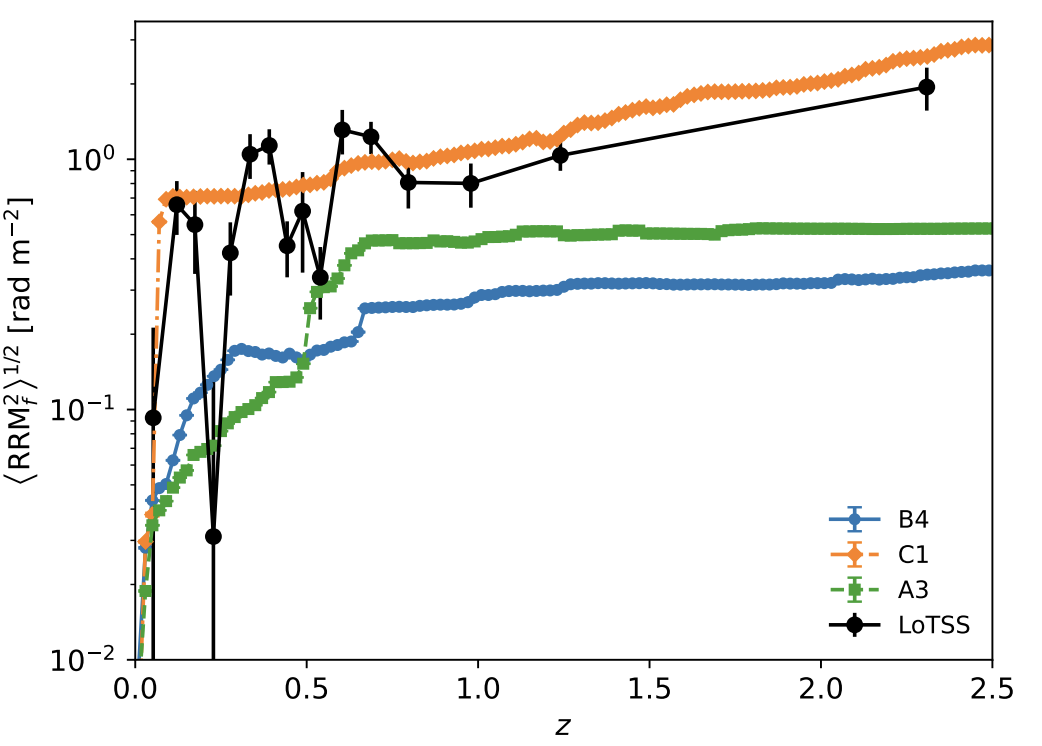}
\caption{Trend of simulated rms of residual Faraday Rotation Measure, observed at $z=0$ and for integrating out to increasingly larger redshift (up to $z=2.5$), for our A3, { B4 and C1 runs. The lines give the mean dispersion of the $RRM_f$ for 100 randomly selected lines of sight through the simulated volume, while the black line with larger error bars shows the observed RRM trend by LOFAR, taken from \citet{Carretti24}}.}
\label{fig:RMz}
\end{figure}

\begin{figure*}
\begin{center}
\includegraphics[width=0.33\textwidth]{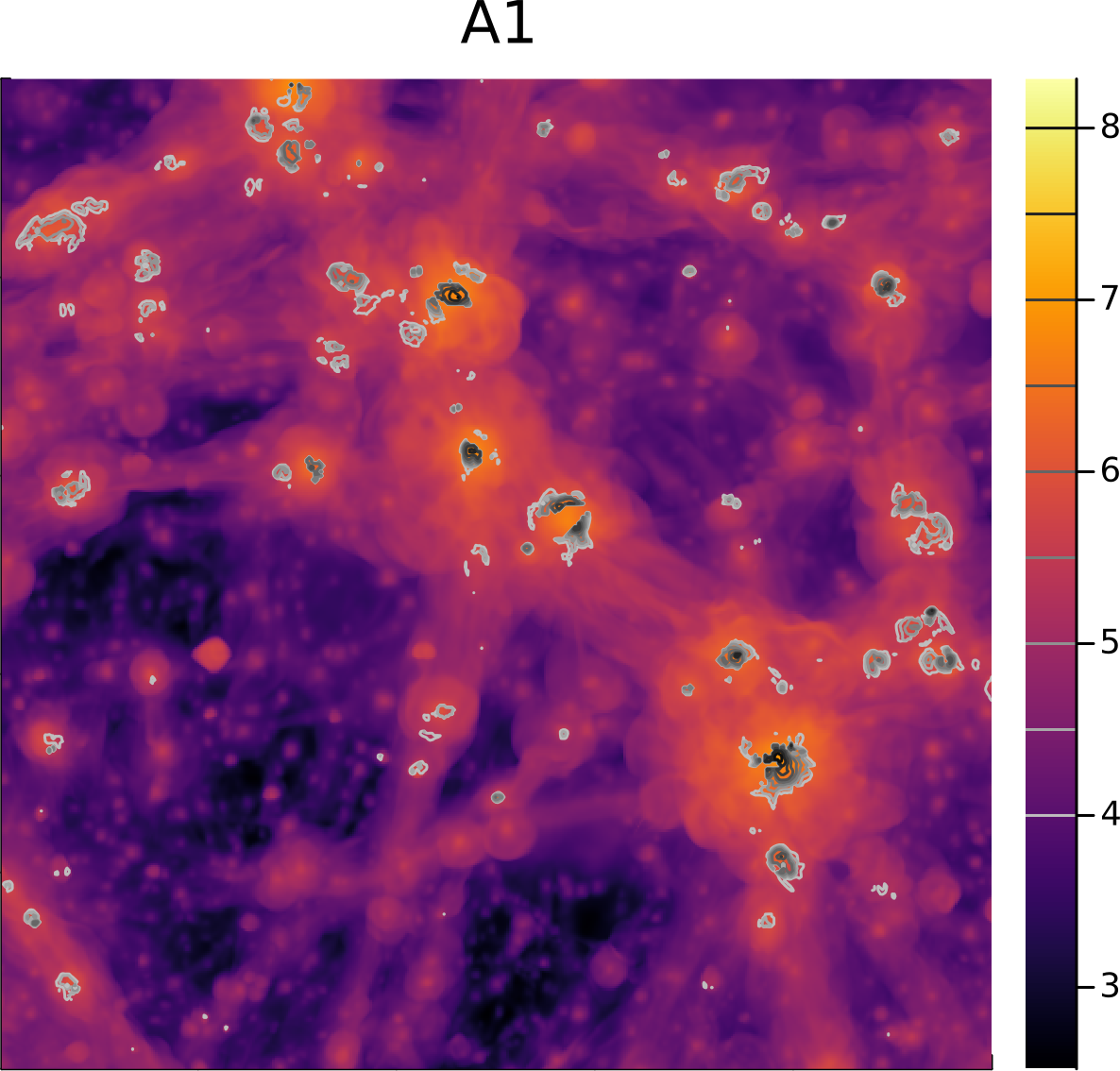}
\includegraphics[width=0.33\textwidth]{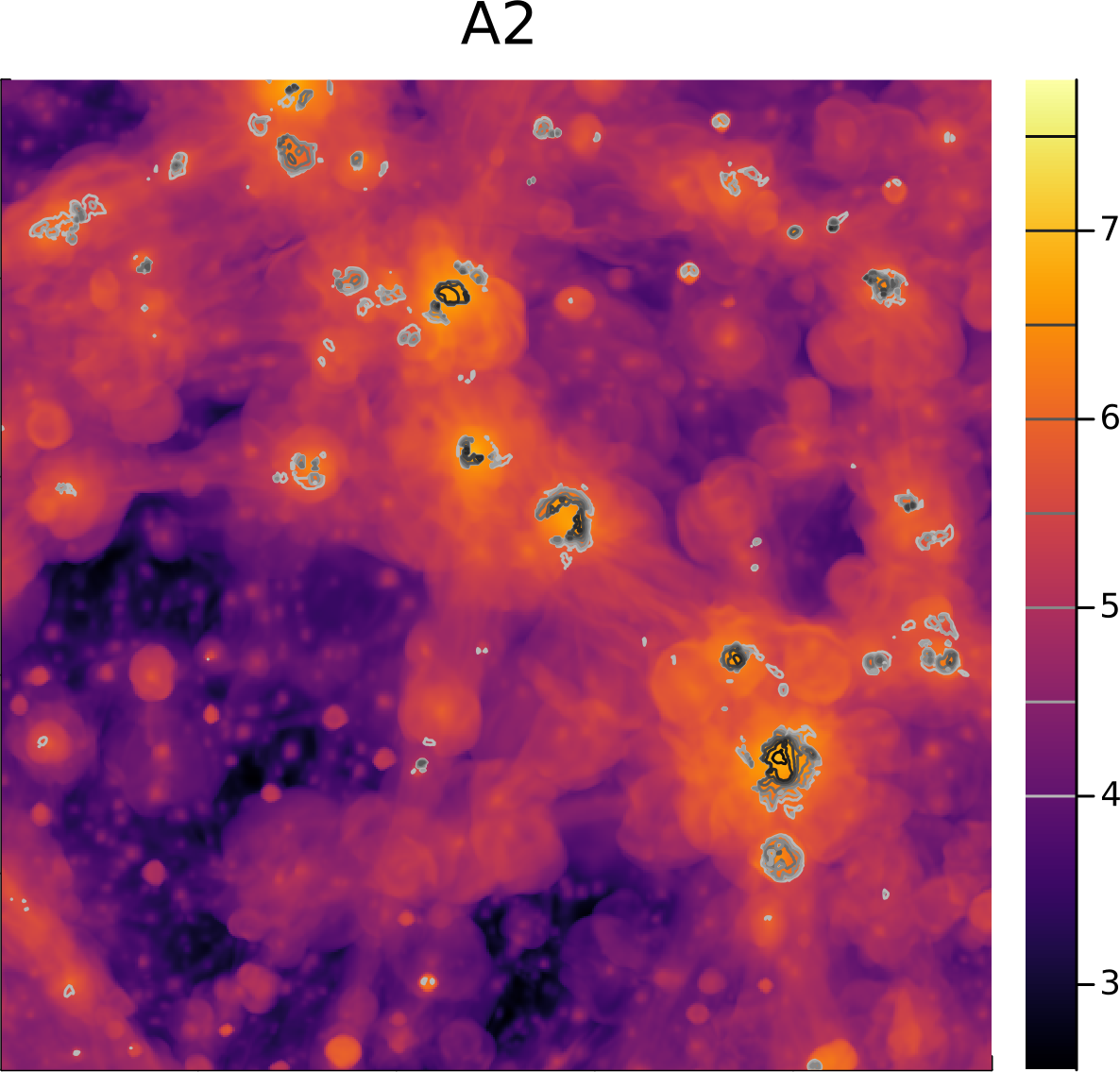}
\includegraphics[width=0.33\textwidth]{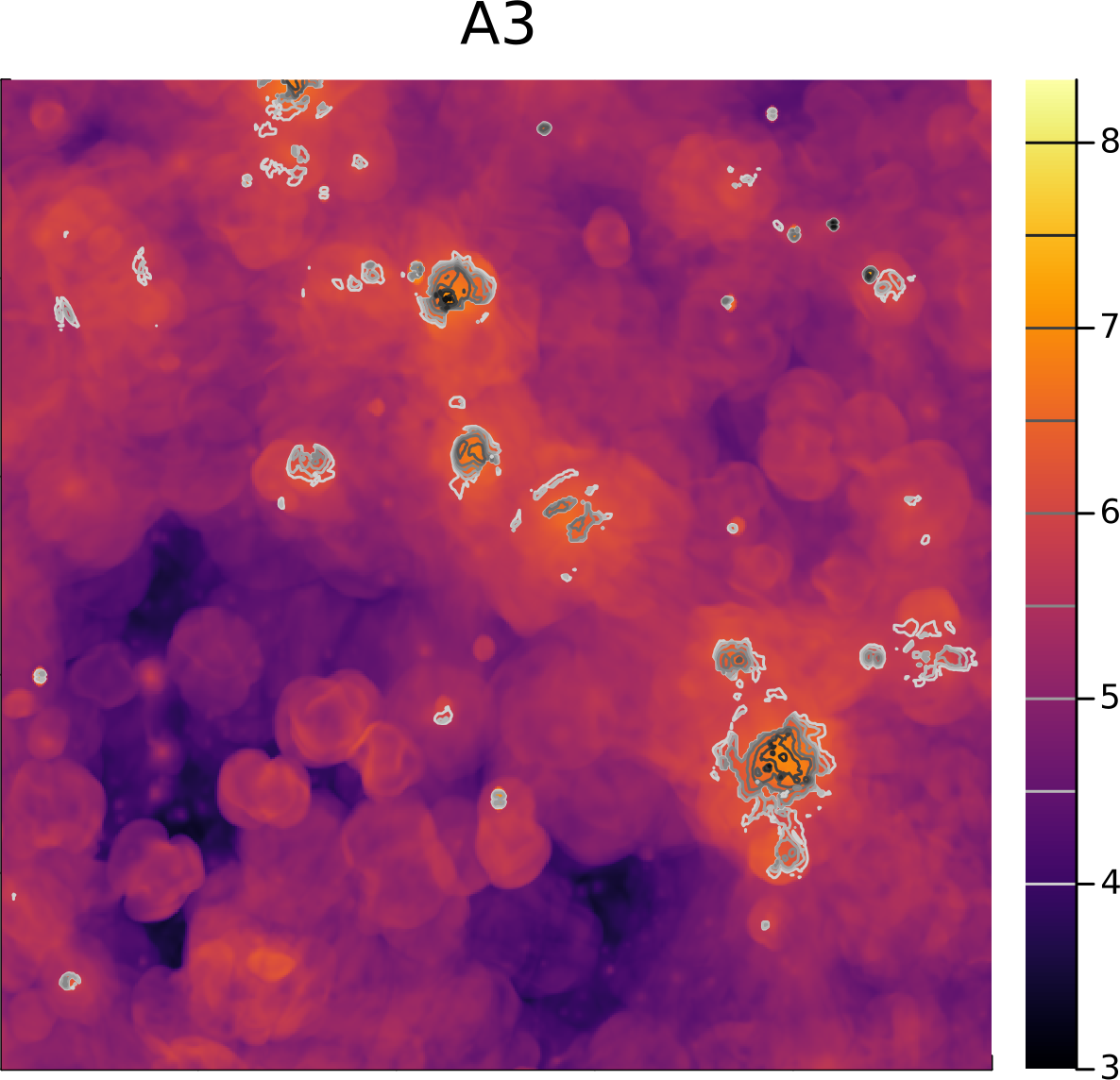}
\includegraphics[width=0.33\textwidth]{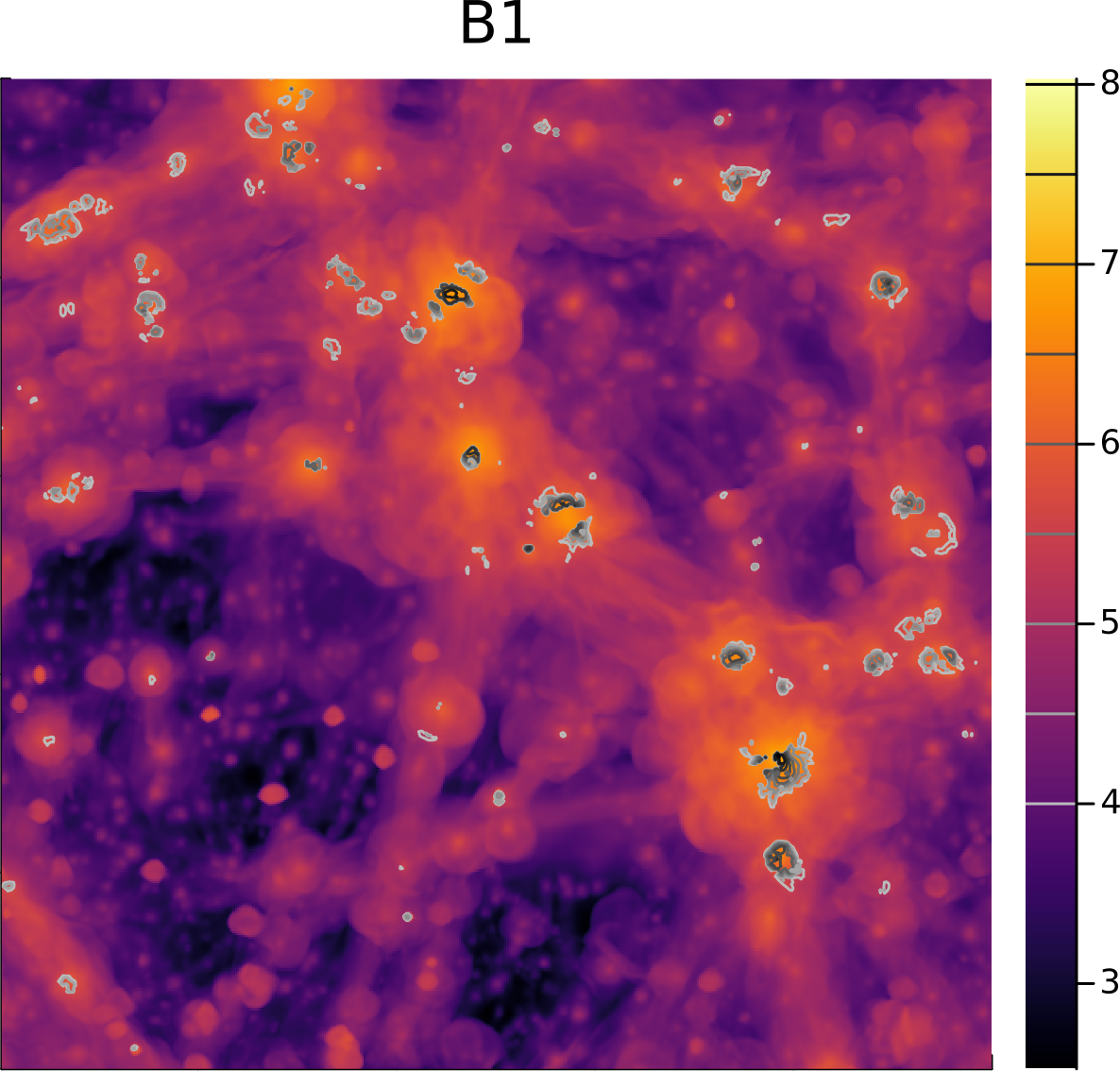}
\includegraphics[width=0.33\textwidth]{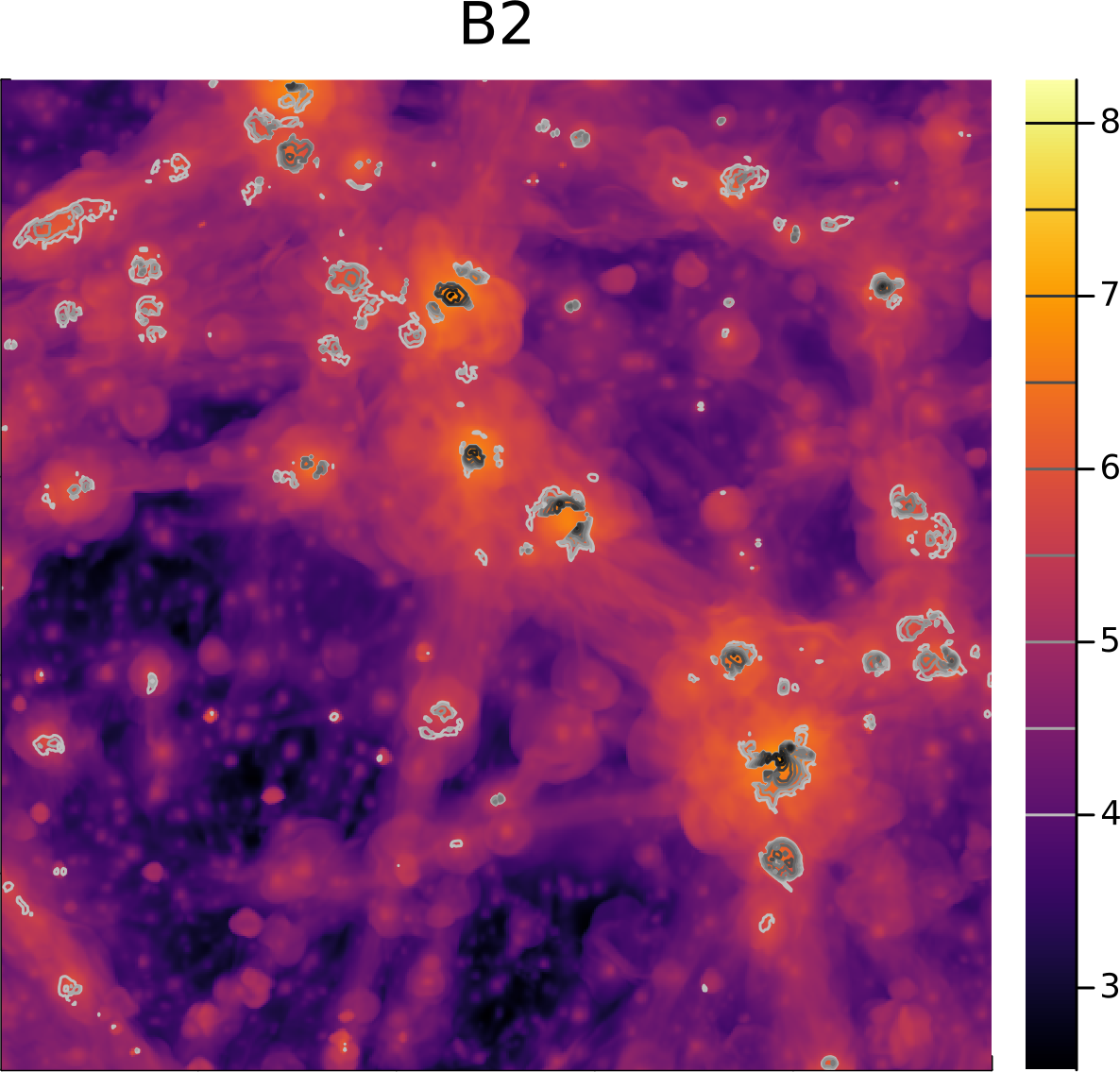}
\includegraphics[width=0.33\textwidth]{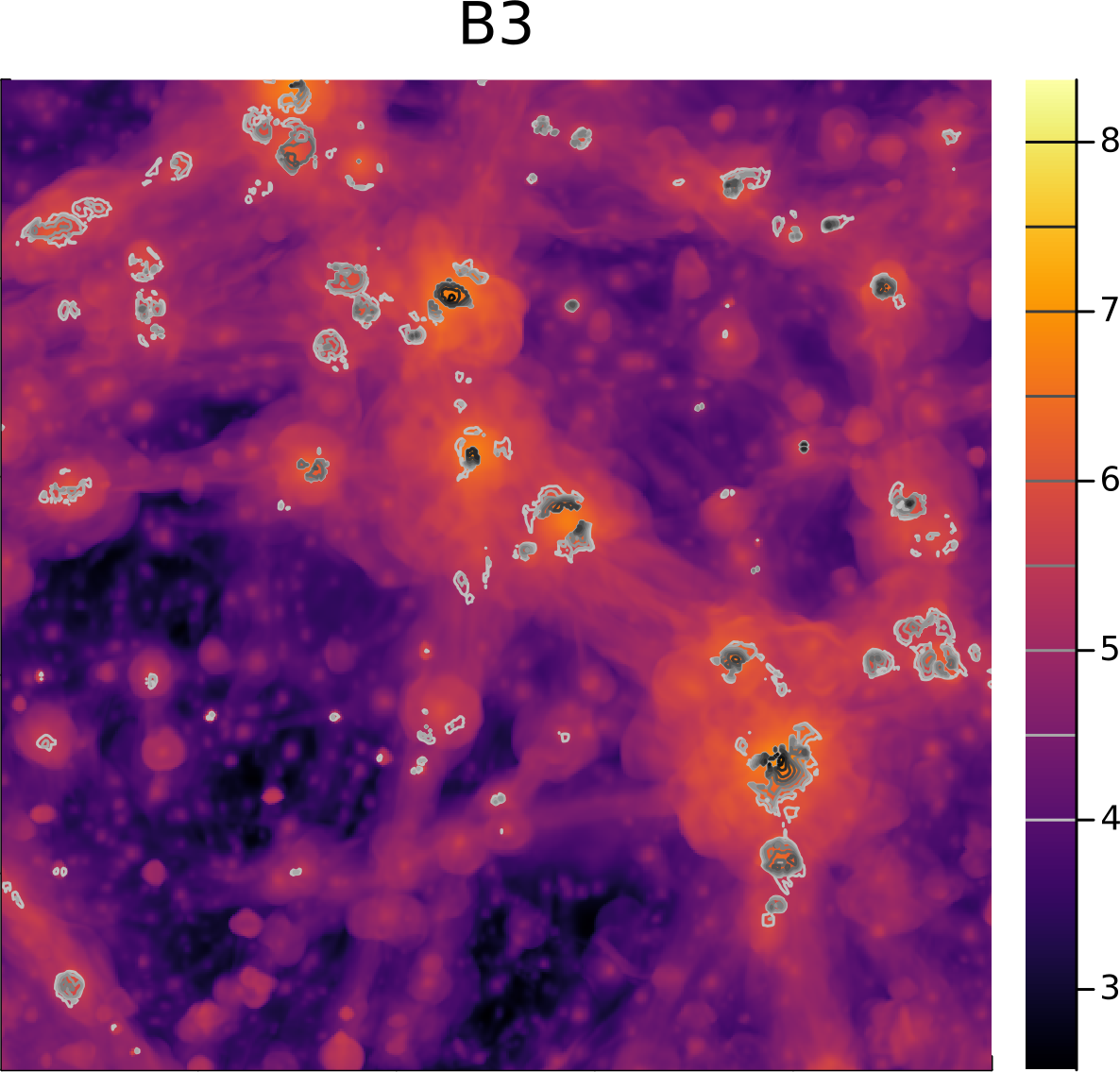}
\includegraphics[width=0.33\textwidth]{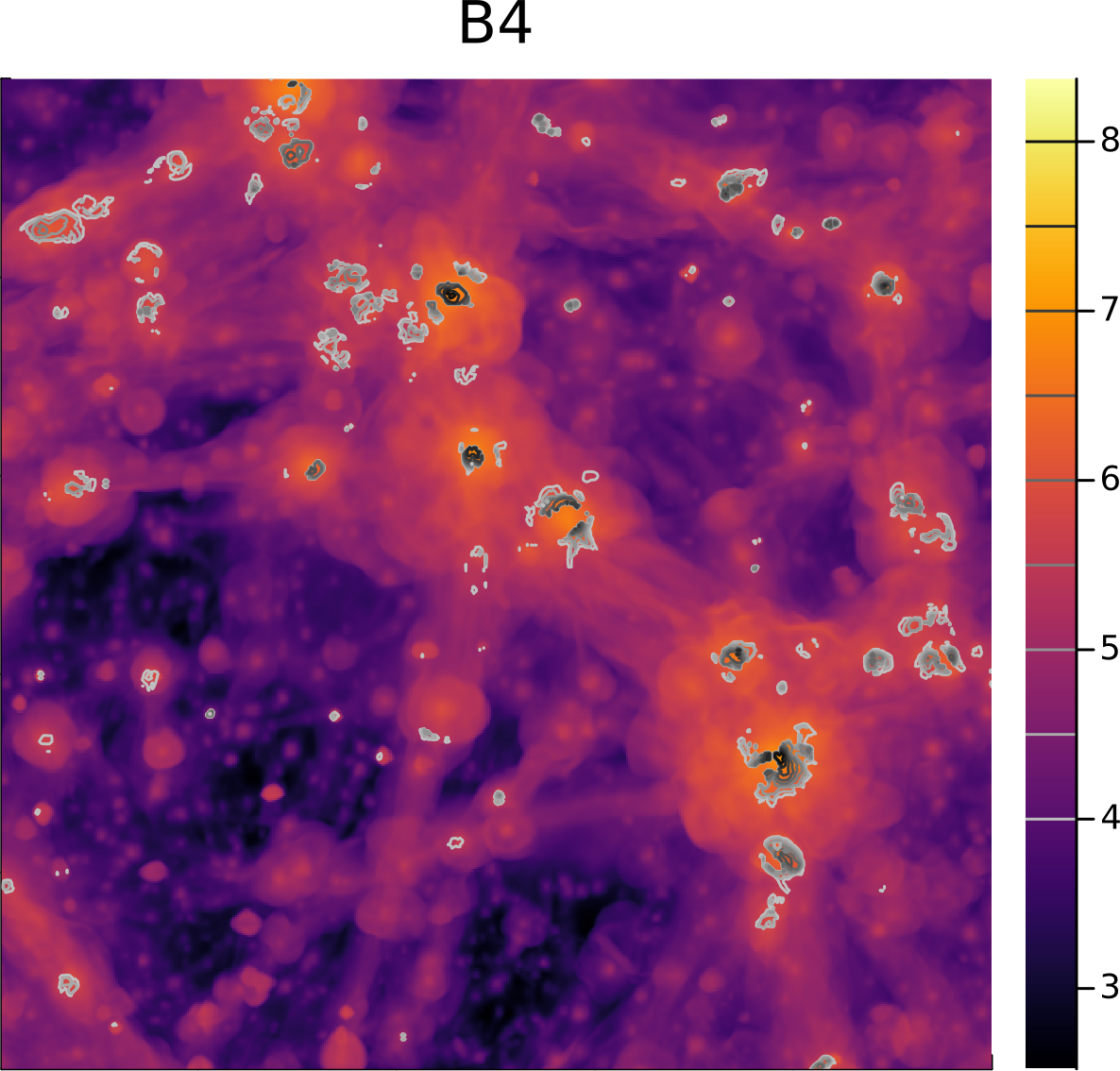}
\includegraphics[width=0.33\textwidth]{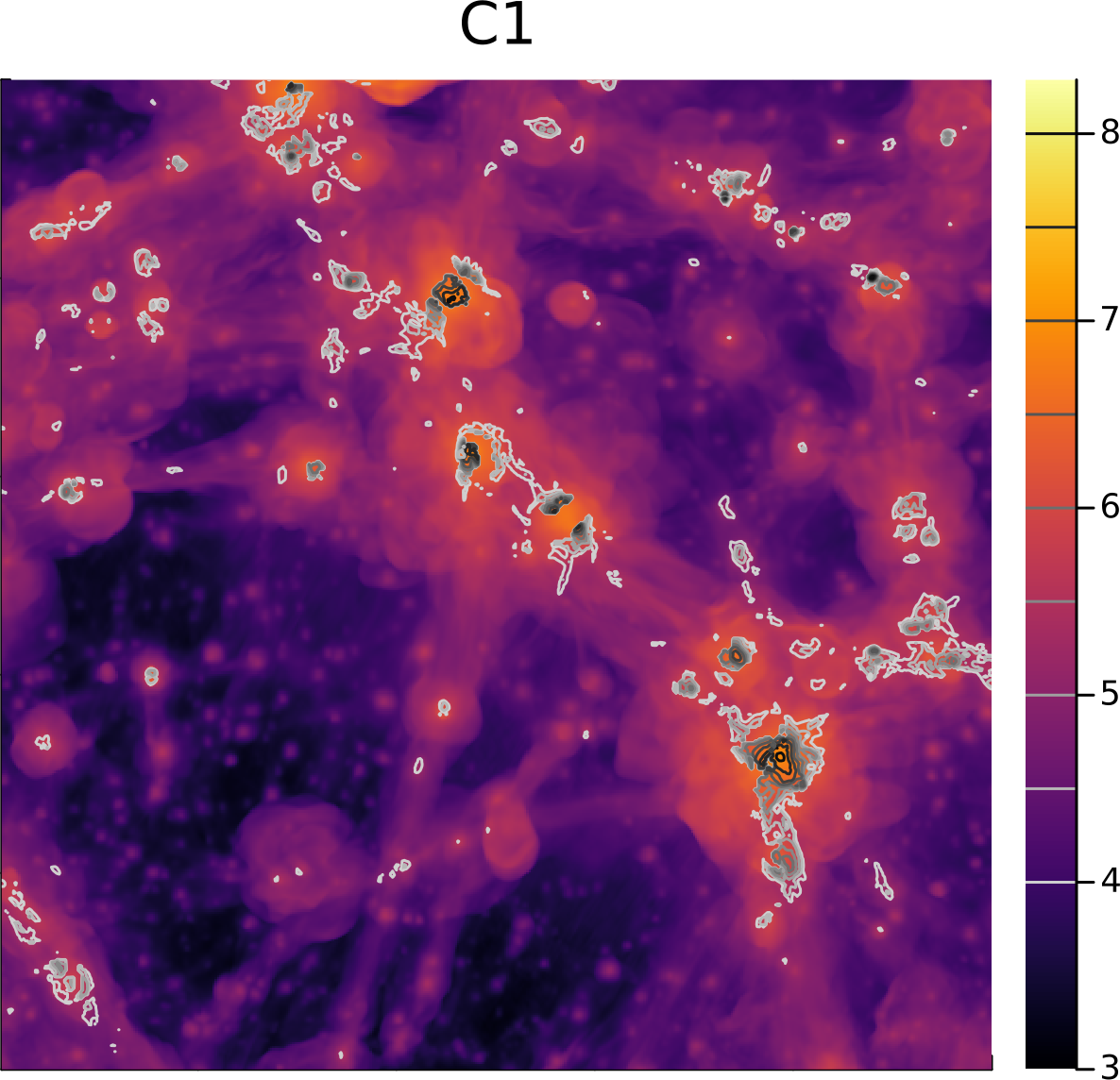}
\includegraphics[width=0.33\textwidth]{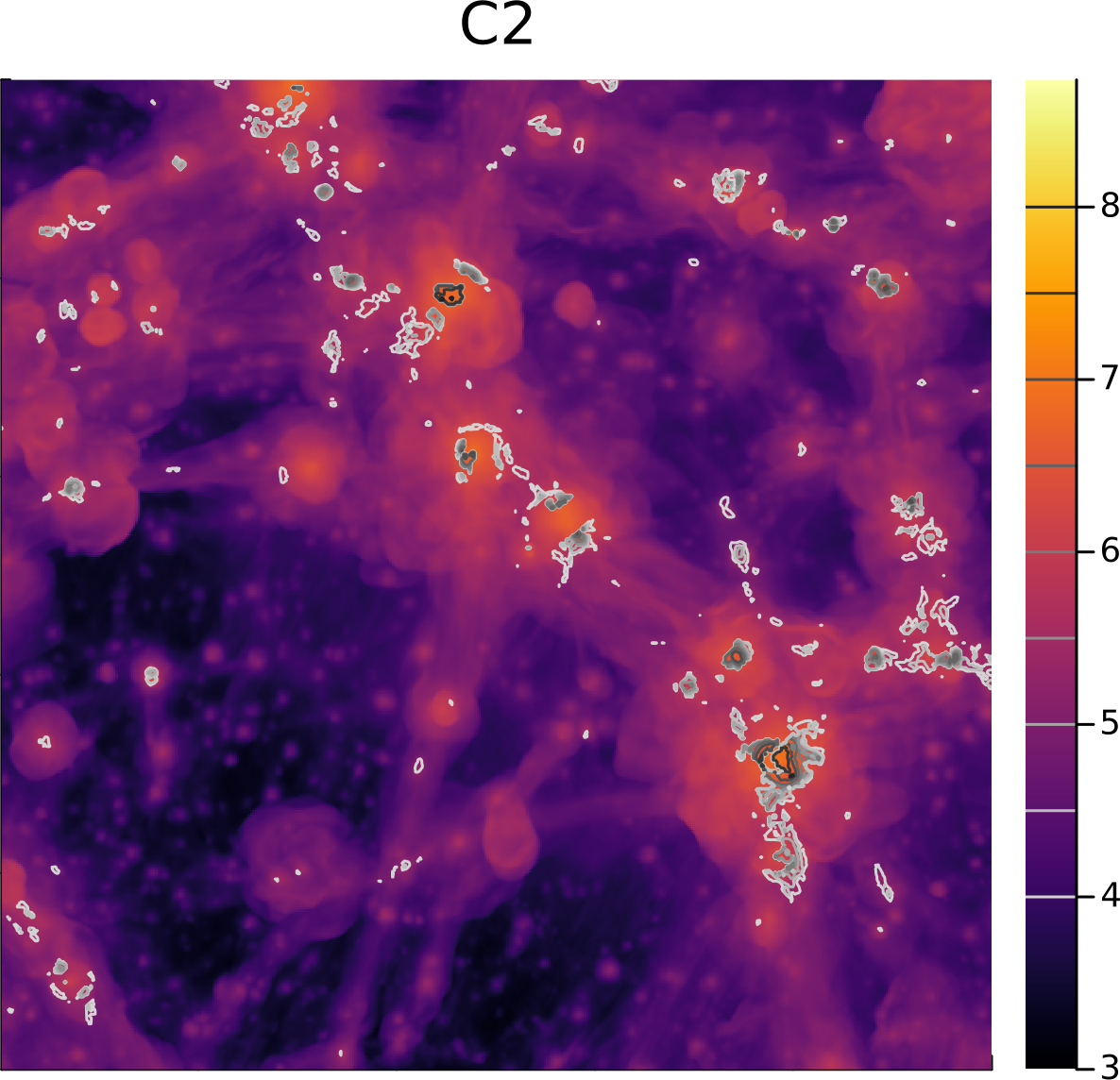}
\caption{Projected mass weighted gas temperature (colors, in units of $\rm log_{\rm 10}[K]$) and over imposed contours of detectable synchrotron radio emission at 150 MHz from CRe accelerated by shocks
in a subselection of our volume (about $20 \rm ~Mpc$ across) in our nine runs at  $z=0.02$.}
\end{center}
\label{fig:mapRadio1}
\end{figure*}

\begin{figure}
\includegraphics[width=0.45\textwidth]{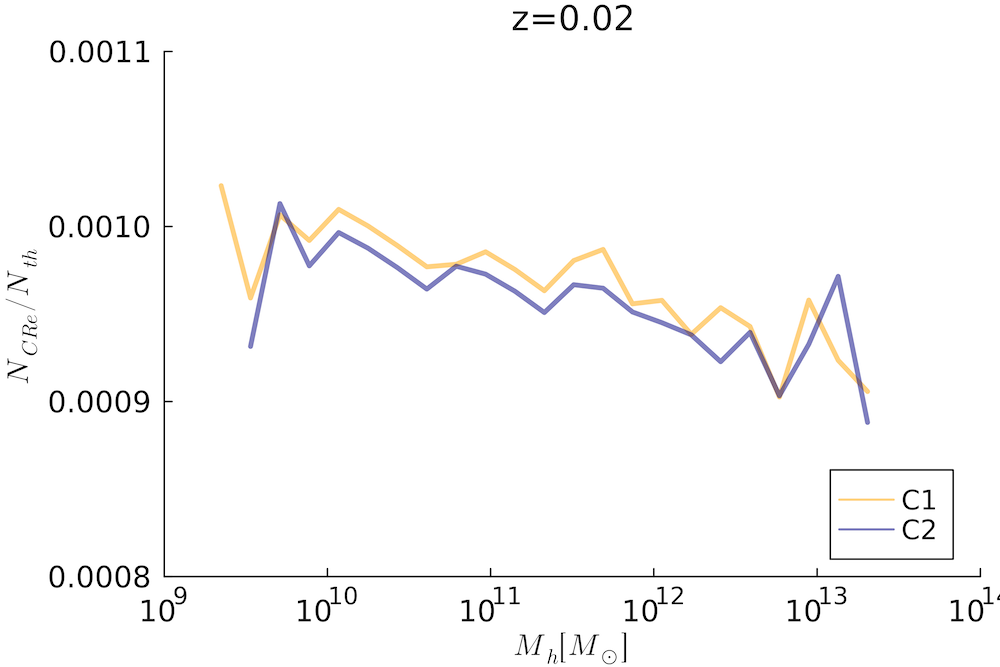}
\caption{Distribution of the ratio between shock-injected CRe density and baryon density (bottom row) within $R_{500}$ of our halos for the  C1 and C2 runs at the end of the simulation.}
\label{fig:obliq}
\end{figure}

\subsection{Radio properties of the cosmic web}

We studied some potentially detectable signatures of the diffuse magnetic fields and CRe injected by the different mechanisms in the radio band. 
First, the Faraday Rotation measurement (RM) of polarised extragalactic sources, integrated over long ($\geq  \rm ~Gpc$) cosmological lines of sight, and removed by the contamination from the Galactic foreground, is a powerful probe of cosmic magnetism, thanks to the sensitivity of modern radio surveys \citep[e.g.][]{vern19,os20,2021MNRAS.505.5038A,2022MNRAS.512..945C,2022MNRAS.515..256P,2023MNRAS.518.2273C,2024arXiv240616230M}.
We thus produced mock RM statistics of cosmological lines of sight (LOS) through the simulated volume of our runs, to constrain the observable effects of our implemented feedback physics on the magnetisation of large scale structures, and compare it to the contribution from purely primordial magnetic fields. Following a procedure already introduced elsewhere in the comparison with real LOFAR data \citep[][]{2023MNRAS.518.2273C,Carretti24}, we built  100 LOS through the simulation, by replicating the simulated volume 153 times, from a set of 17 snapshots saved at nearly equally spaced redshifts, in order to sample the distribution of thermal electron density and of the magnetic field component parallel to the LOS, from $z=2.98$ to $z=0$ . Each full LOS is $\approx 6.502$  comoving $\rm ~Gpc$ (containing $156,672$ cells in total) long and it is assembled by randomly varying the volume-to-volume crossing position to minimise the production of periodic artefacts, moving along the same cartesian axis of the simulation. 
Additionally, all cells in the simulation with a gas matter over-density larger than $50$ times the cosmic average gas density were flagged and excluded from the integration to minimise the RM contamination by cluster halos. This is meant to mimic the more complex procedure applied to real 
LOFAR observations, where the information of spectroscopic sample of more than 150,000 halos from \citep[][]{2015ApJ...807..178W} is used to excise LOS crossing too close to the possible contamination by galaxies \citep[see][and discussion therein]{2023MNRAS.518.2273C}.

For all LOS we integrated 
\begin{eqnarray}
    RRM_f [\rm rad/m^2] &=& 0.812 \int_z^0 \frac{n_e\,B_\parallel}{(1+z)^2}\, dl,
\end{eqnarray}
 where $n_e$ [cm$^{-3}$] is the density of thermal electrons, $dl$ [pc] is the differential path length and $B_\parallel[\rm \mu G]$ is the magnetic field component along the integration path. 

The panels of Figure \ref{fig:mapRM} show the example of the full RRM$_f$ integrated within the snapshot of a $42.5^3 \rm Mpc^3$ box at $z=0.02$ for all our runs. 
With the exception of the C1 and C2 runs, which included a primordial stochastic magnetic field (with spectrum $P_B \propto k^{-1}$ and normalisation $B_{\rm 1Mpc}=0.37 \rm ~nG$) in the initial conditions,  all models produce a similar amplitude of Faraday Rotation, in all cases strongly enhanced in the proximity of dense halos. 
More subtle threads of enhanced Faraday Rotation can be seen also in filaments in runs C1 and C2, highlighting the larger volume filling factor of magnetic fields in primordial models and the fact that only primordial magnetic field models are capable of significantly magnetise cosmic filaments. 
Such differences are best highlighted when integrated for the full long LOS, and after excising the contamination by halos, as in Fig.~\ref{fig:RMz}, in which we computed the full  RRM$_f$ integrated up to $z=3$ limited to the interesting comparison of { models A3, B4 and C1}. 
{ This is motivated by the fact that, in all previous tests, model A3 stood out as the one producing the largest volume filling factor of astrophysical magnetic fields in voids. Hence it is the one in our models which can be most competitive with primordial magnetic fields. However, for reasons discussed above, the amount of star formation and feedback developed in A3 are unrealistically high, while they are more realistic in B4.
The latter, featuring a more prolonged activity by star formation and AGN at low redshifts, produces higher values of RRM$_f(z)$ for $z \leq 0.5$, while the more bursty A3 model dominates the rest of  RRM$_f(z)$ for $0.5 \leq z \leq 2.5$.  In Fig.\ref{fig:RMz} we compare our simulations  to the recent measurements of residual Rotation Measure obtained using LOFAR by \citet{Carretti24}. The latter include the subtraction, at all redshifts, of a local intervening astrophysical contribution at the location of  polarised sources (which cannot be modelled in our simulations), with an assumed, $\propto 1.1/(1+z)\rm [rad/m^2]$ redshift trend as motivated in \citet{Carretti24}. 
In all cases, both purely astrophysical scenarios for the seeding of extragalactic magnetic fields produce values of  RRM$_f(z)$ much lower, at all redshifts, than LOFAR observations. Model C1, which also features the stochastic primordial magnetic field, yields instead values in the $\sim 1-3\, \rm rad/m^2$ range for $z \geq 0.5$, and thus it provide the best reproduction of observed LOFAR among our models (also including the additional primordial models tested in \citealt{Carretti24}).}

Next, we compute the  synchrotron radio emission from CRe accelerated by cosmic shocks, which gives another promising way to detect the cosmic web, especially relevant for the next generation of radio observations, culminating with the Square Kilometre Array  \citep[e.g.][]{brown11,va15radio,2020PASA...37....2W,2021PASA...38...47H,2022A&A...662A..87O}.

The panels in Figure \ref{fig:mapRadio1} show the projected synchrotron radio emission from the CRe accelerated by shocks in all our runs at $z=0.02$, with additional red contours showing the detectable radio emission, assuming as before $\geq 3 \sigma_{\rm noise}$ detection where $\sigma_{\rm noise}=1.05 \mu Jy/\rm arcsec^2$ is of the order of the sensitivity of LOFAR for the LOTSS survey for a $\theta=25"$ beam. 
Once more, the much more volume filling distribution of magnetic fields predicted for the primordial models in run C1 and C2 greatly increases the average synchrotron emissivity of filaments, which produce a nearly fully connected radio cosmic web across scales of tens of Megaparsecs.
However, the bulk of the emission from CRe accelerated by shocks around filaments remains a factor $\sim 10$ or more below the LOFAR HBA detection threshold, with the occasional exception of the peripheral regions of some halos, and of a few very few short-range intra-group bridges, which are barely detectable with LOFAR HBA here. 
This is consistent with present constraints from recent radio observations: the stacking detection of (likely) cosmic filaments by \citet{vern21,vern23} have been obtained using up to $\sim 6 \cdot 10^5$ pairs of physically connected halos of luminous red galaxies. Based on this, the imaging detection of filaments with LOFAR HBA shoud be possible only with a $\sim 25$ better sensitivity, i.e. 
$\sigma'_{\rm noise}=0.014 \mu Jy/\rm arcsec^2$ at 150MHz for a $\theta=25"$ beam. This estimate is also in line with the latest limits from the non-detection of filaments in LOFAR HBA using a smaller sample of massive filaments, by \citet{2023MNRAS.523.6320H}.

\subsection{Role of shock obliquity}
\label{subsec:obliq}

Finally, we also tested the impact of an obliquity-dependent acceleration of CRe in shocks, which can be assessed by comparing our runs C1 (no obliquity dependence) and C2 (injection of CRe by shocks only for $\theta_B \geq 45^\circ$). 

Both the projected maps of radio emission (not shown) and the properties of CRe distribution within halos (Fig.~\ref{fig:obliq}) show that in general the differences of the two models are extremely small, with the C2 model having only a tiny ($\leq 10\%$) reduction in the amount of injected CRs within halos of all masses, compared to the C1 model.  This is consistent with previous results from the few numerical simulations which investigated the effect of shock obliquity in cosmological simulations. They reported that most of the energy dissipation in the cosmic web happens through the oblique shocks \citep[][]{Banfi20,boss},  due to the fact that, within halos, merger shocks sweep a quasi-random distribution of magnetic field orientations \citep[][]{wi17}, while the magnetic field around accretion shocks in filaments and outside clusters is predominantly aligned with the shock surface, as an effect of the local gas dynamics  \citep[][]{2021MNRAS.503.4016B}. { Finally, very recent kinetic plasma simulations by \citet{Gupta24} have reported large levels of acceleration of relativistic electrons even by quasi-parallel shocks, due to the onset of resonant and non-resonant streaming instabilities induced by the acceleration of cosmic ray protons. Even though they limited to one dimension, the implications of these models is that the acceleration of electrons by cosmic shocks might be larger than expected, solely based on diffusive shock acceleration. In this case, our approach is flexible to allow us to test different CRe injection efficiencies than the one tested here (Sec.~\ref{subsec:dsa}).}

\section{Discussion}
\label{sec:discussion}

We comment here on a few important physical processes which are not included in our models. 

First, our simulations neglect { the effects of cosmic-ray diffusion or cosmic ray streaming in our simulation}.  Predicting the diffusion of CRs in a tangled magnetic field is a non trivial problem in astrophysics \citep[e.g.][for reviews]{2000PhR...327..109B,10.3389/fspas.2023.1154760,PhysRevD.89.123001}. 
However,  even the most energetic CR electrons contributing to the energy density of relativistic particles in the cosmic web ($\sim 10 \rm ~GeV$) can only diffuse over a spatial scale that is several orders of magnitude smaller than the length scales typically covered by advection during the same time.  A diffusion coefficient of cosmic rays in the range of  $D \leq 10^{30}-10^{31} \rm cm^2/s$ has been derived for the intracluster medium \citep[e.g.][and discussion therein]{bj14}. this implies that the diffusion timescale over a $L_D \approx 1 \rm ~Mpc$ scale is much longer than the Hubble time, and longer than any other relevant energy loss process for these cosmic rays: $\tau_{\rm diff} \sim L_D^2/(4D) = 7.5-75 \rm ~Gyr$.   Moreover, considering that CRs sourced by galactic evolution processes are injected in  highly magnetised regions (AGN or star forming regions), their diffusion length scale is very likely to be even smaller than this (because $L_D  \propto 1/|B|$) indicating that the effect of CR diffusion can be reasonably neglected compared to the one of advection, for the sake of our analysis. { However, on the scale of galaxies the effects CR diffusion and streaming can be relevant to correctly reproduce their radio emission. Our simulations are not designed to match observations on those spatial scales ($\leq 40 \rm ~kpc$). There the modelling of CR transport, diffusion and streaming in the multi-phase interstellar medium, as well as the interplay between primary and secondary cosmic rays e.g. \citep[e.g.][]{2018PhRvL.121b1102E,2021ApJ...922...11A,hopkins22,2022MNRAS.510.3917G} adds complexity to our simplistic treatment and will deserve more investigation in the future.}

Second,  the recipes of CRe injection adopted in this work are rather simplistic, and can made more sophisticated in several ways and based on the (ongoing) exploration of shocks in collisionless plasmas which particle-in-cell simulations make possible. 
While we already explored the role of shock obliquity in the budget of accelerated CRe (Sec.~\ref{subsec:obliq}), different efficiencies and injection spectra from strong shocks have been proposed in the literature \citep[e.g.][]{2012JCAP...07..038C,2020ApJ...897L..41X}. 
Moreover, more complex and realistic theoretical recipes to model the injection of electrons from low-Mach number and oblique shocks, in which shock-drift acceleration can operate to pre-accelerate electron into DSA,can be considered \citep[e.g.][]{2022ApJ...927..132A}. While this is not expected to change our estimates of radio emission from filaments and cluster outskirts, where the direct injection of shocks should be dominant and the magnetic obliquity should be close to perpendicular \citep[e.g.][]{Banfi20}, it might increase the level of emission from weak internal shocks driven my mergers in the intracluster medium, where the expected DSA efficiency is low \citep[][]{guo14a}.

Our simulations also neglected (see discussion in Sec.~\ref{radiogalaxies}) the continuous injection of secondary CRe by hadronic collisions, which are surely important within galaxies \citep[e.g.][]{2021MNRAS.508.4072W}. Although this mechanism is crudely taken care of in our estimate of radio emission (Sec.~\ref{radiogalaxies}), and it is deemed to be overall subdominant for the seeding of CRe on large scales, compared to AGN and shocks, future work may include also this additional seeding mechanism for CRe at run-time.

{ Finally, another caveat is the role of spatial resolution in our predictions for CRe and magnetic field-related statistics. While with previous work we tested already that the statistics of cosmic rays  by accretion-driven shocks is  well converged already for $\leq 100 \rm kpc$ resolution \citep[e.g.][]{va11comparison,scienzo}, it remains to be assessed with further studies whether the volume filling factor of magnetic fields and CRe injected by galaxies are equally well converged, or not. }

\section{Conclusions}
\label{sec:conclusions}

We have introduced new {\enzo} cosmological simulations  to study the  injection and the evolution of relativistic electrons in the cosmic web.
We focused on their number density and on their average age and we simulated their injection by three main processes: structure formation shocks, AGN feedback and star formation. { Our procedure tracks the evolution of relativistic electrons with typical minimum energies of $\sim 1-10\, \rm MeV$ and accelerated up to $\sim 1-10 \rm\, GeV$ or beyond. The spectral energy distribution  can be reconstructed in post-processing, thanks to the combination of our spectral ageing model (Sec.~\ref{subsec:sync}) and of the additional tracking of the average time elapsed since the injection of cosmic rays in any particular cell (Sec~\ref{subsec:tau}.}

Given the spatial and mass resolution limits of our simulations, as well as of the approximations used to mimic the growth and feedback by SMBH, we have calibrated sub-grid physics to reproduce known observable relations, rather than developing a full physical model to solve galaxy-scale physics, or the formation and growth of SMBH. 
We showed that it is possible to calibrate this procedure to well reproduce several key observations (like the stellar mass distribution in galaxies, the correlation between SMBH and host galaxy masses,  the radio power distribution function of galaxies), and to use this information to robustly predict the amount of CRe injected in large-scale structures by cosmic evolution. 
In particular, we report the first and very good reproduction of the luminosity distribution of radio emission from galaxies in the simulation and up to $z \sim 2$, using a distribution of magnetic fields and CRe evolved at run time ({ focusing on the number density of CRe and tracking their average age with our simple exponential-decay age tracing model}) and using ideal MHD simulations.

Our main results concerning the evolving distribution of these two important non-thermal components can be so summarised:

\begin{itemize}
    \item {\it structure formation shocks dominate the injection of CRe in the majority of the Universe}. Even for conservative assumptions on the CRe acceleration efficiency from Diffusive Shock Acceleration (calibrated to reproduce known radio emission on the scale of clusters of galaxies), shocks dominate the volume-filling population of CRe in the cosmic web, as well as the total amount of injected CRe in all investigated models. This finding is not altered by including the obliquity-dependent injection efficiency by shocks. 
    \item { {\it AGN feedback dominates the injection of CRe with respect to stellar feedback}, both within halos and also when volume averaged statistics are considered. While stellar feedback starts to enrich the Universe earlier with CRe and magnetic fields, the more powerful activity by AGN until late cosmic times end up dominating the overall production of CRe, both in high- and low-mass halos.} 
    \item {\it known astrophysical sources can only magnetise a tiny fraction of the Universe.} By computing the filling factor of magnetic fields in the different runs, we estimate $f \leq 15 \%$ of the Universe at all redshift can be realistically magnetised above $10^{-15} \rm G$ by galaxies (both including AGN feedback and star formation-driven outflows), so any primordial magnetic field larger than this would dominate observations. Even our most extreme feedback model (A3) the produced filling factor ($\sim 37\%$) is not enough to account for the non detection of Inverse Compton Cascade emission from blazars, or explain the trends of Faraday Rotation Measurements obtained with LOFAR. 
    
\end{itemize}

Our new numerical framework (which is scalable to larger, or more resolved simulations) represents a first important step towards the self-consistent coupling of galaxy formation processes to their observable, or invisible, non-thermal output.
{ While this approach can still be subject to improvements on the side of the implemented galaxy evolution physics,} this is key to investigate the important role played by several sources in seeding of  large-scale emissions with a "fossil" population of mildly relativistic electrons, which Fermi I and Fermi II processes can further re-energise \citep[e.g.][for a review]{bj14}. 

Based on the results of this work, we can now quantify the total budget of CRe injected within halos by all mechanisms considered in our simulations by $z=0$. The budget of CRe injected by each mechanism scales with the assumed acceleration efficiencies: $\xi_{\rm e, \mathcal{M}\geq 5}$ is the CRe injection efficiency for strong shocks, $\xi_{\rm AGN}$ is the CRe injection efficiency assumed for AGN feedback and $\xi_{\rm SF}$ is our assumed CRe injection efficiency from star formation (averaged over our $41.5^3\rm ~kpc^3$ resolution).  
There are of course several other dependencies on the assumed physical implementations of AGN physics and SMBH growth, star formation and feedback, which cannot be easily prescribed in this way due to the non-linearity of their couplings. 

In order to 
{ give an order of magnitude estimate of the various contribution to CRe from all considered mechanisms within matter halos, we use our best model B4, and based on Fig.~\ref{fig:halosCRe} we can parametrise the most important model dependencies as follows :

\begin{equation}
\frac{N_{\rm CRe}}{N_{\rm th}}= 5.51 \cdot 10^{-4}\frac{\xi_{\rm e, \mathcal{M}\geq 5}}{4.6\cdot 10^{-4}}+4.9 \cdot 10^{-5}\frac{\xi_{AGN}}{2 \cdot 10^{-4}}+6 \cdot 10^{-6}\frac{\xi_{SF}}{10^{-5}} , 
\end{equation}

for our largest, $\sim 5 \cdot 10^{13} M_{\odot}$ halos at $z=0$, and

\begin{equation}
\frac{N_{\rm CRe}}{N_{\rm th}}= 6.26 \cdot 10^{-4}\frac{\xi_{\rm e, \mathcal{M}\geq 5}}{4.6\cdot 10^{-4}}+2.1 \cdot 10^{-5}\frac{\xi_{AGN}}{2 \cdot 10^{-4}}+2.1 \cdot 10^{-6}\frac{\xi_{SF}}{10^{-5}} , 
\end{equation}

for the smallest $\sim 5 \cdot 10^{9} M_{\odot}$ halos at $z=0$.}
While the details of their star formation rate density, or galaxy luminosity functions, may differ significantly at specific redshifts, their final integrated budget of CRe only differs by $\sim 60\%$. The total amount of injected CRe in large-scale structures appears therefore robust prediction of the model, despite the several assumptions and open theoretical problems related to galaxy physics and to the evolution and formation of SMBH.  

It is now important to ask whether these amount of CRe are sufficient to account for the budget of fossil radio emitting electrons required by the modelling of diffuse radio emission in clusters of galaxies. 
In a recent review \citep[][]{vb24} we
computed the number of CRe that must be contained within the volume of observed radio halos at the centre of clusters of galaxies, as implied by simplistic assumptions of energy equipartition between cosmic rays and magnetic field, and for a variety of observed radio spectra and (unknown) ratio between the energy density of CR protons and electrons.  For a  $M=10^{15} M_{\odot}$ cluster of galaxies, from $N_{\rm CRe}\sim 5 \cdot 10^{66}$ to 
$N_{\rm CRe}\sim 10^{68}$  electrons within a $\rm Mpc^3$ volume  are required to explain the observed radio emission of a medium power radio halo, depending on the spectra energy distribution of the electron population, on their minimum energy as well as on the amount of CR protons (which enters the derivation of the equipartition magnetic field). 

Our new simulations show that the amount of CRe in massive  halos at low redshift is generally in line with the above estimates, and thus that the combination of the CRe seeding mechanisms explored in this work is more than enough to fuel Fermi processes in their production of diffuse radio emissions in clusters. Moreover, the same approach will also allow us to better understand the acceleration physics in more peripheral and less dense regions of the cosmic web, which started being probed by recent radio observations \citep{2019Sci...364..981G,2020MNRAS.499L..11B,vern21,2022Natur.609..911C,2022SciA....8.7623B,vern23}, considering that it is presently unclear how much fossil CRe, or magnetic fields,  are present at such large scales. 
This appears timely and crucial, considering that a systematic survey of these regions will be one of the future main goals of the Square Kilometre Array \citep[e.g.][]{2020PASA...37....2W}.

\section*{Acknowledgements}
{ We thank our anonymous reviewer for the valuable help in improving the quality of this work.} F.V. acknowledges the CINECA award  "IsB28\_RADGALEO" and "IsCc4\_FINRADG"  under the ISCRA initiative, for the availability of high-performance computing resources and support. 
F.V. also gratefully acknowledges the usage of computing  at the  CSCS-ETH supercomputing centre (Lugano, Switzerland) on Piz-Daint under project ID s1096. F.V. and S. M.'s research has been supported by Fondazione Cariplo and Fondazione CDP, through grant n° Rif: 2022-2088 CUP J33C22004310003 for "BREAKTHRU" project.
F.V. also acknowledges financial support from the European Union's Horizon 2020 program under the ERC Starting Grant "MAGCOW", no. 714196. MB acknowledges funding by the Deutsche Forschungsgemeinschaft (DFG) under Germany's Excellence Strategy -- EXC 2121 ``Quantum Universe" --  390833306 and the DFG Research Group "Relativistic Jets".
 M. T. was supported by the National Research Foundation of Ukraine under Project No. 2020.02/0073. M. T. would like to thank the Armed Forces of Ukraine for providing security to perform this work.
 We acknowledge fruitful scientific discussions with Y. Dubois, G. Brunetti, V. Heesen, A.Botteon, M. S. S. L. Oei and H. Rottgering. 

 \bibliographystyle{aa}

 \bibliography{franco3}

\appendix

\section{Test resimulations with physical variations}

\begin{table*}[bht]
    \centering
    \begin{tabular}{cccccccc}
                  ID & $t_{*}[\rm Myr]$ & $\rho_{\rm AGN}$ & $\alpha_{B,cold}$& $\alpha_{B,hot}$&$f_{\rm k,cold}$ &$f_{\rm k,hot}$ & comment\\ \hline
                   { testA}& 30&20 &1 & 10 & 0.81& 0.1 & reference\\ \hline
           test\_time001& 60&20 & 1 & 10 & 0.81& 0.1& $\tau=10$ Myr\\
           test\_time1& 60&20 & 1 & 10 & 0.81& 0.1& $\tau=1$ Gyr\\ \hline  test\_AGN\_sep& 60&20 & 1 & 10 & 0.81& 0.1& 4 cells AGN separation\\
        testB& 30&20  &1 & 5 & 0.81& 0.1 & AGN variation\\
        testC& 30&20 & 1 & 1 & 0.81& 0.1 & AGN variation\\
        testD& 30&40 & 1 & 10 & 0.81& 0.1& AGN variation\\
        testE& 30&60 & 1 & 10& 0.81& 0.1& AGN variation\\ \hline
        testF& 1&20 & 1 & 10 & 0.81& 0.1& SF variation\\
        testG& 100&20 & 1 & 10 & 0.81& 0.1& SF variation\\
        testH& 20&20 & 1 & 10 & 0.81& 0.1& SF variation\\
        testI& 60&20 & 1 & 10 & 0.81& 0.1& SF variation\\ \hline 
         test\_noAGN& 60&- & - & - & -& -& no AGN\\
          test\_noSF& -&20 & 1 & 1 & 0.81& 0.1& no SF \\
           test\_noSFnoAGN& -&- & - & - & -& -& no SF, no AGN\\ \hline 
         
    \end{tabular}
    \caption{Important parameters varied in our additional tests, run on a $21.25^3 \rm Mpc^3$ volume simulated with $512^3$ cells and dark matter particles, as explained in the Appendix. All other parameters are as in the B3 model of Tab.\ref{tab:mhd} (run testA, in particular, is identical to  model B3  in the main paper).}
    \label{tab:mhd1}
\end{table*}

Here we present several important numerical tests in which we resimulated the same cosmic volume (smaller than the one used in the main paper) to compare the outcome of important physical or numerical variations of our algorithms. This is meant to assess the robustness of the conclusions of the main paper, and their dependence on the many sub-grid physical prescriptions for galaxy formation.

In detail, we used a $21.25^3 \rm Mpc^3$ volume using $512^3$ cells and $512^3$ dark matter particles, hence these test simulations cover the same spatial and mass resolution of the runs employed in the main paper, but have a $8$ times less volume.  All explored variations of parameters are given in Tab.\ref{tab:mhd1}, and in the next Subsections we comment on each of these variations in more detail.

\subsection{Tests on the relativistic electrons modelling}
\label{sec:A1}
\begin{figure}
\includegraphics[width=0.49\textwidth]{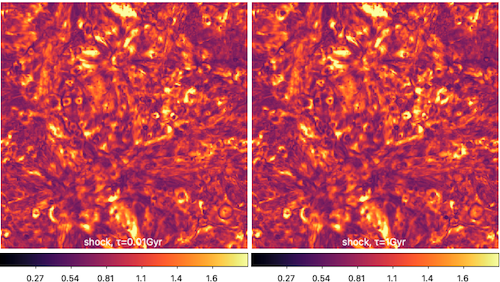}
\includegraphics[width=0.49\textwidth]{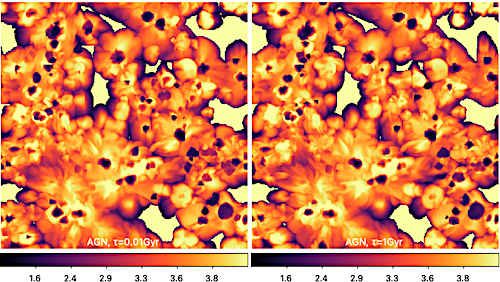}
\includegraphics[width=0.49\textwidth]{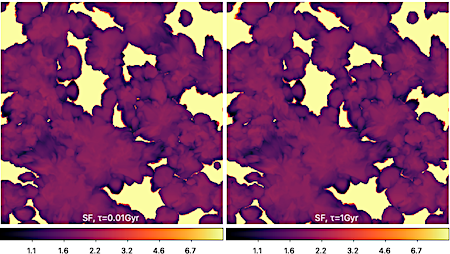}
\caption{Projected mean (CRe-weighted) age of CRe since their last injection (in units of [Gyr]) by shocks (top row), AGN (middle row) or star formation (bottom row) for two control resimulations at $z=0.02$, which employed $\tau=0.01 \rm ~Gyr$ (model test\_001, left panels) or $\tau=1~\rm Gyr$ (model test\_time1, right panels) to solve Eq.\ref{eq:age}. Each panel is $21.25 \rm ~Mpc$ across. No significant differences in the reconstructed age of the different CRe population are found.}
\label{fig:appendix1}
\end{figure}

First, we tested the robustness of our new proposed method to track the advection and ageing of relativistic electrons in {\enzo} cosmological simulations (Sec.~\ref{sec:methods}).
While the specific choice of the timescale $\tau$ used for the age reconstruction in Eq.\ref{eq:age} is arbitrary and, a single injection of CRe would always allow for perfect recovery with this technique, multiple injections complicate the process as older or younger population of CRe may in principle differently weight the resulting age, depending on the choice of $\tau$. 
We thus tested the use of $\tau=10 \rm ~Myr$ (model test\_time001 in Tab.\ref{tab:mhd1}), $\tau=100 \rm ~Myr$ (as in the main paper, model testA) and $\tau=1 \rm ~Gyr$ (model test\_time1), in all cases using the same physical model of run B3 of the main paper. Figure \ref{fig:appendix1} gives the projected CRe-weighted maps for the time elapsed since the last injection of CRe, separately for the three different families of CRe and for the two  extreme 10 ~Myr and 1~Gyr cases. Beside small random differences (due to our implementation of random axis orientations for feedback events), the distribution of ages look extremely similar in both cases, for all considered CR species. We thus conclude that, for the distribution of ages concerned with the physical evolution of our simulations, the choice of this parameter is not critical for a correct a-posteriori reconstruction of the time elapsed since the injection of CRs.

\subsection{Tests on variations of the AGN modelling}
\label{sec:A2}

Our run-time procedure for AGN feedback, without relying on actual SMBH Lagrangian sink particles, is based on a few assumptions to generate realistic injection site of feedback events. As detailed in Sec.\ref{subsec:bh}, we identify at run time local high baryonic density peaks ($100$ times larger than the baryonic mean density), and there we  measure the total mass contained within a fixed comoving radius (2 cells, i.e. $83 \rm ~kpc$). We also adopt a local exclusion criterion to avoid generating multiple SMBH within a radius of 3 cells ($124.5 ~\rm kpc$) from the central gas density peak. 

 With a test run (test\_AGN\_sep in Tab.\ref{tab:mhd1}),  we resimulated the same  $21.5^3 \rm Mpc^3$ volume using the same local exclusion criterion of $3$ cells as in the baseline simulation of the main paper, or using $4$ cells instead (so $166  ~\rm kpc$ from the central gas density peak). The top panel of Figure \ref{fig:appendix2} shows the integrated star formation history in this volume, which does not show significant differences in the two cases. Other quantities related to this (i.e. injection of CRe by star formation or AGN, halo scaling relations, etc) equally show negligible differences, suggesting the adopted minimum separation between SMBH host cells in our run-time procedure is not particularly critical here.

 The same Figure also shows the effect of other variations related to our SMBH implementation and AGN feedback: we varied the values of the $\alpha_B$ "boost" parameters used in the Bondi formalism to compute the matter accretion from the hot and the cold gas phase (Sec.~\ref{subsec:bh}) in models testB and testC, or we increased the minimum density ($\rho_{\rm AGN}$) to form SMBH at run time (testD and testE). The effects of all variations in the star formation history is minimal, while our fiducial choice of parameters $\rho_{\rm AGN}=20$, $\alpha_{\rm B,cold}=1$ and $\alpha_{\rm B,cold}=10$ (e.g. in the B4 model of the main paper) overall gives the best results when coupled to the prediction of luminosity functions of radio galaxies (see Sec.\ref{radiogalaxies} in the main paper). 
 
\subsection{Tests on variations of the stellar formation modelling}
\label{sec:A2b}

\begin{figure}
\includegraphics[width=0.49\textwidth]{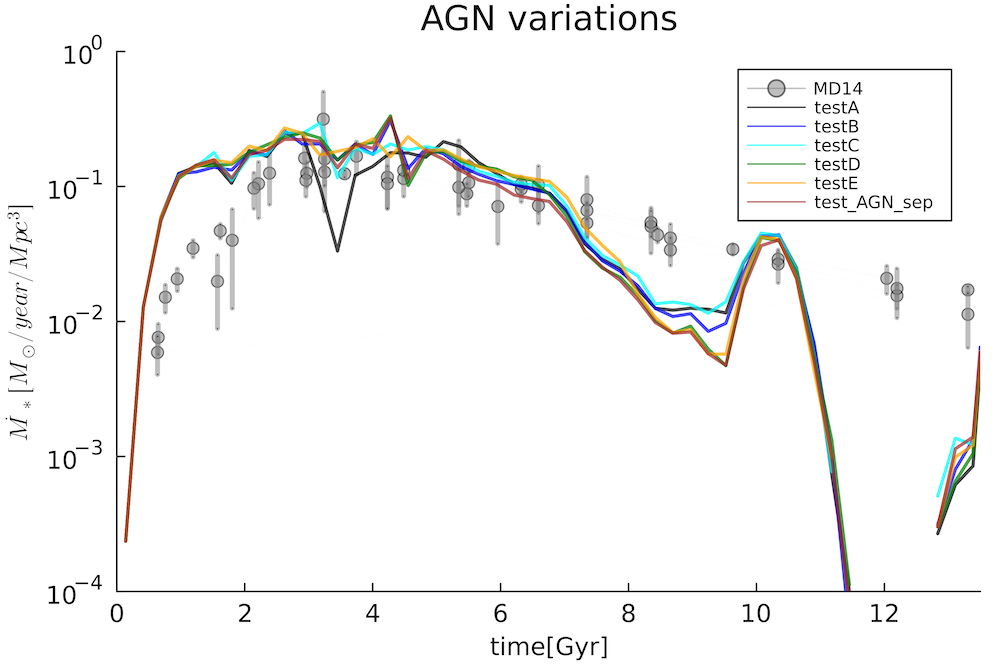}
\includegraphics[width=0.49\textwidth]{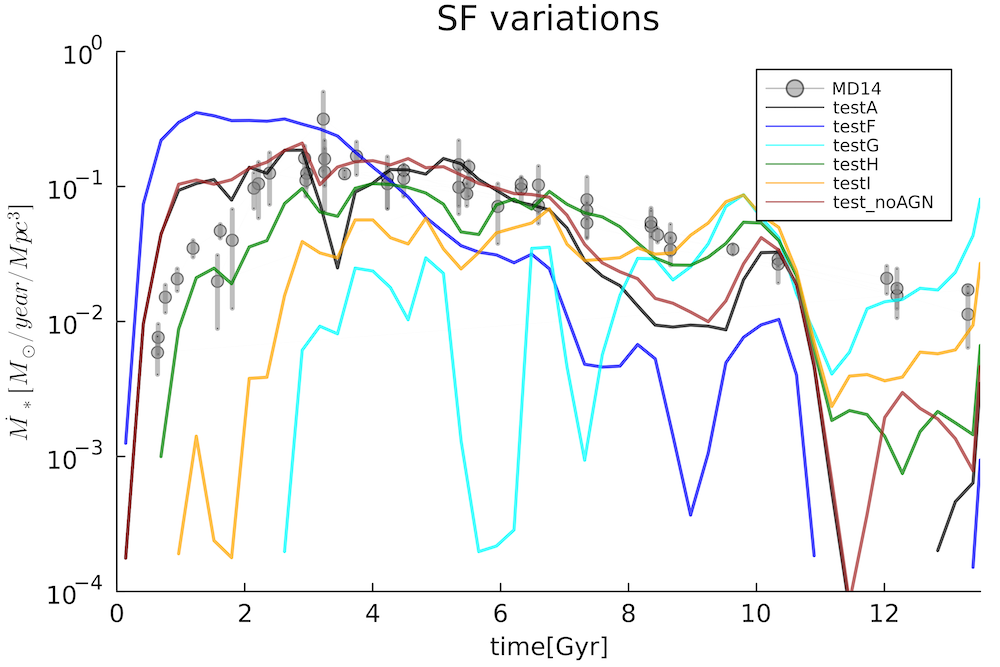}
\caption{History of the simulated cosmic star formation density (i.e. normalised for a $\rm Mpc^3$ comoving volume) for several resimulations of a $21.5^3 \rm ~Mpc^3$ volume, in which we varied the parameters of our star formation feedback implementation (top panel), or of our prescriptions for AGN feedback (bottom panel). See Sec.~\ref{sec:A2} for explanations.}
%The grey points with error bars show the observed cosmic star formation derived in \citet{2014ARA&A..52..415M}. }
\label{fig:appendix2}
\end{figure}

With other tests, we varied the dynamical timescale associated with the formation of stellar particles, which is a fundamental parameter in our model (Sec.~\ref{subsec:sfr}), while keeping instead the AGN feedback prescription fixed. In this case the effects are more evident: a short $t_*=1 ~\rm Myr$ timescale allows the fast formation of a large quantity of stars, reaching a peak already $\sim 1 \rm ~Gyr$ after the start of the simulation, followed by a steady decline. Conversely, a much longer $t_*=100 ~\rm Myr$ timescale delays the formation of the first stars to $\sim 3 \rm ~Gyr$ after the star of the simulation. As shown by the bottom panel of Fig.~\ref{fig:appendix2} both trends of cosmic star formation are quite far from the observed ones, and the best combination of parameters in our resolution regime is given by the testH model ($t_*=20 ~\rm Myr$, as in the B4 model of the main paper).

\subsection{The interplay between AGN, star formation and shocks}
\label{sec:A4}
\begin{figure*}
\includegraphics[width=0.99\textwidth]{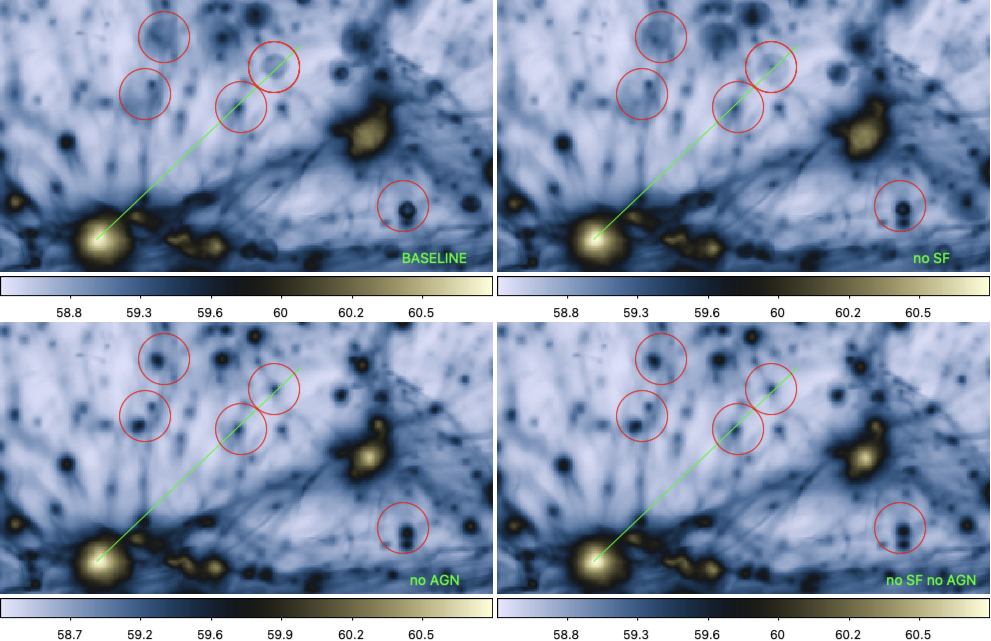}
\caption{Maps of the total number of CRe (injected by shocks only, in units of [$\log_{\rm 10}(N_{\rm CRe})$]) along the line of sight of a $10 \times 10 \rm Mpc^2$ zoomed region in our test runs, in which we included all sources of feedback in the simulation ("baseline" model, as in the main paper), or we switched off AGN feedback ("no AGN"), or stellar feedback ("no SF") or both ("no SF no AGN"). The red circles identify regions in which obvious differences are visible between models, while the green lines show the direction for which we generated radial profiles of the number of CRe in Fig.~\ref{fig:appendix5}.}
\label{fig:appendix4}
\end{figure*}

With other three variations, we resimulated the  same  $21.5^3 \rm Mpc^3$ volume alternatively by switching off AGN, star formation or both, in order to assess the contribution by galaxy evolution processes to CRe and magnetic fields, as well as the interplay between feedback events and the additional generation of shocks. The reference galaxy formation model is again testA here ("baseline" in the Figure). 
Figure \ref{fig:appendix3} shows the projected maps of the number of CRe only injected by shocks in the four relevant models (testA, test\_noAGN, test\_noSF and test\_noSFnoAGN) at $z=0$. The general distribution of CRe is very similar in all cases, with a few notable exceptions if AGN feedback is included or not, marked by the red circles: AGN release powerful shocks around their host halo, which both inject new CRs via diffusive shock acceleration, as well spread already existing CRe from previous shocks on a larger volume. At low redshift, the stellar feedback is in general too weak, in our simulations, to produce similar effects.
Figure \ref{fig:appendix4} give the line profile for the projected CRe distribution in four cases (for the green lines in the previous Figure), showing the subtle modulation of the by AGN, which typically smoothens the distribution of CRe outside from the centre of halos. 

We conclude that shocks driven by structure formation are, by far, the main contributor of CRe from the DSA mechanism. However, the feedback from AGN can, beside directly injecting new CRe through their jets/outflows, also drive powerful shocks in the intergalactic medium, which further add CRe in coccoon-like structures surrounding halos.

\subsection{The contribution of different astrophysical sources to magnetic fields}
\label{sec:A5}

By comparing the same above four models (testA, test\_noAGN, test\_noSF and test\_noSFnoAGN) we can also better visualise the role of each mechanism to the magnetisation of the cosmic web.
Figure \ref{fig:appendixB} shows the mean (mass-weighted) magnetic field strength along the line of sight for these models at three different epochs ($z=3$, $=1$ and $=0$). 
Above $z=3$, the magnetisation in and around halos is dominated by star formation-driven outflows. If both AGN and star formation are off, only the compression and (sub-grid) amplification of magnetic fields from structure formation are active. 
In a second stage ($z \sim 1$)  AGN feedback takes over as it generates large outflows of baryons, which carry magnetic fields with them.  Finally, in the the present day Universe ($z=0$) the cosmic web is magnetised at all scales but, regardless of the magnetisation from AGN-driven outflows, 
most of the cosmic volume remains dominated by its initial primordial magnetic field. 
The typical magnetic field strength at the virial radius of our most massive halos here ($\sim 5 \cdot 10^{13} M_{\odot}$) is $\sim 0.1\mu G$ if AGN feedback is activated (with or without stellar feedback), $\sim 0.01 \mu G$ if only stellar feedback is active, and $\sim 0.001 \mu G$ if no feedback sources are active.

In filaments, on the other hand, the  magnetic field amplitude by the end of our runs are always similar to the test\_noSFnoAGN run, meaning that their value is mostly anchored to the evolution of primordial magnetic fields, and that no contamination from astrophysical fields can cover outshine them.

\begin{figure}
\includegraphics[width=0.49\textwidth]{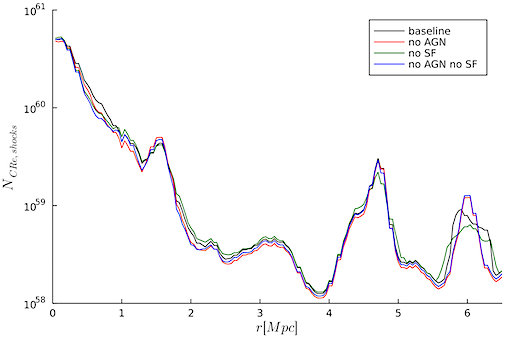}
\caption{Line profile of the total number of CRe (injected by shocks only) for the four variations and the same line in Fig.~\ref{fig:appendix4}.}
\label{fig:appendix5}
\end{figure}

\begin{figure}
\includegraphics[width=0.49\textwidth]{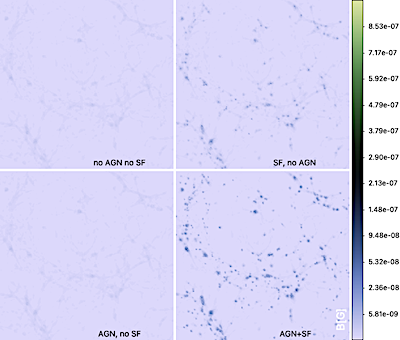}
\includegraphics[width=0.49\textwidth]{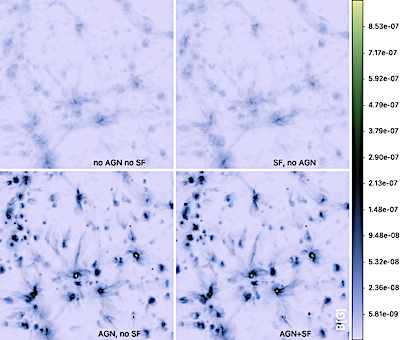}
\includegraphics[width=0.49\textwidth]{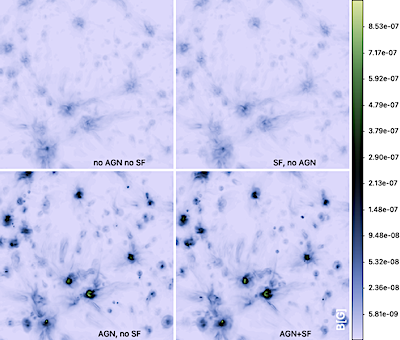}
\caption{Projected mass-weighted magnetic field strength for runs  test\_noSFnoAGN, test\_noAGN, test\_noSF and testA, at three different epochs:$z=3$ (top), $z=1$ (middle) and $z=0$ (bottom).}
\label{fig:appendixB}
\end{figure}

\begin{figure}
\includegraphics[width=0.49\textwidth]{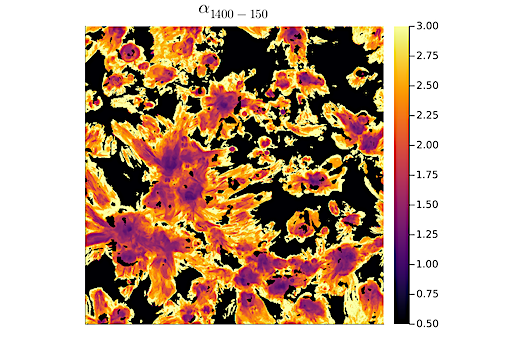}
\includegraphics[width=0.49\textwidth]{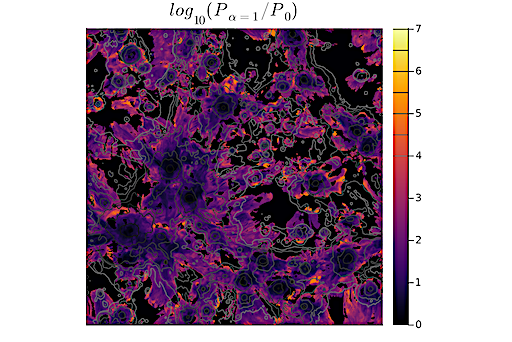}
\caption{Top panel: distribution of radio spectral index between  $1400$ and $150$ MHz for CRe injected by shocks, for our fiducial post-processing treatment of CRe spectra, for a $21.25^3 \rm ~Mpc^3$ volume at $z=0$. 
Bottom panel: for the same volume and epoch, 
map of the ratio between the synchrotron radio power at $150$ MHz computed assuming a fixed $\alpha=-2$ for the momentum spectrum of CRe injected by shocks, and the power from CRe assuming their realistic spectrum as in Sec.~\ref{subsec:sync}. }

\label{fig:appendix3}
\end{figure}

\section{Tests on the spectral modelling of CRe}
\label{sec:A3}

We tested the validity of our procedure to age the spectra of injected CRe (which combines our treatment of the time elapsed since the injection of CRe in cells in Sec.~\ref{subsec:tau}  with the use of template CRe momentum spectra depending on the local thermo-dynamical values in cells, in Sec. \ref{subsec:sync}) against simpler approaches with a fixed spectrum for CRe. 
The top panel of Fig.~\ref{fig:appendix3} gives the radio spectral index of the synchrotron radio emission from shock accelerated CRe, measured between $1400$ and $150$ MHz at for our test volume at $z=0$. 

The bottom panel shows instead the ratio between the radio power at $150$ MHz obtained with a simpler constant spectral index for all cells (i.e. $\alpha=2$ in momentum spectrum, yielding a $I(\nu) \propto \nu^{-1}$ radio emission spectrum in DSA), and the power computed with our procedure to evolve CRe spectra as in the main paper. 

If one adjust a fixed spectrum to $\alpha=2$, the simplified calculation can reasonably reproduces the radio emission from recent merger shocks internal to mass halos and filaments, but it would dramatically exceed, by orders of magnitude, the radio emission in less dense environments, 
 dominated by older populations of CRe. As is obvious from the distribution of spectral indices, no single value $\alpha$ can reproduce the realistic range of spectral indices of CRe across the cosmic web. 
 We remark that assuming a fixed spectrum is not the standard procedure used in many works in the literature \citep[e.g.][]{sk11,va15radio,2017MNRAS.470..240N}. However, even in such work only the prompt emission from shock injected CRe can be estimated. 
 Our simulations show instead that different populations of can dominate the budget of CRe in different regions of the cosmic web, and hence that a  realistic spectral modelling of the emission (like the one proposed here), is necessary.

\end{document}